\documentclass[final,3p]{elsarticle}

\usepackage{amssymb,mathrsfs,amsmath}
\usepackage{listings}
\usepackage{xcolor}
\usepackage{bbold}
\usepackage{color}
\usepackage{algorithm}
\usepackage{algpseudocode}
\usepackage {tabularx}
\usepackage[colorlinks=true, linkcolor=blue, citecolor=red, urlcolor=magenta]{hyperref}
\usepackage[utf8]{inputenc}
\usepackage[T1]{fontenc}

\definecolor{codegreen}{rgb}{0,0.6,0}
\definecolor{codegray}{rgb}{0.5,0.5,0.5}
\definecolor{codepurple}{rgb}{0.58,0,0.82}
\definecolor{backcolour}{rgb}{0.95,0.95,0.92}

\lstdefinestyle{mystyle}{
    backgroundcolor=\color{backcolour},   
    commentstyle=\color{codegreen},
    keywordstyle=\color{magenta},
    numberstyle=\tiny\color{codegray},
    stringstyle=\color{codepurple},
    basicstyle=\ttfamily\footnotesize,
    breakatwhitespace=false,         
    breaklines=true,                 
    captionpos=b,                    
    keepspaces=true,                 
    numbers=left,                    
    numbersep=5pt,                  
    showspaces=false,                
    showstringspaces=false,
    showtabs=false,                  
    tabsize=2
}

\newcounter{bla}

\journal{arXiv}

\newcommand{\NESSi}{{\tt NESSi}{}}
\newcommand{\NESSione}{{\tt NESSi$\,$1.0}{}}
\newcommand{\NESSitwo}{{\tt NESSi$\,$2.0}{}}
\newcommand{\libcntr}{{\tt libcntr}{}}

\begin{document}

\begin{frontmatter}

\title{\textbf{NESSi 2.0}: The \textbf{N}on-\textbf{E}quilibrium \textbf{S}ystems \textbf{Si}mulation package version 2.0}

\author[Hamburg]{Fabian K\"unzel\corref{cor1}} \cortext[cor1]{Corresponding author.} \ead{fabian.kuenzel@uni-hamburg.de}
\author[PSI,Fribourg]{Michael Sch\"uler}
\author[JSI,Ljubljana]{Denis Gole\v{z}}
\author[Tohoku,RIKEN]{Yuta Murakami}
\author[Fribourg]{Sujay Ray}
\author[Erlangen]{Christopher Stahl}
\author[Erlangen]{Jiajun Li}
\author[Orebro]{Hugo U. R. Strand}
\author[Fribourg]{Philipp Werner}
\author[Hamburg,CUI]{Martin Eckstein}

\affiliation[Hamburg]{organization={Institute of Theoretical Physics, University of Hamburg, 20355 Hamburg, Germany}}
\affiliation[Fribourg]{organization={Department of Physics, University of Fribourg, CH-1700 Fribourg, Switzerland}}
\affiliation[CUI]{organization={The Hamburg Centre for Ultrafast Imaging, Hamburg, Germany}}
\affiliation[PSI]{organization={PSI Center for Scientific Computing, Theory and Data,
Paul Scherrer Institute, 5232 Villigen PSI, Switzerland}}
\affiliation[JSI]{organization={Jo\v{z}ef Stefan Institute, SI-1000 Ljubljana, Slovenia}}
\affiliation[Ljubljana]{organization={Faculty of Mathematics and Physics, University of Ljubljana, 1000 Ljubljana, Slovenia}}
\affiliation[Tohoku]{organization={Institute for Materials Research, Tohoku University, Sendai 980-8577, Japan}}
\affiliation[RIKEN]{organization={Center for Emergent Matter Science, RIKEN, Wako, Saitama 351-0198, Japan}}
\affiliation[Orebro]{organization={School of Science and Technology, \"Orebro University, SE-701 82 \"Orebro, Sweden}}
\affiliation[Erlangen]{organization={Department of Physics, University of Erlangen-N\"urnberg, 91058 Erlangen, Germany}}

\begin{abstract}
Nonequilibrium Green's functions provide a powerful framework for studying quantum many-body dynamics including the laser-induced dynamics in solids. The {\bf N}on-{\bf E}quilibrium {\bf S}ystems {\bf Si}mulation package (\NESSi{}) offers an efficient platform for such simulations, ranging from perturbative approaches like nonequilibrium $GW$ to nonequilibrium dynamical mean-field theory. However,  simulations based on nonequilibrium Green's functions become computationally demanding when the dynamics span a large temporal range, such as from sub-femtosecond electron dynamics to the picosecond dynamics of collective modes.  Due to the memory integral in the Kadanoff-Baym equations, which serve as equations of motion for nonequilibrium Green's functions, the computational cost scales as $\mathcal{O}(N_t^3)$ with the number of timesteps $N_t$, and the memory requirement scales as $\mathcal{O}(N_t^2)$. In this work, we extend \NESSi{} by incorporating techniques that aim to overcome this bottleneck: (i) By truncating the memory integrals in the KBE to a maximum of $N_c$ timesteps, the computational complexity is reduced to $\mathcal{O}(N_tN_c^2)$, and the memory requirement to $\mathcal{O}(N_c^2)$. Provided that the results converge with respect to the cutoff $N_c$, memory truncation allows to extend the simulations to significantly longer times.  (ii)  We introduce functionalities to describe nonequilibrium steady states, i.e. time-translationally invariant nonequilibrium states. Such states are relevant for transport settings, and they provide an approximate  description of slowly evolving (prethermal) nonequilibrium states.
\\

\noindent \textbf{NEW VERSION PROGRAM SUMMARY}

\begin{small}
\noindent
{\em Program Title:}               NESSi.                           \\
{\em Developer's repository link:} https://github.com/nessi-cntr/nessi \\
{\em Licensing provisions:} MPL-2.0.  \\
{\em Programming language:}             C++, python.                      \\
{\em External routines/libraries:} cmake, eigen3, fftw3 (optional), hdf5 (optional), mpi (optional), omp (optional). \\
{\em Journal reference of previous version:} \cite{1}.                  \\
{\em Does the new version supersede the previous version?:}  Extension of the existing library.  \\
{\em Reasons for the new version:} Reduce computational and memory cost to extend simulation times and directly simulate nonequilibrium steady states. \\
{\em Summary of revisions:} Extension to memory-truncated and steady-state Kadanoff-Baym equations. \\
{\em Nature of problem:} Solves equations of motion of time-dependent Green’s functions on the Kadanoff-Baym contour for a memory-truncated self-energy and in the steady state. \\
{\em Solution method:} Higher-order solution methods of integral and integro-differential equations on the Kadanoff-Baym contour with memory-truncation in the integral kernel and evaluation of Fourier integrals in the steady state.  \\
 \\

\end{small}

\end{abstract}


\end{frontmatter}

\section*{List of abbreviations}
\begin{tabbing}
\hspace{3cm}\= \textbf{Notation} \hspace{1cm} \= \textbf{Description} \\
\> DMFT \> dynamical mean-field theory\\
\> DOS \> density of states\\
\> FFT\> fast Fourier transform\\
\> FFTW\> Fastest Fourier Transform in the West algorithm\\
\> HDF5\> Hierarchical Data Format version 5\\
\> KB \> Kadanoff-Baym\\
\> KBE \> Kadanoff-Baym equation\\
\> NEGF \> nonequilibrium Green's function\\
\> NESS \> nonequilibrium steady state
\end{tabbing}

\section{Introduction}

Nonequilibrium Green's function (NEGF) techniques based on the Keldysh formalism provide a versatile field-theoretical approach to investigate quantum many-particle systems out of equilibrium, with applications from condensed matter physics to plasma and particle physics \cite{KadanoffBaym_1962, Keldysh_1964, KamenevBook, stefanucci_nonequilibrium_2013}. In the condensed matter context, they are  particularly useful to study the dynamics in solids induced by short laser pulses, which drive collective processes on the  intrinsic timescale of the electronic motion \cite{Sentef2021, Giannetti2016, Murakami2025}. Numerical simulations using NEGFs  rely on correlation functions (Green's functions, response functions) which depend on two time arguments. Their equations of motion are the so-called Kadanoff-Baym equations (KBEs) \cite{KadanoffBaym_1962}, which correspond to a real-time formulation of the Dyson equation. The KBEs are integral equations, where the forward propagation in time depends on an integration of the past evolution, with a memory kernel determined by the self-energy. The nonequilibrium systems simulation package \NESSi{}  provides a general-purpose framework for performing such simulations, with data structures to store one- and two-time correlation functions, as well as routines to solve their equations of motion and to evaluate elementary diagrammatic expressions \cite{NESSi}. It can be used within perturbative frameworks,  such as nonequilibrium variants of Hedin's $GW$ method \cite{Aryasetiawan_1998, Golez2016} or the fluctuation exchange approximation \cite{Bickers1989, Sayyad2019, Stahl2021}, and it provides a framework for nonequilibrium dynamical mean-field theory (DMFT) \cite{Aoki_2014}, where the effective model to be solved represents an impurity atom embedded into a self-consistently determined host.   As a result, \NESSi{} has been widely used to simulate condensed matter dynamics in photo-excited solids, see Ref.~\cite{Murakami2025} for a recent review. 

Correlation functions within the Keldysh formalism are defined with time arguments on a closed time contour $\mathcal{C}$ \cite{Keldysh_1964, stefanucci_nonequilibrium_2013}. In order to describe a system which is in thermal equilibrium with temperature $T$ at a given initial time $t_0$ before the perturbation, one can choose a contour which consists of a forward real-time branch $\mathcal{C}_1$ from $t_0$ to a final time $t_{\text{max}}$, a backward real-time branch  $\mathcal{C}_2$, and an imaginary-time branch  $\mathcal{C}_3$ from $t_0$ to $t_0-i\beta$, with $\beta=1/k_BT$. For contour-ordered correlation functions including general NEGFs, the time evolution along the imaginary branch is used to prepare the equilibrium initial state.  NEGF simulations can be computationally expensive because the KBEs require a memory integral over the entire past evolution of the system at each timestep. Most implementations are based on  an equidistant discretization of the real-time contour \cite{NESSi,stan_time_2009, Freericks2006, balzer_nonequilibrium_2012}. With $N_t$ real-time points, the computational cost  scales as $\mathcal{O}(N_t^3)$, while the memory requirement scales as $\mathcal{O}(N_t^2)$. However, in many applications, the dynamical range of the simulation can span several orders of magnitude in time, from the fastest electronic timescales (sub-femtosecond regime) over the pico-second dynamics of order parameters in photo-induced phase transitions, to the classical dynamics which can be described using phenomenological theories such as time-dependent Ginzburg-Landau theory.  Several strategies have been explored to address the computational challenges of long-time NEGF simulations. One can compress the two-time function using hierarchical matrix structures \cite{Kaye2021} or quantics tensor trains \cite{Shinaoka2023, Sroda2024}. Alternative methods include adaptive timestepping schemes \cite{Meirinhos2022,Lang2025}, exponential fitting procedures of the long-time tails \cite{Yin2022}, and machine learning based compression \cite{Zhu_2025}.
 Notable progress has also been made within approximate schemes, particularly the $\mathcal{O}(N_t)$ algorithm within the generalized Kadanoff-Baym approximation \cite{Lipavski1986, Schluenzen2020}. A strategy that most straightforwardly integrates with the existing \NESSi{} implementation, \NESSione{}, relies on  truncating memory integrals in the KBEs after a cutoff time $t_c$ \cite{Schueler2018, Stahl2022}. The latter then serves as a numerical control parameter. If $N_c$ represents the number of timesteps corresponding to $t_c$, this method reduces the computational complexity to $\mathcal{O}(N_t N_c^2)$, and memory usage to an amount $\mathcal{O}(N_c^2)$, which is independent of the propagation time. This approach has already been successfully applied to nonequilibrium DMFT simulations, where it allowed to extend $t_{\text{max}}$ by up to two orders of magnitude  (see, e.g., Ref.~\cite{Picano2021b, Dasari2020}).

In this manuscript, we present an extension of the \NESSi{} library, \NESSitwo{}, which incorporates techniques for addressing long-time and slow nonequilibrium dynamics based on the memory-truncated KBEs. \NESSitwo{} includes a new data structure (a moving Green's function window) which stores Green's functions on a two-time window ranging up to timestep $n$, but with a restricted memory depth of $N_c$ timesteps. A typical real-time simulation then begins with a conventional \NESSione{} simulation up to time $N_c$. After that, it is sufficient to shift the memory-restricted window forward in the timestepping procedure, maintaining a constant memory cost and computation time at each timestep.   In this manuscript, we explain the theoretical background of the memory-truncated KBEs and provide details on its implementation in \NESSitwo{}.  The usage of the library is also explained in an independent online manual on the webpage 
{\tt https://nessi.readthedocs.io/en/latest/}, which can be generated locally  following the compilation instructions in Sec.~\ref{subsec:installation_libcntr}. 

In the extreme limit of a slow nonequilibrium evolution, a system can reach a nonequilibrium steady state (NESS). A NESS is characterized by time-translationally invariant correlation functions that however do not obey universal equilibrium fluctuation-dissipation relations \cite{KamenevBook}. Typically, NESSs arise in open quantum systems under external bias, as studied in charge or energy transport settings. Quasi-steady states can also emerge in the dynamics of quantum many-particle systems:  In particular, if the evolution of a system is constrained by nearly conserved quantities, the system can reach a prethermal state which can persist as an almost stationary state over a long period of time.  Such quasi-steady prethermal states can in some cases be  approximately described as a NESS,  by weakly coupling the system to suitable reservoirs \cite{lange2017}. For example, after the photo-excitation of large-gap insulators, recombination bottlenecks can sustain a long lifetime of photo-carriers. The resulting slowly evolving photo-doped state can be approximated as a NESS by maintaining the non-thermal carrier population through the weak coupling of suitable charge reservoirs to the system \cite{Jiajun2021PRB, Kuenzel2024}. 

Within the NEGF formalism, the treatment of NESSs is simpler than that of the full two-time dynamics, since the memory of the initial state is lost, and one can reduce the time arguments of Green's functions to the two-branch contour $\mathcal{C}_1\cup\mathcal{C}_2$. Moreover, in the steady state all correlation functions depend  only on the difference between the two time arguments, such that the solution of the KBEs can be formulated in frequency space.  In the context of the quantum impurity models relevant for DMFT, NESSs also offer  perspectives for non-perturbative treatments, using  quantum Monte Carlo methods \cite{Profumo2015, Erpenbeck2023}, tensor-train representations of diagrammatics \cite{Eckstein2024, Kim2024}, and Lindblad master equations \cite{Arrigoni2013}. To complement the memory-truncation approach for real-time simulations, \NESSitwo{} also includes data structures for storing Green's functions in a NESS, which integrate with the real-time structures. For example, NESS Green's functions can be initialized using data from real-time or memory-truncated Green's functions at a given time slice, or the memory truncated time-evolution can be initialized in a time-translationally invariant way using a NESS Green's function.   We provide routines to solve Dyson equations and elementary diagrammatic expressions in the NESS formalism, expanding the capabilities of  \NESSi{} to problems such as transport and quasi-steady descriptions of slowly evolving states. 

This paper is organized as follows. Section \ref{secmemKBE} explains the memory truncated KBEs.  We start with 
a brief recapitulation of the basic functionalities of the previous version of  \NESSi{}, \NESSione{}, in Sec.~\ref{secnessi1} and then discuss the implementation of the memory-truncation scheme in the remaining subsections. Section~\ref{sec:steadystate} is devoted to the steady-state approach. In Secs.~\ref{sec:steadystategf} and \ref{section:steady_state_dsyon} we provide some background on steady-state Green's functions and the solution of the Dyson equation, and then detail the numerical implementation in Secs.~\ref{sec:ness:impl} and \ref{sec:steadystate_dyson_num}. Section~\ref{subsec:installation_libcntr} explains the compilation of \verb|libcntr|, while Sec.~\ref{sec:examples} presents benchmarks and results obtained with the example programs. 

\section{Memory-truncated KBEs}
\label{secmemKBE}

\label{trunc}
\subsection{Review of \NESSione{}: Solution of the full KBEs}
\vspace*{2mm}

\label{secnessi1}

\subsubsection*{Green's functions}\vspace*{2mm}

In general, we deal with two-time functions $G(t,t')$ with time arguments on the KB contour $\mathcal{C}=\mathcal{C}_1\cup\mathcal{C}_2\cup\mathcal{C}_3$, which represent contour-ordered correlation functions.\footnote{For an overview over the Keldysh formalism, see Refs.~\cite{KamenevBook, stefanucci_nonequilibrium_2013}, as well as the original \NESSione{} paper \cite{NESSi}: Section 3 of Ref.~\cite{NESSi} contains a more detailed list of relevant relations involving two-time Green's functions and the KBE, while the following section gives only a brief summary.}  An example is  the Green's function $G_{ab}(t,t') = -i\langle T_\mathcal{C} c_a(t) c_b^\dagger(t')\rangle $ for fermionic or bosonic particles, where $c_a$ ($c_a^\dagger$) represents the annihilation (creation) operator for a particle in an orbital $a$, and $T_\mathcal{C}$ is the contour-ordering operator.  Other important  examples include correlation functions $G_{ab}(t,t')=-i \langle T_\mathcal{C} u_a(t)u_b(t') \rangle $ of real fields $u$, such as phonon displacement fields. Hence, if not indicated otherwise, Green's functions $G(t,t')$ are understood as matrices carrying orbital, spin, and/or spatial (site) indices. 
To parametrize the contour-ordered Green's functions, one can use the Matsubara component, with two arguments $-i\tau_1$ and $-i\tau_2$ on the imaginary branch,
\begin{align}
G^M(\tau_1-\tau_2) = i G(-i\tau_1,-i\tau_2),
\end{align}
the greater and lesser real-time functions,
\begin{align}
G^<(t,t') &= G(t_+,t'_-), \label{eq:lesser_nessi}
\\
G^>(t,t') &= G(t_-,t'_+), \label{eq:greater_nessi}
\end{align}
where $t_+$ ($t_-$) denotes a time argument on the
forward 
contour $\mathcal{C}_1$
(backward 
contour $\mathcal{C}_2$), as well as the functions with mixed arguments,
\begin{align}
G^\lceil(t,\tau) &=G(t_\pm,-i\tau),
\\
G^\rceil(\tau,t) &=G(-i\tau,t_\pm).
\end{align}
Because of their physical relevance as response functions or spectral functions, we also introduce the retarded and advanced Green's functions 
\begin{align}
G^R(t,t') &= \theta(t-t')[G^>(t,t')-G^<(t,t')], \label{eq:retarded_nessi}
\\
G^A(t,t') &= \theta(t'-t)[G^<(t,t')-G^>(t,t')]. \label{eq:advanced_nessi}
\end{align}
The physical meaning of the Green's functions becomes more obvious through their Fourier transforms in the steady state formalism (Sec.~\ref{sec:steadystategf}).

\begin{figure}[tbp]
\includegraphics[width=0.99\textwidth]{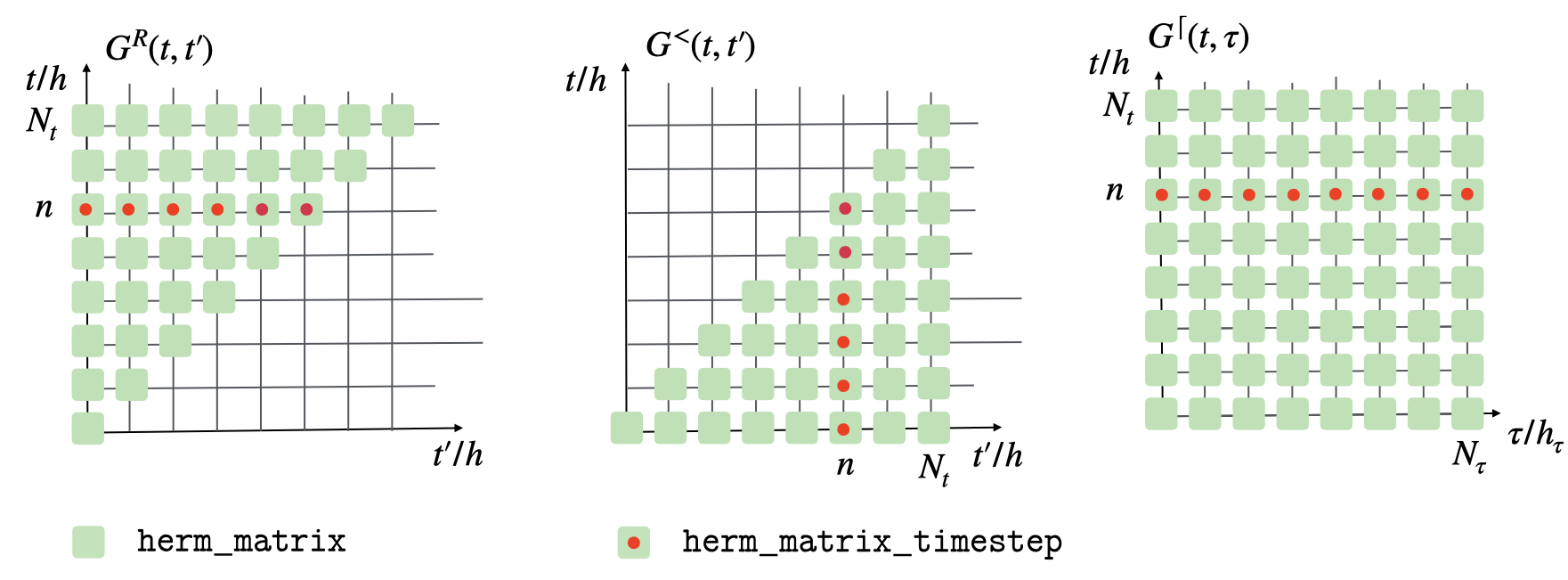}\\
\caption{
Green's function domain and data structures on a discrete time grid. The Hermitian domain \eqref{eq:storage_herm} for a two-time Green's function $G(t,t')$ is shown by the filled squares (the Matsubara component is not shown here). The class {\tt herm\_matrix} stores the values of a Green's function on the Hermitian domain up to a maximum timestep $N_t$, which is sufficient to represent a Hermitian Green's function with the symmetry $G=G^\ddagger$. Filled squares with dots indicate the time slice $n$. The data on one time slice $n$ of a Green's function can be stored using the class {\tt herm\_matrix\_timestep}. The timestep on the real axis is denoted by $h$ and $h_{\tau}$ is the timestep on the Matsubara axis.}
\label{fig01}
\end{figure}

All Green's functions satisfy a periodic (antiperiodic) boundary condition on $\mathcal{C}$ for bosonic (fermionic) correlation functions, the so-called Kubo-Martin-Schwinger boundary condition, as well as a causality constraint, which implies that $G(t,t')$ is fully determined by  $G^M, G^R, G^A, G^<, G^\rceil, G^\lceil$. Moreover, for each Green's function it is convenient to define a Hermitian conjugate $G^\ddagger$ by 
\begin{subequations}
  \label{eq:hermconjg}
  \begin{align}
    \label{eq:hermconjg_gtrless}
   [G^\ddagger]^\gtrless(t,t^\prime) &=    -\left( G^\gtrless(t^\prime,t) \right)^\dagger ,\\
    \label{eq:hermconjg_ret}
    [G^\ddagger]^\mathrm{A}(t,t^\prime) &= \left(  G^\mathrm{R}(t^\prime,t) \right)^\dagger,\\
    [G^\ddagger]^\mathrm{R}(t,t^\prime) &= \left(  G^\mathrm{A}(t^\prime,t) \right)^\dagger,\\
       [G^\ddagger]^\lceil(\tau,t) &= - \xi \left(  G^\rceil(t,\tau + \beta) \right)^\dagger,\\
       [G^\ddagger]^\rceil(t,\tau) &= - \xi \left(  G^\lceil(\tau + \beta, t) \right)^\dagger,\\
       [G^\ddagger]^\mathrm{M}(\tau) &= \left(  G^\mathrm{M}(\tau) \right)^\dagger,\
  \end{align}
\end{subequations}
where $\dagger$ denotes the conventional matrix adjoint and $\xi=-1$ ($\xi = +1$) is the fermionic (bosonic) sign. Green's functions of bosonic and fermionic particles, contour-ordered correlations functions of real fields, and their self-energies have a Hermitian symmetry $G^\ddagger=G$. In \NESSione{}, the basic container {\tt herm\_matrix} for two-time functions therefore stores only the elements needed to define a Hermitian Green's function in a non-redundant manner.  The domain of storage for the Green's function is shown in Fig.~\ref{fig01}; it includes the minimal set of entries
\begin{subequations}
  \label{eq:storage_herm}
  \begin{align}
    &G^M(m h_\tau) \ , \ m=0,\dots,N_\tau \ , \\
    & G^<(j h, n h) \ , \ n=0,\dots,N_t, j=0,\dots,n \ , \\
    &G^R(n h, j h) \ , \ n=0,\dots,N_t, j=0,\dots,n \ , \\
    &G^\rceil(n h, m h_\tau) \ , \ n=0,\dots,N_t, m=0,\dots,N_\tau,
  \end{align}
\end{subequations}
where $h$ ($h_\tau$) denote a timestep in real (imaginary) time, and $N_t$ ($N_\tau$) is the number of real (imaginary) time points (i.e., $\beta=N_\tau h_\tau$ and $t_\text{max}=hN_t$). The storage domain of the {\tt herm\_matrix} object will also be referred to  as the Hermitian domain of the contour function.

\subsubsection*{Kadanoff-Baym equations}\vspace*{2mm}

Typical NEGF simulations involve two tasks: The first is the  solution of the Dyson equation
\begin{align}
    G=G_{0}+G_{0} \ast\Sigma *G,
    \label{Dyson01}
\end{align}
i.e. the calculation of the interacting Green's function from the noninteracting Green's function $G_0$ and the self-energy $\Sigma$. Here, all functions are understood as two-time functions on $\mathcal{C}$, and $\ast$ denotes the convolution 
\begin{align}
    [A\ast B] (t,t')= \int_\mathcal{C} d\bar t\,A(t,\bar t) B(\bar t,t')
    \label{convl}
\end{align}
on the Keldysh contour $\mathcal{C}$. The second task is the evaluation of the self-energy. In diagrammatic perturbation theory, $\Sigma$ is itself expressed either in terms of $G_0$ or $G$. The expression of the self-energy $\Sigma$ in terms of $G$ is therefore often a self-consistent equation, making the KBEs nonlinear.   In a wide class of approximations, including $GW$, FLEX, and the 2nd Born approximation, the basic building block of the expression for $\Sigma$ are 
products
of the form 
\begin{align}
\label{bubble1}
C_1(t,t') = A_{a_1,a_2} (t,t')B_{b_2,b_1}(t',t),
\\
\label{bubble2}
C_2(t,t') = A_{a_1,a_2} (t,t')B_{b_1,b_2}(t,t').
\end{align}
Due to their diagrammatic representation, the first is called a ``particle-hole bubble'' and the second a ``particle-particle bubble''. These are in essence point-wise products of Green's functions, but the $R,<,\rceil$ and $M$ components of $C$ must be properly expressed in terms for the corresponding components of $A$ and $B$ according to the respective Langreth rules \cite{NESSi}. 

The \NESSione{} package provides routines for the solution of the Dyson equation \eqref{Dyson01}, the convolution \eqref{convl}, and 
the computation of bubble products \eqref{bubble1},\eqref{bubble2}. The Dyson equation can be solved in two forms: If the noninteracting Green's function satisfies a differential equation with single-particle Hamiltonian $\epsilon(t)$ 
(understood as matrix carrying orbital, spin, and/or spatial (site) indices).
\begin{eqnarray}
    G_{0}^{-1}(t,t^{\prime})=\delta_{\mathcal{C}}(t,t^{\prime})\left[i\partial_t-\epsilon(t)\right],
\end{eqnarray}
one arrives at the integral-differential form of the Dyson equation
\begin{eqnarray}
    \left[i\partial_t-\epsilon(t)\right]G(t,t^{\prime})-[\Sigma \ast G](t,t')=\delta_{\mathcal{C}}(t,t^{\prime}),
\label{intdif_eq}
\end{eqnarray}
where $\delta_{\mathcal{C}}(t,t^{\prime})$ is the Dirac delta function defined on the KB contour. Below, we will frequently use the abstract notation 
\begin{eqnarray}
    G^{-1} = i\partial_t-\epsilon-\Sigma 
\label{intdif_eq_general_notation}
\end{eqnarray}
to denote this integral-differential equation. Alternatively, by defining  $F=-G_0\ast\Sigma$, one arrives at an integral equation
\begin{align}
(1+F)\ast G = G_0.
\label{vie2}
\end{align}
The two respective equations are called {\tt dyson} and {\tt vie2}  in \NESSi, where the name {\tt vie2}  indicates that the numerical solution is obtained by a mapping to Volterra integral equations of the 2nd type. A solution is provided under the assumption that $G_0$, $G$, and $\Sigma$  are Hermitian. (Note, that for $F=-G_0\ast\Sigma$,  $F^\ddagger=-\Sigma \ast G_0 \neq  F$ is not Hermitian, but the solution of Eq.~\eqref{Dyson01} is.)

\subsubsection*{Timestepping}\vspace*{2mm} 

Due to causality, both Eqs.~\eqref{intdif_eq} and \eqref{vie2} can be solved using a timestepping procedure. 
To make this transparent, we denote by the time slice $\mathcal{T}[G]_n$ of a Hermitian two-time function 
all elements for which the larger of the two time arguments is $n$. \NESSione{} provides the class {\tt herm\_matrix\_timestep}, which  stores 
\begin{subequations}
  \begin{align}
    (\mathcal{T}[G]_n)^<_{j} &= G^<(j h, n h) \ ,  j=0,\dots,n \ , \\
    (\mathcal{T}[G]_n)^R_{j} &= G^R(n h, j h) \ ,  j=0,\dots,n \ , \\
    (\mathcal{T}[G]_n)^\rceil_{m} &= G^\rceil(n h, m h_\tau) \ ,  m=0,\dots,N_\tau \,,
  \end{align}
\end{subequations}
for
$n\ge 0$ ($\mathcal{T}[G]_{-1}$ refers to the Matsubara component). The corresponding values for one timestep $n$ are indicated by the red dot in Fig.~\ref{fig01}. The key causal property of Eqs.~\eqref{intdif_eq} and \eqref{vie2} implies that, given the input $\Sigma$ (or $F$ and $G_0$), the output Green's function $G$ can be computed on a timestep $\mathcal{T}[G]_{n+1}$ from its value on timesteps  $\mathcal{T}[G]_{m}$ with $m\le n$. Similarly, in self-consistent theories, the dependence of $\Sigma$ on $G$ is causal, i.e., $\mathcal{T}[\Sigma]_{n}$ can be obtained from  $\mathcal{T}[G]_{m}$ with $m\le n$.  
All operations in \NESSione{} (convolution,  the solution of Eqs.~\eqref{intdif_eq} and \eqref{vie2}, simple algebra operations on  Green's functions, the evaluation of the particle-hole and particle-particle bubbles, etc.) are therefore implemented for a given timestep. An exception is the solution of Eqs.~\eqref{intdif_eq} and \eqref{vie2} for the first few real-timesteps $n=0,...,k$, where $k$ is the given order of the quadrature rules for the memory integrals ($k\le 5$).  The solution on these first $k$ timesteps is done simultaneously in a startup (or ``bootstrapping'') routine \cite{NESSi}.  A typical simulation thus proceeds in the timestepping manner sketched in the pseudocode algorithm~\ref{alg:matsubara}. A worked out example is found in Sec.~\ref{sec:movingexample}.

\begin{algorithm}[tbp]
\caption{Solving the Dyson equation with a self-consistent self-energy.}
\label{alg:matsubara}
\begin{algorithmic}[1]
        \State Start with a guess for $\mathcal{T}[\Sigma]_{-1}$.
        \While{not converged}
        \State \textbf{Solve} Dyson equation for $\mathcal{T}[G]_{-1}$ (Matsubara).
         \State \textbf{Compute} $\mathcal{T}[\Sigma]_{-1}$ from $\mathcal{T}[G]_{-1}$.
        \EndWhile  
        \State Generate a guess for $\mathcal{T}[\Sigma]_{m}$, $m=0,...,k$ (e.g., $\Sigma=0$).
        \While{not converged}
        \State \textbf{Solve} Bootstrapping of the Dyson equation for $\mathcal{T}[G]_{m}$, $m=0,...,k$.
         \State \textbf{Compute} $\mathcal{T}[\Sigma]_m$ for $m=0,...,k$.
        \EndWhile      
    \For{$n = k + 1$ to $N_t$}
        \State \textbf{Extrapolate} $\Sigma$ by one timestep to estimate $\mathcal{T}[\Sigma]_n$.
        \While{not converged}
        \State \textbf{Solve} Dyson equation for $\mathcal{T}[G]_n$.
        \State \textbf{Compute} $\mathcal{T}[\Sigma]_n$ from $\mathcal{T}[G]_m$ for $m\le n$.
        \EndWhile
    \EndFor
\end{algorithmic}
\end{algorithm}

\subsection{Memory-truncated Green's functions and KBE}

\begin{figure}[tbp]
\includegraphics[width=0.99\textwidth]{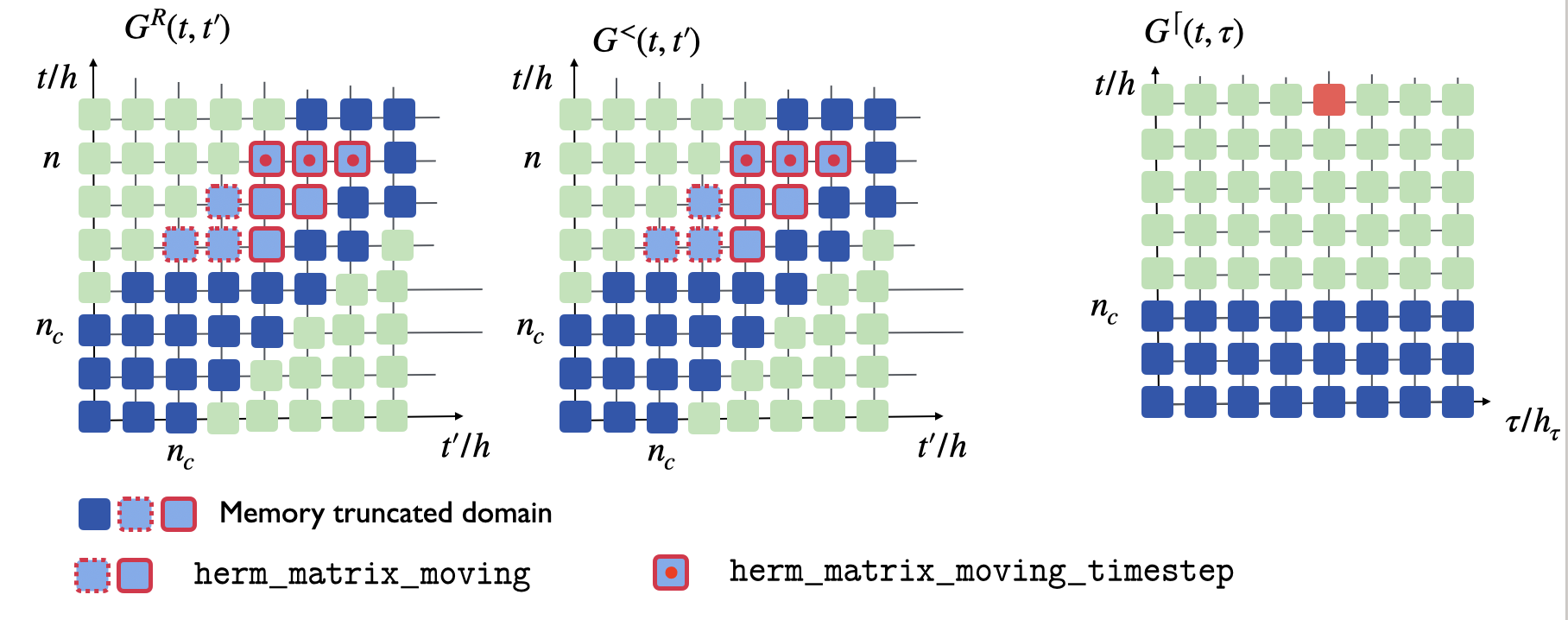}
\caption{
Domains for data structures on a discrete time grid. The memory truncated domain, defined by Eqs.~\eqref{sgshsjwkqqma01} and \eqref{sgshsjwkqqma02}, is indicated by dark squares, for a cutoff $n_c=2$. The domain of a memory truncated time slice 
$\mathcal{T}[G]^{n_c}_n$ (Eq.~\eqref{time slice}) is shown by the squares with a red dot. Squares with a solid boundary denote the triangular window $\Delta[G]^{n_c}_n$, while squares with a dashed boundary show the extension of the triangular window \eqref{time slice-tri} to the moving window $\mathcal{M}[G]^{n_c}_n$ [Eq.~\eqref{time slice-moving}]. }
\label{fig02}
\end{figure}

Because of the convolution integrals in the timestepping, the computational effort to solve the above equations scales like $\mathcal{O}(N_t^3)$ in CPU time and $\mathcal{O}(N_t^2)$ in memory. In certain cases, this limitation can be circumvented by adopting the following  truncation scheme which is included in \NESSitwo{}. The truncation scheme can be applied when the self-energy $\Sigma(t,t')$ decays to zero sufficiently fast  as a function of the time difference,  such that the following approximation can be made:
\begin{eqnarray}
    &&\Sigma^{R}(t,t^{\prime})=\Sigma^{<}(t,t^{\prime})=0 \hspace{1cm} \text{for} \hspace{1cm} |t-t^{\prime}|> t_{c}, 
        \label{sgshsjwkqqma01}
        \\
    &&\Sigma^{\rceil}(t,\tau)=0 \hspace{2.86cm} \text{for} \hspace{1cm} t> t_{c}.
    \label{sgshsjwkqqma02}
\end{eqnarray}
Here $t_{c}=hn_c$ is the memory cutoff. We will refer to the remaining region as the memory-truncated domain,
\begin{subequations}
  \label{momorybtruncatde}
  \begin{align}
    G^<(j h, n h) \ , \ |n-j|\le n_c,\\
    G^R(n h, j h) \ , \ |n-j|\le n_c,
  \end{align}
\end{subequations}
which is shown by the dark shaded squares in Fig.~\ref{fig02}.  We can define a partial time slice\footnote{Note, that for truncated memory Green's functions, the Hermitian domain for the lesser component  is defined with swapped time arguments compared to the original implementation, cf. Fig.~\ref{fig01} and Fig.~\ref{fig02}.} of the contour function $G$ as
\begin{eqnarray}
\mathcal{T}[G]_{n}^{n_c} = \{ G^{R}(nh, (n-m)h), G^{<}(nh, (n-m)h), 0\le m\le n_c \}.
\label{time slice}
\end{eqnarray}
 The key observation in Ref.~\cite{Stahl2022} was that, under the assumption that the self-energy satisfies the constraints \eqref{sgshsjwkqqma01} and \eqref{sgshsjwkqqma02}, the Dyson equation \eqref{intdif_eq} can be solved for the partial time slice $\mathcal{T}[G]_{n}^{n_c}$ if we know $\Sigma$ on the partial time slice $\mathcal{T}[\Sigma]_{n}^{n_c}$, and $G$ on a  triangular domain 
\begin{eqnarray}
\Delta[G]_{n}^{n_c} = \bigcup_{m=0}^{n_c} \mathcal{T}[G]_{n-m}^{n_c-m},
\label{time slice-tri}
\end{eqnarray}
as shown by the squares with the solid red boundary in Fig.~\ref{fig02}. The argument for this is given in Sec.~\ref{kbetheory}.

For the numerical implementation, it is favorable to extend the triangular domain $\Delta[G]_{n}^{n_c}$ to a moving window 
\begin{eqnarray}
\mathcal{M}[G]_{n}^{n_c} = \bigcup_{m=0}^{n_c} \mathcal{T}[G]_{n-m}^{n_c},
\label{time slice-moving}
\end{eqnarray}
represented by the squares with solid and dashed red boundaries in Fig.~\ref{fig02}. The values of the Green's function in the moving window \eqref{time slice-moving} can be stored in a new data structure {\tt herm\_matrix\_moving}. The advantage of the  moving window \eqref{time slice-moving} over the minimal triangular domain is that the window can be easily shifted forward in time: To advance $\mathcal{M}[G]_{n}^{n_c} $ to $\mathcal{M}[G]_{n+1}^{n_c} $, the values on the earliest step $\mathcal{T}[G]_{n-n_c}^{n_c}$ are eliminated, while a new  leading step $\mathcal{T}[G]_{n+1}^{n_c}$ is added. Because of the rectangular data alignment, only the internal references to the time slices $\mathcal{M}[G]_{m}^{n_c}$ have to be recalculated, and no copying of remaining data is required.  With this, a time propagation within the memory truncated KBEs can proceed as described in the pseudocode algorithm \ref{alg:moving}. 

\begin{algorithm}[tbp]
\caption{Memory truncated KBEs with a self-consistent self-energy.}
\label{alg:moving}
\begin{algorithmic}[1]
        \State Use the conventional algorithm \ref{alg:matsubara} to compute $\mathcal{T}[G]_{n}$ and $\mathcal{T}[\Sigma]_{n}$ for $n\le n_c$.
        \State 
        Initialize moving windows $\mathcal{M}[G]^{n_c}_n$ and $\mathcal{M}[\Sigma]^{n_c}_n$ at $n=n_c$ with the input from timesteps $m\le n$.        
        \For{$n = n_c + 1$ to $N_t$}
        \State \textbf{Advance} $\mathcal{M}[X]^{n_c}_m$ for $X=G,\Sigma$ from $m=n-1$ to $m=n$.
        \State {\bf Predict} timestep  $\mathcal{T}[G]^{n_c}_n$ within $\mathcal{M}[G]^{n_c}_n$  by extrapolation from $\mathcal{T}[G]^{n_c}_m$, $m<n$.        
        \While{not converged}
        \State \textbf{Compute} $\mathcal{T}[\Sigma]^{n_c}_n$ from $\mathcal{M}[G]^{n_c}_n$.
        \State \textbf{Solve} Dyson for $\mathcal{T}[G]^{n_c}_n$, with input from $\mathcal{M}[G]^{n_c}_n$ and  $\mathcal{M}[\Sigma]^{n_c}_n$.
        \EndWhile
    \EndFor
\end{algorithmic}
\end{algorithm}

\subsection{Memory-truncated integral equations: Details}
\label{kbetheory} 
\vspace*{2mm}

In this section, we demonstrate that the knowledge of $\Sigma$ on the memory-truncated domain allows to compute $G$ on the same domain \cite{Stahl2022}, and give details of the numerical  implementation.  We  aim to determine $G$ on the restricted timestep $\mathcal{T}[G]_n^{n_c}$ with $n>n_c$.  First, the Dyson equation \eqref{intdif_eq} can be written in terms of the components $X^M,X^R,X^<$, and $X^\rceil$ for $X=G,\Sigma$.  For $\mathcal{T}[G]_n^{n_c}$, we only need the equations for the retarded and lesser components \cite{NESSi},
\begin{eqnarray}
\label{KB_eq1}
    &&\left[i\partial_t-\epsilon(t)\right]G^{R}(t,t^{\prime})-\int_{t^{\prime}}^{t}d\bar{t}\,\,\Sigma^{R}(t,\bar{t})G^{R}(\bar{t},t^{\prime})=0, \\
    \label{KB_eq2}
    &&\left[i\partial_t-\epsilon(t)\right]G^{<}(t,t^{\prime})-\int_{0}^{t}d\bar{t} \,\, \Sigma^{R}(t,\bar{t})G^{<}(\bar{t},t^{\prime})  =
        \nonumber\\
        &&\hspace{10mm}=\int_{0}^{t^{\prime}}d\bar{t}\,\, \Sigma^{<}(t,\bar{t})G^{A}(\bar{t},t^{\prime})
    - i\int_{0}^{\beta}d\tau \,\, \Sigma^{\rceil}(t,\tau)G^{\lceil}(\tau,t^{\prime}).
\label{KB_eq3}
\end{eqnarray}
These equations are solved iteratively with a timestepping procedure by discretizing the time axis with a constant timestep $h$. Here, we will use subscripts $X_{n,m}=X(nh,mh)$ to indicate discrete time arguments. For $t=nh$ with $n>n_c$ ($t>t_c$), the last integral vanishes together with $\Sigma^{\rceil}$ [Eq.~\eqref{sgshsjwkqqma02}]. 

To solve the integral-differential equation, we proceed as in Ref.~\cite{NESSi} and combine a $(k+1)$th order backward approximation for the derivative
\begin{align}
\label{BDdiff}
[\partial_tf(t)]_{n} = \frac{1}{h}\sum_{l=0}^{k+1} \alpha_l f_{n-l},
\end{align}
with $k$th order Gregory quadrature rules for the integral,
\begin{align}
\label{Gquadrature}
\int_{0}^{nh} \!\!dt \,f(t) = h \sum_{m=0}^{\text{max}(n,k)} w^{(n)}_m  f_{m}.
\end{align}
The weights $\alpha_l$ and $w^{(n)}_m$ are given in Ref.~\cite{NESSi}, with the simplest case ($k=0$) being $\alpha_0=-\alpha_1=0.5$ for the derivative and the trapezoidal rule for the integral ($w^{(n)}_n=w^{(n)}_0=0.5$,  $w^{(n)}_m=1$  for $0<m<n$).  The Gregory integration with $k$ points has an error of order $\mathcal{O}(h^{k+2})$. An important thing to note is that the quadrature requires at least $k+1$ points, such that the integral  \eqref{Gquadrature} needs an extension of the function $f$ outside the domain $[0,nh]$ if $n<k$.  In the following, we therefore assume a  truncation at $n_c>k$. 

\subsubsection*{Retarded component}
\vspace*{2mm}

We can now apply the discretization \eqref{BDdiff} and \eqref{Gquadrature} to Eq.~\eqref{KB_eq1} at $(t,t')=(nh,mh)$ on the truncated time slice $\mathcal{T}[G]^{n_c}_n$, i.e., with $n-m\le n_c$:
\begin{align}
\label{KB_eq1A}
\sum_{l=0}^{k+1}
\frac{i\alpha_l}{h} 
G^{R}_{n-l,m}
-\epsilon_n G^{R}_{n,m}
-
h\sum_{l=0}^{\text{max}(n-m,k)} w^{(n-m)}_{l}
\Sigma^{R}_{n,n-l} G^{R}_{n-l,m}=0.
\end{align}
For $m<n-k$, $\text{max}(n-m,k)$ becomes $n-m$, and one can immediately see that all four functions $X_{ab}$ in this equation  satisfy the condition $n\ge a \ge n-n_c$ and $0\le a-b \le n_c$, i.e., the values are part of the moving windows $\mathcal{M}[X]^{n_c}_n$.  Hence one can simply solve this equation for $G^R_{n,m}$, which yields the  proposed time evolution algorithm to determine $G^R_{n,m}$ based on the moving windows $\mathcal{M}[X]^{n_c}_n$ for $X=G,\Sigma$. 

The points $G^R_{n,m}$ with $n-k\le m\le n$ need special consideration. For this case, the second sum in Eq.~\eqref{KB_eq1A} extends to $l=k$, such that there are points in the integral which depend on $ G^{R}_{n-l,m}$ with $n-l<m$. In order to obtain a $k$th order accurate algorithm, one would need a differentiable extension of $ G^{R}(t,t')$ to the domain $t<t'$. Using Eqs.~\eqref{eq:retarded_nessi}  and \eqref{eq:hermconjg_gtrless}, the most straightforward approach appears to be the extension 
\begin{align}
\label{hvmappring}
X^{R}_{n-l,m} \to  - \big[X^{>}_{m,n-l} - X^{<}_{m,n-l}\big]^\dagger =  - \big[X^{R}_{m,n-l} \big]^\dagger,
\end{align}
which for $n-l<m$, $n-m\le k$  and $k<n_c$ again depends only on values of $G$ in  $\mathcal{M}[G]^{n_c}_n$. Interestingly, however, the algorithm based on the extension \eqref{hvmappring} for $G$ within the retarded time propagation  turns out to be unstable. To determine $G^R_{n,m}$ with $n-k\le m\le n$, we instead discretize the conjugate equation to Eq.~\eqref{KB_eq1},
\begin{eqnarray}
\label{KB_eq1cc}
    &&-i\partial_{t'}G^{R}(t,t^{\prime})-G^{R}(t,t^{\prime})\epsilon(t')-\int_{t^{\prime}}^{t}d\bar{t}\,\,G^{R}(t,\bar{t})\Sigma^{R}(\bar{t},t^{\prime})=0,
\end{eqnarray}
using an approximation for the derivative $\partial_{t'}G^{R}(t,t^{\prime})$ on the first $k+1$ steps $m=n,...,n-k$ that is based on polynomial interpolation of $G^R_{n,m}$ on these points. The discretized equation then leads to a simultaneous  linear equation for all  $G^R_{n,m}$, $m=n,...,n-k$, where now $\Sigma$  instead of $G$ needs to be extended using Eq.~\eqref{hvmappring}. This concludes the timestepping for $G^R$. 

\subsubsection*{Lesser component}
\vspace*{2mm}

We now apply the discretization to the second equation \eqref{KB_eq3}, 
\begin{align}
\label{KB_eq3A}
\sum_{l=0}^{k+1}
\frac{i\alpha_l}{h} 
G^{<}_{n-l,m}
-\epsilon_n G^{R}_{n,m}
=
I_{n,m} + J_{n,m},
\end{align}
where $I_{n,m}$ and $J_{n,m}$ denote the memory integrals,
\begin{align}
I_{n,m}
&=\int_{0}^{t}d\bar{t}\,\Sigma^{R}(t,\bar{t})G^{<}(\bar{t},t^{\prime})\Big|_{t=nh,t'=mh},
\\
\label{Jn}
J_{n,m}&=
\int_{0}^{t^{\prime}}d\bar{t}\,\Sigma^{<}(t,\bar{t})G^{A}(\bar{t},t^{\prime})\Big|_{t=nh,t'=mh} \nonumber \\
&=
\int_{0}^{t^{\prime}}d\bar{t}\,\Sigma^{<}(t,\bar{t})G^{R}(t^{\prime},\bar{t})^\dagger\Big|_{t=nh,t'=mh}.
\end{align}
In the first integral, one can use the constraint \eqref{sgshsjwkqqma01} to restrict  the integration range to $[t-t_c,t]$,
\begin{align}
I_{n,m}
&=\int_{t-t_c}^{t}d\bar{t}\,\Sigma^{R}(t,\bar{t})G^{<}(\bar{t},t^{\prime})
=
h\sum_{l=0}^{n_c} w^{(n_c)}_l \Sigma^{R}_{n,n-l} G^{<}_{n-l,m}.
\end{align}
Since $n-m\le n_c$ and $l\le n_c$, the time difference of the arguments of $G$ satisfies $|(n-l)-m|\le n_c$. Hence either $G^{<}_{n-l,m}$ itself or $G^{<}_{m,n-l}$ is in the window $\mathcal{M}[G]^{n_c}_n$; in the latter case, we can use the Hermitian symmetry 
\begin{align}
\label{hvmappringles}
G^{<}_{n-l,m} =   - \big[G^{<}_{m,n-l} \big]^\dagger.
\end{align}
For the second integral \eqref{Jn}, the memory constraint \eqref{sgshsjwkqqma01}  implies
\begin{align}
J_{n,m}&
=
\int_{t-t_c}^{t^{\prime}}d\bar{t}\,\Sigma^{<}(t,\bar{t})G^{A}(\bar{t},t^{\prime})\Big|_{t=nh,t'=mh}.
\label{,xblaz}
\end{align}
We can again choose a discretization for which the result depends only on elements of $\mathcal{M}[X]^{n_c}_n$ with $X=G,\Sigma$:
If $t'-(t-t_c)= (m-n+n_c)h \ge kh$ we take
 \begin{align}
J_{n,m}
&=
h
\sum_{l=n-n_c}^{m}
w^{(m-(n-n_c))}_{l-(n-n_c)}
\Sigma^{<}_{n,l}G^{A}_{l,m}
\nonumber \\ &=
h
\sum_{l=n-n_c}^{m}
w^{(m-(n-n_c))}_{l-(n-n_c)}
\Sigma^{<}_{n,l}(G^{R}_{m,l})^\dagger.
\end{align}
If $t'-(t-t_c) < kh$, the integrand in \eqref{,xblaz} must be extended outside the interval. By choosing a forward extension
 \begin{align}
J_{n,m}&
=
h
\sum_{l=n-n_c}^{(n-n_c)+k}
w^{(m-(n-n_c))}_{l-(n-n_c)}
\Sigma^{<}_{n,l}G^{A}_{l,m}
\end{align}
together with the Hermitian extension \eqref{hvmappring} for $G$, again all points within the integrand fall within the moving window $\mathcal{M}[X]^{n_c}_n$ with $X=G,\Sigma$
(assuming that $n_c>k$). 
Hence we can again solve Eq.~\eqref{KB_eq3A} with the discrete expressions for $I_{n,m}$ and $J_{n,m}$ for $G^<_{n,m}$, which concludes the timestepping for the lesser component.

\subsubsection*{Volterra equation {\tt vie2}}
\vspace*{2mm}

Because the memory integrals in the Volterra equation \eqref{vie2} have the same structure as for the Dyson equation \eqref{intdif_eq}, the  timestepping on the restricted window can  be performed in an analogous fashion, and the explicit equations will not be reproduced here. The most important application of Eq.~\eqref{vie2}  is the  RPA equation
\begin{align}
\chi = \chi_0 +  \chi_0 \ast U \ast \chi,
\label{rpa}
\end{align}
where $\chi$ and $\chi_0$ denote a susceptibility, and $U$ a time-local interaction. This equation is mapped to an equation of type \eqref{vie2}, with 
\begin{align}
F(t,t')=-\chi_0(t,t')U(t'),\,\,\,F^\ddagger(t,t')=-U(t)\chi_0(t,t').
\end{align}
If $\chi_0$ decays in time as in Eqs.~\eqref{sgshsjwkqqma01} and  \eqref{sgshsjwkqqma02}, the RPA equation \eqref{rpa} can then be solved within the memory-truncated framework.

At this point, one should note a technical issue related to the convolution \eqref{convl}: As the analysis of the memory integrals in the Dyson equation has demonstrated, the convolution $C=A\ast B$ on the 
memory-truncated slice
$\mathcal{T}[C]^{n_c}_n$ can be calculated from the windows $\mathcal{M}[A]^{n_c}_n$ and $\mathcal{M}[B]^{n_c}_n$ if $A$ satisfies the constraints \eqref{sgshsjwkqqma01} and  \eqref{sgshsjwkqqma02}, while there is no restriction on $B$. However, if $B$ does not satisfy the constraints \eqref{sgshsjwkqqma01} and  \eqref{sgshsjwkqqma02}, the reverse convolution $C^\ddagger= B\ast A$ cannot be computed from the windows  $\mathcal{M}[A]^{n_c}_n$ and $\mathcal{M}[B]^{n_c}_n$. For this reason, we currently do not provide a general convolution routine for memory-truncated Green's functions. This also implies that in the present implementation Eq.~\eqref{Dyson01} cannot be solved by mapping it to the integral equation \eqref{vie2} with $F=-\Sigma\ast G_0$, because this would require us to compute both $F=-\Sigma\ast G_0$ and $F^\ddagger=- G_0\ast \Sigma$, although only $ \Sigma$ satisfies the memory constraints \eqref{sgshsjwkqqma01} and  \eqref{sgshsjwkqqma02}. Instead the Dyson equation should be solved  in the integral-differential form. 

\subsubsection*{Density matrix}
\vspace*{2mm}

Given a Green's function $A$, we define the density matrix by the equal time contribution 
\begin{align}
\label{density_matrix_A}
\rho_{A}(t) = \xi_A i A^<(t,t), 
\end{align}
where $\xi_A=\pm1$ for a bosonic (fermionic) Green's function $A$. Furthermore, one frequently needs the density matrix of the convolution of two Green's functions $A$ and $B$. We hence define the equal--time lesser component of their convolution 
\begin{align}
\label{density_matrix_AB}
\rho_{AB}(t) = 
\xi_A
i (A\ast B)^<(t,t),
\end{align}
which allows to compute for example the interaction energy from the equal-time convolution $\rho_{\Sigma G}(t)$ of the self-energy and the Green's function ($\xi_X$ refers to the sign of the function $X$, 
and both $A$ and $B$ have the same sign).
While we do not provide a general convolution routine, the computation of the convolution at equal times can be done safely. Using the same Langreth rules as in the right-hand side of Eq.~\eqref{KB_eq3A}, and the decay of either $A$ or $B$ outside the memory-truncated domain, we have 
\begin{align}
\rho_{AB}(nh)
&=i\xi_A 
\int_{(n-n_c)h}^{nh}d\bar{t}
\left(A^{R}(nh,\bar{t})B^{<}(\bar{t},nh)
+
A^{<}(nh,\bar{t})B^{R}(nh,\bar{t})^\dagger\right)
\\
&=i\xi_A 
h
\sum_{l=0}^{n_c}
w^{(n_c)}_{l}
\left(-A^{R}_{n,n-l} (B^{<}_{n,n-l})^\dagger
+
A^{<}_{n,n-l} (B^{R}_{n,n-l})^\dagger\right).
\label{density_matrix_AB_sum}
\end{align}
In the second equation, the Hermitian symmetry $B=B^\ddagger$ was assumed. 
All values needed to compute $\rho_{AB}(nh)$ are therefore in the Hermitian domain of $A$ or $B$.

\subsection{Main classes and routines for the memory-truncated KBE}
\vspace*{2mm}

\begin{table}[tbp]
    \centering
    \begin{small}
\begin{tabularx}{\textwidth}{|p{0.43\textwidth}|>{\raggedright\arraybackslash}X|}
        \hline
        \texttt{herm\_matrix\_moving<T>(int tc, int size1, int sig)} 
        &      
       Green's function on a memory-truncated moving window $\mathcal{M}[G]^{n_c}_\circ$ [Eq.~\eqref{time slice-moving}].
        \\ \hline
        \texttt{herm\_matrix\_timestep\_moving <T>(int tc, int size1, int sig)} 
        &         
       Green's function on a memory-truncated time slice $\mathcal{T}[G]^{n_c}_\circ$ [Eq.~\eqref{momorybtruncatde}].
        \\ \hline
        \texttt{function\_moving<T>(tc, size1)} 
        &         
        leading $n_c+1$ elements of a function $f(t)$.
        \\ \hline
\hline
\texttt{C.set\_[les|ret](i,j,M)} & Sets \( C^{<|R}(\circ-i,\circ-i-j) \) to \texttt{M}. \\
\hline
\texttt{C.get\_[les|ret](i,j,M)} & \texttt{M} is set to \( C^{<|R}(\circ-i,\circ-i-j) \). \\
\hline
\texttt{tC.set\_[les|ret](j,M)} & Sets \( C^{<|R}(\circ,\circ-j) \) to \texttt{M}. \\
\hline
\texttt{tC.get\_[les|ret](j,M)} & \texttt{M} is set to \( C^{<|R}(\circ,\circ-j) \). \\
\hline 
\texttt{F.set\_value(j,M)} & Sets \( f(\circ-j) \) to \texttt{M}. \\
\hline
\texttt{F.get\_value(j,M)} & \texttt{M} is set to \( f(\circ-j) \). \\
\hline        
    \end{tabularx}
    \end{small}
    \caption{{Upper part:} Explicit constructor for new main classes in \NESSitwo{}; {\tt size1} is the orbital matrix dimension of the objects, {\tt sig} the statistics sign of the Green's functions ($\pm 1$ for bosons/fermions). The template parameter {\tt T} denotes the precision ({\tt double} or {\tt float}). The constructors allocate the memory and initialize all elements with zero.  {Lower part:}   Read/write components of a  \texttt{herm\_matrix\_moving} object {\tt C} which stores the moving window $\mathcal{M}[G]^{n_c}_\circ$ with leading physical timestep $\circ$, a  \texttt{herm\_matrix\_timestep\_moving} object {\tt tC} which stores the moving window $\mathcal{T}[C]^{n_c}_\circ$ with leading physical timestep $\circ$, 
    and a \texttt{function\_moving}   object {\tt F} which stores elements of a time-dependent function $f(t)$. Indices must satisfy $0<i,j< ${\tt C.tc}. {\tt M} is a complex square matrix; in \texttt{get} routines, {\tt M} is resized to the correct dimensions \texttt{size1}, in \texttt{set} routines, {\tt M}  must have dimension \texttt{size1}. }
    \label{tab:nessi_trunc_cons}  
\end{table}

\subsubsection*{Main classes} 
\vspace*{2mm}

The functionality to treat memory-truncated Green's functions  in \NESSitwo{} is defined under the same namespace {\tt cntr} as the full Green's functions. The new main classes in \NESSitwo{} are  {\tt herm\_matrix\_moving}, which can store a memory-truncated moving window $\mathcal{M}[G]^{n_c}_n$, {\tt herm\_matrix\_timestep\_moving}, which stores  memory-truncated time slice  $\mathcal{T}[G]^{n_c}_n$, as well as {\tt function\_moving} to store a function $f(t)$ on times $t=nh,...,(n-n_c)h$ (Table \ref{tab:nessi_trunc_cons}). In the description below, we will refer to $n$ as the leading physical timestep of the objects. The classes in Table \ref{tab:nessi_trunc_cons} do not store $n$ explicitly. Instead, this information is assumed to be taken care of by the environment, such as a timestepping routine.  In the expressions defining the action of the routines, such as in Table \ref{tab:nessi_trunc_cons}, we therefore often indicate the leading timestep simply by the placeholder $\circ$. Times within these objects are instead addressed relative to the leading time: For example, if a \texttt{herm\_matrix\_moving} object {\tt G} stores the data $\mathcal{M}[G]^{n_c}_n$ attached to leading timestep $n$, the ``$j$th slice of  {\tt G}'' refers to the data at the physical time slice $\mathcal{T}[G]^{n_c}_{n-j}$.  This relative time access is in particular also used in the access to individual entries of the functions, see lower part of Table \ref{tab:nessi_trunc_cons}.

\subsubsection*{Forward move and initialization of the moving window} 
\vspace*{2mm}

To control the timestepping procedure, the most elementary routines are the forward move of the moving window, and the initialization. Initialization can be done via the member function {\tt set\_from\_G\_backward}, which initializes the moving window $\mathcal{M}[C]^{n_c}_{\circ}$ from timesteps $m,m-1,...,m-n_c$ of a given full Green's function $A$ (see Table \ref{tab:forward_move}), i.e.,
\begin{align}
\label{wegshakasl}
C^{R,<}(\circ-i,\circ-i-j)\leftarrow 
\begin{cases}
A^{R,<}(m-i,m-i-j) & \text{if} \quad m-i-j\ge 0,
\\
0 & \text{if} \quad m-i-j< 0.
\end{cases}
\end{align}
Here we omit the discretization step  $h$ for simplicity of notation, e.g. $C^<(a,b)\equiv C^<(ah,bh)$.
The initialization will usually be called with $m=n_c=$\,{\tt C.tc\_}, to start a timestepping based on a truncated time window from an non-truncated simulation up to time $m=n_c$. Note, that in this case only the triangle $\Delta [C]^{n_c}_n$ [Eq.~\eqref{time slice-tri}] is set, which is however sufficient to start the timestepping: 
\begin{lstlisting}[language=c++,numbers=none]
// Typical start of timestepping with truncated window tc=nc*h:
int size1 = 1, nc = 10;
int nt = nc, ntau = 100;
cntr::herm_matrix<double> G(nt, ntau, size1, FERMION);
// Perform initialization of G up to time nt=nc ...
cntr::herm_matrix_moving<double> G_trunc(nc, size1, FERMION);
G_trunc.set_from_G_backward(G, G, nc);
\end{lstlisting}

\begin{table}[tbp]
\centering
\begin{small}
\begin{tabularx}{\textwidth}{|p{0.3\textwidth}|X|}
\hline
\texttt{C.set\_from\_G\_backward (herm\_matrix\& A, herm\_matrix\& Acc, int m)} & Initialize the moving window $\mathcal{M}[C]^{n_c}_{\circ}$ represented by {\tt C} with the full Green's function $A$, see Eq.~\eqref{wegshakasl}; {\tt Acc} represents the Hermitian conjugate $A^\ddagger$; {\tt C.tc\_ <= m <= A.nt} required. 
\\
\hline
\texttt{C.forward()} &  If  {\tt C} stores the data $\mathcal{M}[C]^{n_c}_n$ with leading timestep $n$, after  the call to {\tt forward()}, ${\tt C}$ will represent $\mathcal{M}[C]^{n_c}_{n+1}$, with the new leading timestep $\mathcal{T}[C]^{n_c}_{n+1}$ replaced by the previous last time slice $\mathcal{T}[C]^{n_c}_{n-n_c}$.
 \\
\hline
\end{tabularx}
\end{small}
\caption{Initialization and forward move (referred to as advancing in Algorithm \ref{alg:moving}) of the moving window. Similar routines {\tt F.forward()} exist for the {\tt function\_moving} class.}
\label{tab:forward_move}
\end{table}

An alternative way to initialize memory-truncated Green's functions in a  time-translationally invariant way from steady-state Green's functions is discussed in Sec.~\ref{sec:ness:impl}.

If  a \texttt{herm\_matrix\_moving} object {\tt C} stores the data $\mathcal{M}[C]^{n_c}_n$ attached to leading timestep $n$, the member function \texttt{C.forward()}  is used to shift the time window forward by one step, such that the new leading timestep corresponds to the physical time $n+1$. For example, if {\tt C} represents a moving window with leading timestep $n$, the {\tt forward()} move implies
\begin{lstlisting}[language=c++,numbers=none]
C.get_les(i, j, M1);  // a matrix M1 set to C^<(n-i, n-i-j)
C.forward();
C.get_les(i+1, j, M2);  // M2 now equals M1, assuming 0 <= i < tc.
\end{lstlisting}

The forward shift is performed by cyclically reassigning pointers rather than by copying data. Through this, the data at the new leading timestep $\mathcal{T}[C]^{n_c}_{n+1}$ are effectively replaced by data at the previous last time slice $\mathcal{T}[C]^{n_c}_{n-n_c}$. In a timestepping algorithm, however, these data are usually replaced immediately after the forward move. A worked out example is given in Sec.~\ref{sec:impl:kbe}.

\subsubsection*{File access} 
\vspace*{2mm}

For the classes \texttt{herm\_matrix\_moving} and \texttt{herm\_matrix\_timestep\_moving}, we provide member functions \texttt{print\_to\_file} and \texttt{read\_from\_file} to store objects into human readable text files, and read from them (see Table \ref{tab:fileio}). For more efficient storage, the binary HDF5 format should be used, similar as for the full Green's functions of type {\tt herm\_matrix} in \NESSione{} (see Ref.~\cite{NESSi} for an explanation of the HDF5 file format). Particularly useful is the function {\tt write\_timeslice\_to\_hdf5} in Table \ref{tab:fileio}, which allows to store selected time slices of a Green's function during a memory-truncated evolution.  A worked out example is given in Sec.~\ref{sec:impl:kbe}. For post-processing analyses, we provide the Python module  \texttt{ReadCNTRhdf5}, which allows to read HDF5 Green's functions and time slices into standard arrays (see the example in Sec.~\ref{sec:impl:kbe}). HDF5 files can be created and opened using the HDF5 interface (see Ref.~\cite{NESSi}, as well as the online documentation). Use of HDF5  requires compilation of \texttt{libcntr} using the \texttt{hdf5=ON} flag (see Sec.~\ref{subsec:installation_libcntr}). 

\begin{table}[tbp]
\centering
\begin{small}
\begin{tabularx}{\textwidth}{|p{0.4\textwidth}|X|}
\hline
\texttt{C.print\_to\_file(const char *filename, int precision=16)} & 
Create a text file named \texttt{filename} and write the content of \texttt{C} into it. \\
\hline
\texttt{C.read\_from\_file(const char *filename)} & 
Initialize \texttt{C} with data from a text file previously written with \texttt{print\_to\_file}.  \\
\hline
\texttt{C.write\_to\_hdf5(ARGS)} &
Store \texttt{C}  in a HDF5 file defined by {\tt ARGS}. \\
\hline
\texttt{C.read\_from\_hdf5(ARGS)} &
Initialize \texttt{C}  from a HDF5 file defined by {\tt ARGS}. \\
\hline
\texttt{C.write\_timeslice\_to\_hdf5(int i, ARGS)} &
Store timeslice {\tt i} (relative to the leading time) of \texttt{C}  in a HDF5 file, in the format of a timestep object
(only for \texttt{herm\_matrix\_moving}).
 \\
\hline
\end{tabularx}
\end{small}
\caption{Selected member functions of \texttt{herm\_matrix\_moving} and  \texttt{herm\_matrix\_timestep\_moving} for writing to (reading from) text or HDF5 files. In the HDF5 variants, if {\tt ARGS } is {\tt (hid\_t fid}), the data is written to (read from) a HDF5 group with handle {\tt fid}. If {\tt ARGS} is { \tt(hid\_t fid, const char *g\_name}), the data is written to (read from) a sub-group with name {\tt g\_name} within the group {\tt fid} (for  {\tt write}, the sub-group is first created). If {\tt ARGS } is { \tt(const char *f\_name, const char *g\_name)}, the file {\tt f\_name} is created ({\tt write}) or opened for reading ({\tt read}), and the data is written to (read from) a sub-group with name {\tt g\_name}.
}
\label{tab:fileio}
\end{table}

\subsubsection*{Timestep-wise manipulation and access}
\vspace*{2mm} 

The {\tt libcntr} library provides a number of routines which allow to manipulate data on a whole time slice of the moving window $\mathcal{M}[C]_{n}^{n_c}$ or $\mathcal{T}[C]_{n}^{n_c}$, using data from another Green's function. The  basic syntax is 
\begin{align}
\text{\tt C.do\_something(int i,...,A,int j, ... )}
\end{align}
to perform an action on time slice $i$ (relative to the leading time slice) of $C$ using the data of $A$ at  time slice $j$ (relative to the leading time slice). For example, if {\tt C} and {\tt A} represent $\mathcal{M}[C]_{n}^{n_c}$ and $\mathcal{M}[A]_{m}^{n_c}$ at leading physical times $n$ and $m$ of $C$ and $A$ respectively, {\tt C.set\_timestep(int i,A,int j)} will copy the data such that $C^{R,<}(n-i,n-i-l)\leftarrow A^{R,<}(m-j,m-j-l)$ for the full slice ($l=0,...,n_c$), assuming that the size of the two objects is consistent ({\tt A.tc=\tt C.tc} and {\tt A.size1=\tt C.size1}).  More operations are listed in Table \ref{tab:copy_timeslices}. The syntax is similar to that for manipulations of full Green's functions of type {\tt herm\_matrix} in \NESSione{}.

In order to evaluate the difference between Green's functions, we provide a function {\tt distance\_norm2} to compute the $L_2$ norm distance between the time slices $i$ and $j$ (relative to the leading time slice) of two Green's functions (Table~\ref{tab:copy_timeslices})
\begin{align}
\label{distancenorm}
||A-B||_2^2 = \sum_{X=R,<} \sum_{l=0}^{n_c} \left|A^{X}(\circ-i,\circ-i-l)-B^{X}(\circ-j,\circ-j-l)\right|^2.
\end{align}
Moreover, routines to measure the density matrix \eqref{density_matrix_A} and the convolution density matrix \eqref{density_matrix_AB} can be convenient. 

\begin{table}[tbp]
\centering
\begin{small}
\begin{tabularx}{\textwidth}{|p{0.48\textwidth}|X|}
\hline 
\texttt{A.set\_timestep(i,B,j)} & Copy time slice {\tt j} of \texttt{B} into time slice \texttt{i} of  \texttt{A}.
$A^{<,R}(\circ-i,\circ-i-l)$ set to $ B^{<,R}(\circ-j,\circ-j-l)$ for $l=0,...,n_c$.  \\
\hline
\texttt{A.set\_matrixelement (i,i1,i2,B,j,j1,j2)} & Set matrix element {\tt (i1,i2)} of \texttt{A} at time slice \texttt{i}  to  element {\tt (j1,j2)} of \texttt{B} at time slice {\tt j}.
\\
\hline 
\texttt{A.incr\_timestep(i,B,j,alpha)} & 
Increment \texttt{A} at timestep \texttt{i} by \texttt{ alpha\,$\cdot$B} at time slice {\tt j} (scalar \texttt{alpha}).
\\
\hline
\texttt{A.set\_timeslice\_zero(i)} & Set the entire time slice \texttt{i} of \texttt{A} to zero.
\\ 
\hline
\texttt{A.smul(i,alpha)} & Multiply time slice \texttt{i} of \texttt{A} by scalar \texttt{alpha}.
\\
\hline
\texttt{distance\_norm2(A,i,B,j)}&
Return difference norm \eqref{distancenorm} between time slice \texttt{i} of \texttt{A} and time slice \texttt{j} of \texttt{B}.
\\
\hline
\texttt{density\_matrix(rho,A)} &
{\tt rho} (complex matrix) is set to $\rho_{A}$ (Eq.~\eqref{density_matrix_A}), at the leading timestep of $A$.
\\
\hline
\texttt{convolution\_density\_matrix(rho,A,B,h)} &
{\tt rho} $\to$ $\rho_{AB}$ (Eq.~\eqref{density_matrix_AB}), at the leading timestep of $A$ and $B$; $h$ is the time-discretization.
\\ 
\hline
\end{tabularx}
\end{small}
\caption{Selected routines for simple manipulations of Green's functions at a given time slice. Both objects {\tt A} and {\tt B} can be of type \texttt{herm\_matrix\_moving}, \texttt{herm\_matrix\_timestep\_moving}. If {\tt A} ({\tt B}) is of type \texttt{herm\_matrix\_timestep\_moving}, the respective argument {\tt i} ({\tt j}) is omitted. {\tt i} and {\tt j} are understood relative to the leading timestep, as shown for the first line in the table.
}
\label{tab:copy_timeslices}
\end{table}

\subsubsection*{Diagram utilities} 
\vspace*{2mm}

We provide two basic functions that compute particle-particle and particle-hole bubbles on a given timestep, and multiply Green's functions with time-dependent functions, similar as for {\tt herm\_matrix} objects in \NESSione{} (Table \ref{tab:diagram}). The provided diagram utilities follow the syntax for the corresponding routines for types {\tt herm\_matrix} in \NESSione{}, omitting only the timestep argument, since the operations are always performed on the leading time slice. For example, {\tt C.right\_multiply(f)} will set $C^{<,R}(\circ,\circ-l)$ to  $C^{<,R}(\circ,\circ-l)f(\circ-l)$ for the full slice $l=0,\ldots,n_c$. A worked out example, which evaluates a second-order self-energy for a general time-dependent
interaction can be found in the example Sec.~\ref{sec:impl:res}. 

\begin{table}[tbp]
\centering
\begin{small}
\begin{tabularx}{\textwidth}{|p{0.45\textwidth}|X|}
\hline
\texttt{cntr::Bubble1(C,c1,c2,A,Acc, a1,a2, B,Bcc,b1,b2)} &
Particle-hole bubble \eqref{bubble1}: $C_{c1,c2}(t,t')$ set to $ i A_{a1,a2}(t,t') \, B_{b2,b1}(t',t)$. \\
\hline
\texttt{cntr::Bubble2(C,c1,c2,A,Acc, a1,a2, B,Bcc,b1,b2)} &
Particle-particle bubble \eqref{bubble2}: $C_{c1,c2}(t,t') $ set to $ i A_{a1,a2}(t,t') \, B_{b1,b2}(t,t')$. \\
\hline \hline
\texttt{A.left\_multiply (f)} & Set $A(t,t') \rightarrow f(t) A(t,t')$. \\
\hline
\texttt{A.right\_multiply(f)} & Set $A(t,t') \rightarrow A(t,t') f(t')$. \\
\hline
\texttt{A.left\_multiply\_hermconj(f)} & Set $A(t,t') \rightarrow f(t)^\dagger A(t,t')$. \\
\hline
\texttt{A.right\_multiply\_hermconj(f)} & Set $A(t,t') \rightarrow A(t,t') f(t')^\dagger$. \\
\hline
\end{tabularx}
\end{small}
\caption{Diagram utilities: Objects {\tt A}, {\tt B}, {\tt Acc}, {\tt Bcc}, {\tt C}  can be of type \texttt{herm\_matrix\_moving}, \texttt{herm\_matrix\_timestep\_moving}, and {\tt f} is \texttt{function\_moving}.  The operations are always performed on the leading timestep. \texttt{Xcc} contains the Hermitian conjugate \( X^\ddagger \) of \( X \). If \texttt{Xcc} is omitted, \( X = X^\ddagger \) is assumed.
}
\label{tab:diagram}
\end{table}

\subsubsection*{Integral equations} 
\vspace*{2mm}

The timestepping solution of the Dyson equation \eqref{intdif_eq} and the integral equation \eqref{vie2} has been described in Sec.~\ref{kbetheory}. The function calls are summarized in Table \ref{tab:dyson_vie2}. An explicit example is provided in Sec.~\ref{sec:movingexample}.

\begin{table}[tbp]
\centering
\begin{small}
\begin{tabularx}{\textwidth}{|p{0.4\textwidth}|X|}
\hline
\texttt{cntr::dyson\_timestep (herm\_matrix\_moving \&G, herm\_matrix\_moving \&Sigma, function\_moving<T> \&H, T mu, int SolveOrder, T dt)} &
Solves the truncated Dyson equation \eqref{intdif_eq} for the Green's function \texttt{G} at the leading timestep.  \texttt{H} represents the single-particle energy $\epsilon(t)$ in Eq.~\eqref{intdif_eq}.  See Sec.~\ref{kbetheory} for details.  \\
\hline\hline
\texttt{cntr::vie2\_timestep (herm\_matrix\_moving \&G, herm\_matrix\_moving \&F, herm\_matrix\_moving \&Fcc, herm\_matrix\_moving \&Q, int SolveOrder, T dt)} &
Solves the linear Volterra integral equation of second kind, Eq.~\eqref{vie2}, on the leading timestep. 
{\tt F} and {\tt Fcc} represent the kernel $F$ and its Hermitian conjugate $F^\ddagger$, respectively. See Sec.~\ref{kbetheory} for details. \\
\hline
\end{tabularx}
\end{small}
\caption{Routines for solving truncated Dyson and {\tt vie2} equations via timestepping methods. 
 \texttt{SolveOrder} is the accuracy of the integration routines; use {\tt MAX\_SOLVE\_ORDER} ($=5$) by default. \texttt{dt} is the real-timestep size $h$. }
\label{tab:dyson_vie2}
\end{table}

\section{Steady-state NEGFs on the Keldysh contour}
\label{sec:steadystate}

\subsection{Nonequilibrium steady-state Green's functions}
\label{sec:steadystategf}
\vspace*{2mm}

Another framework for solving the KBEs is the Keldysh formalism \cite{Keldysh_1964} for NESSs. In comparison to two-time functions defined on the three-legged KB contour $ \mathcal C = \mathcal C_1 \cup \mathcal C_2 \cup \mathcal C_3$ \cite{NESSi}, the main assumption in the steady-state formalism is the 
absence of correlations with the initial equilibrium state represented by the imaginary time branch 
$\mathcal C_3$, such that the mixed self-energies $\Sigma^{\rceil}$ and $\Sigma^{\lceil}$ vanish. The vertical  branch can then be shifted to $t = - \infty$ and eliminated from the equations, such that time arguments are restricted to the two-branch contour $\mathcal C_K = \mathcal C_1 \cup \mathcal C_2$ \cite{Aoki_2014}. Second, any contour-ordered two-time Green's function $G(t,t') = -i \langle \mathcal{T}_{\mathcal C} c(t) c^{\dagger}(t') \rangle$ in a  NESS exhibits time-translational invariance $G(t,t') = G(t-t')$, so that a  numerical treatment in Fourier representation is possible.

To represent NESS Green's functions, we choose the retarded component $G^R(t)$ and lesser component $G^<(t)$   on an equidistant real-time grid as the two non-redundant components. The Hermitian symmetry $G^\ddagger=G$ (cf.~Eq.~\eqref{eq:hermconjg}) for steady-state Green's functions  implies 
\begin{align}
\label{herm-symmetry-steady}
G^{</>}(t) = -G^{</>}(-t)^{\dagger}, \quad G^R(t) = G^A(-t)^{\dagger},
\end{align} 
and  is assumed for all functions if not stated otherwise. Green's functions in the time and frequency domain are related as follows,
\begin{align}
\label{gw2t01}
G^R(t) &= -i\theta(t) \int
d\omega \, A(\omega) e^{-i\omega t},
\\
\label{gw2t02}
G^R(\omega+i0) &= \int_0^\infty 
dt \, G^R(t) e^{i(\omega+i0) t},
\\
\label{gw2t03}
G^<(t) &= \int 
\frac{d\omega}{2\pi} \, G^<(
\omega) e^{-i\omega t},
\\
\label{gw2t04}
G^<(\omega) &= \int 
dt \, G^<(t) e^{i\omega t},
\end{align}
with the spectral function 
\begin{align}
A(\omega) =-\frac{1}{2\pi i} \left(G^R(\omega+i0)-G^R(\omega+i0)^\dagger\right).
\end{align} 
In equilibrium at temperature $1/\beta$, the fluctuation-dissipation relation implies
\begin{align}
\label{eq:eq:fdt}
G^<(\omega)= -\xi 2\pi i F_\mu(\omega) A(\omega),
\end{align} 
where the sign $\xi$ is $\xi=\pm1$ for bosonic (fermionic) Green's functions, and the distribution function is  $F_\mu(\omega) =1/(e^{\beta(\omega-\mu)}-\xi)$.

\subsection{Steady-state Dyson equation} \label{section:steady_state_dsyon}
\vspace*{2mm}

In order to solve the Dyson equation  \eqref{intdif_eq_general_notation} in the steady state, we restrict Eqs.~\eqref{KB_eq1} and \eqref{KB_eq2} to the time-translationally invariant case. The first equation becomes
\begin{align}
[i \partial_t - \epsilon] G^R(t) - \int_{0}^{t} d\bar t  \, \Sigma^R(t - \bar t)G^R(\bar t) = \delta(t),
\label{dyson-rett}
\end{align}
which implies
\begin{align}
G^R(\omega+i0^{+}) &= \big[\omega+i0^{+}  - \epsilon-\Sigma^R(\omega+i0^{+})\big]^{-1} 
\label{eq:ness_dyson_1} 
\end{align}
in Fourier representation. Equivalently, one could use the conjugate of \eqref{dyson-rett},
\begin{align}
i \partial_t G^R(t) -  G^R(t) \epsilon  - \int_{0}^{t} d\bar t  \, G^R(t - \bar t)\Sigma^R(\bar t) = \delta(t),
\label{dyson-rettconj}
\end{align}
to derive the same equation \eqref{eq:ness_dyson_1}.
Equation \eqref{KB_eq2} becomes (using time translational invariance)
 \begin{align}
\left[i\partial_t-\epsilon\right]G^{<}(t)-\int_{-\infty}^{t}\!\!d\bar{t}\,\Sigma^{R}(t-\bar{t})G^{<}(\bar{t})  =
\int_{-\infty}^{0}\!\!d\bar{t}\,\Sigma^{<}(t-\bar{t})G^{A}(\bar{t}).
\end{align}
Convolution of this equation from the left with $G^R$, together with a partial integration and 
Eq.~\eqref{dyson-rettconj}, results in 
\begin{align}
G^<(t) - \lim_{\tau\to-\infty}iG^R(t-\tau)G^<(\tau)
 =  \int_{-\infty}^{t} \!\!\!ds \int_{-\infty}^{0}  ds' G^R(t-s) \Sigma^<(s-s') G^A(s').
\label{dyson-less-01}
\end{align}
The second boundary term on the left hand side of the equation vanishes if the functions decay at infinite time. The latter does not hold if there is an exact pole $\sim C \delta(\omega-\epsilon_0)$ in the spectrum, which implies an asymptotic behavior  $G^R(t) \sim -i Ze^{-i\epsilon_0 t}$ and  $G^R(t) \sim  iZ^<e^{-i\epsilon_0 t}$ with constants $Z,Z^<$. In the numerical implementation, we must ensure the decay of the functions within the simulation interval, possibly with a proper regularization (see below). With this, the lesser component is determined by the double convolution
\begin{align}
G^<(t)
 =  \int_{-\infty}^{t} \!\!\!ds \int_{-\infty}^{0}  ds' G^R(t-s) \Sigma^<(s-s') G^A(s')
\label{dyson-less}
\end{align}
or its Fourier representation
\begin{align}
G^<(\omega) &= G^R(\omega) \Sigma^<(\omega) G^A(\omega). \label{eq:ness_dyson_2}
\end{align}

\subsubsection*{Regularization}
\vspace*{2mm}

If the functions $G^{R,<}(t)$ do not decay for large times, the equations must be regularized such that the boundary term in \eqref{dyson-less-01} vanishes and the infinite-time integrals in \eqref{dyson-less} become well-defined. This is in particular necessary if $G^{R}(\omega+i0)$ has a pole at some finite $\omega_p$. In general, a physically well-defined regularization corresponds to adding an additional contribution $\Sigma_{\rm reg}$ to the self-energy, which can represent some kind of dissipative environment. Possible simple choices for fermionic self-energy functions are
\begin{align}
\label{bath-const}
\Sigma_{\rm reg}^R(\omega) = -i\eta
\end{align}
or a Gaussian density of states (DOS)
\begin{align}
\label{bath-gauss}
\Sigma_{\rm reg}^R(\omega) = -i\eta e^{-\omega^2/\omega_{\rm c}^2}
\end{align}
with a sufficiently large cutoff $\omega_{\rm c}$.
Together with a lesser component 
\begin{align}
\label{bath-less-fermion}
\Sigma_{\rm reg}^<(\omega) = \frac{-2 i\text{Im} \Sigma_{\rm reg}^R(\omega)}{e^{\beta(\omega-\mu)}+1},
\end{align}
these self-energies represent a coupling to a noninteracting particle reservoir at inverse temperature $\beta$ and chemical potential $\mu$.  For Green's functions of real bosonic fields, a typical regularization can be a self-energy with linear (Ohmic) DOS
\begin{align}
\label{bath-ohmic}
\Sigma_{\rm reg}^R(\omega) = -i\eta \omega e^{-\omega^2/\omega_{\rm c}^2},
\quad
\Sigma_{\rm reg}^<(\omega) = \frac{ 2 i\text{Im} \Sigma_{\rm reg}^R(\omega)}{e^{\beta\omega}-1}.
\end{align}
Due to the Kramers-Kronig relation, a frequency dependent imaginary part Im$\Sigma_{\rm reg}^R(\omega)$ implies a nonzero real part Re$\Sigma_{\rm reg}^R(\omega)$; Eqs.~\eqref{bath-const},  \eqref{bath-gauss} and \eqref{bath-less-fermion} assume that  Re$\Sigma_{\rm reg}^R(\omega)$ is approximately constant over the relevant frequency range, and is compensated by a static renormalization of the bare level energies. This approximation is justified for large cutoff $\omega_c$  (for Eq.~\eqref{bath-const}, the cutoff is given by the extent of the numerical frequency grid). 

\subsection{Numerical implementation }
\vspace*{2mm}
\label{sec:ness:impl}

\NESSitwo{} provides classes to represent steady-state Green's functions, which can be interfaced with the real-time Green's functions. The steady-state functionality is defined under a namespace {\tt ness2}. The namespace {\tt ness} was reserved for an earlier version of the implementation, which has less functionality and does not interface with the two-time functions, and is not described here. A new namespace was chosen for clarity, although there are no major name ambiguities. In the functions and examples below, the namespace {\tt ness2} is not explicitly indicated.

\subsubsection*{Plain Fourier transforms: \tt fft\_array}
\vspace*{2mm}

The class {\tt fft\_array} is a basic data container which contains two arrays {\tt C.time\_} and {\tt C.freq\_} of {\tt Nft\_} square matrices (stored as consecutive
arrays
of complex numbers), as well as plans to compute the  (non-normalized) discrete Fourier transform
\begin{align}
\mathcal{F}
\left\{C\right\}[l]
= 
\sum_{j=0}^{N_{\rm ft}-1} e^{i2\pi jl/N_{\rm ft}} 
C[j],
\label{tofreq}
\\
\bar{\mathcal{F}}
\left\{C\right\}[l]
=
\sum_{j=0}^{N_{\rm ft}-1} e^{-i2\pi jl/N_{\rm ft}} 
C[j],
\label{totime}
\end{align}
 using the FFTW algorithm \cite{FFTW}; $N_{\rm ft}$ should be a power of $2$ to allow for the most efficient FFT.  Routines for standard element access and simple algebra operations are summarized in Table \ref{tab:fftarray}. The {\tt fft\_array} basically represents an array which is periodic in time and frequency, with a period {\tt Nft\_}. This periodic wrapping is automatically taken into account when addressing elements of the array. 
\begin{table}[h]
\centering
\begin{small}
\begin{tabularx}{\textwidth}{|p{0.5\textwidth}|>{\raggedright\arraybackslash}X|}
\hline
\texttt{ fft\_array(Nft, size)} & Construct \texttt{fft\_array} with {\tt Nft} matrices of dimension {\tt size} $\times$ {\tt size}.\\
\hline
\texttt{ C.fft\_to\_time()} & Set $C_{\rm time}$ to $\bar{\mathcal{F}}\{C_{\rm freq}\}$ using Eq.~\eqref{totime} and FFT.
\\
\hline
\texttt{ C.fft\_to\_freq()} & Set $C_{\rm freq}$ to ${\mathcal{F}}\{C_{\rm time}\}$ using Eq.~\eqref{tofreq} and FFT.
\\
\hline
\texttt{C.set\_element(i,M,domain)} & \( C_{\rm time|freq}[i] \) set to matrix \texttt{M}. 
\\
\hline
\texttt{C.get\_element(i,M,domain)} &  Matrix \texttt{M} set to  \( C_{\rm time|freq}[i] \).
\\
\hline
\texttt{C.incr(fft\_array \&B, cplx a, domain)} &
$C_{\rm time|freq}\rightarrow C_{\rm time|freq}+aB_{\rm time|freq}$.
\\
\hline
\texttt{C.smul (cplx a, domain)} & $C_{\rm time|freq}\rightarrow aC_{\rm time|freq}$. \\
\hline
\texttt{C.set\_zero( domain) } & $C_{\rm time|freq}\rightarrow 0$. \ \\
\hline
\texttt{C.set\_matrixelement(i1,i2, fft\_array \&B, j1,j2, domain)} & 
$(C_{\rm time|freq})_{i_1,i_2}\rightarrow (B_{\rm time|freq})_{j_1,j_2}$.\\
\hline
\texttt{C.left\_multiply (M, domain)} & $C_{\rm time|freq} \rightarrow MC_{\rm time|freq}$. \\
\hline
\texttt{C.right\_multiply (M, domain)} & $C_{\rm time|freq} \rightarrow C_{\rm time|freq}M$. \\
\hline
\texttt{C.left\_multiply\_hermconj(M, domain)} & $C_{\rm time|freq} \rightarrow M^\dagger C_{\rm time|freq}$. \\
\hline
\texttt{C.right\_multiply\_hermconj(M, domain)} & $C_{\rm time|freq} \rightarrow C_{\rm time|freq}M^\dagger$. \\
\hline
\end{tabularx}
\end{small}
\caption{
Some member functions of  {\tt fft\_array}. The constructor has an optional third argument  {\tt FFTW\_FLAG}, which chooses the plan for the construction of the plan in the FFTW library used to do the Fourier transforms (default is {\tt FFTW\_ESTIMATE}). The argument {\tt domain} in the element access can be {\tt fft\_domain::time} or {\tt fft\_domain::freq} to indicate whether the respective operation is done on $C_{\rm time}$ or $C_{\rm freq}$. Routines other than {\tt [set|get]\_element} apply to all {\tt Nft} entries. {\tt M} is a Matrix type (complex eigen 
matrices are supported). In the {\tt get} methods, {\tt M} is automatically resized. In  {\tt [set|get]\_element}, time (frequency) arguments $i$ outside the interval $[0,N_{\rm ft}-1]$ are mapped back to this interval by adding an integer multiple of $N_{\rm ft}$.
}
\label{tab:fftarray}
\end{table}

\subsubsection*{Hermitian steady-state Green's functions: \tt herm\_matrix\_ness}
\vspace*{2mm}

The data type {\tt herm\_matrix\_ness} is used to represent a steady-state function $G$ with Hermitian symmetry \eqref{herm-symmetry-steady}. Non-Hermitian Green's functions (which would, e.g., appear in a convolution $A\ast B$ of two matrix-valued objects), can be avoided in a large class of steady-state applications and are therefore not supported in the current extension of {\tt libcntr}. In contrast to two-time functions like  {\tt cntr::herm\_matrix}, a non-Hermitian Green's function cannot simply be represented by two objects of type {\tt herm\_matrix\_ness}, because in the current implementation the two objects would have to be combined when switching between frequency and time representations. 

An object of type {\tt herm\_matrix\_ness} contains a pair of two members  {\tt ret\_} and {\tt les\_} of type {\tt fft\_array}, representing $G^R$ and $G^<$ in time and frequency. Time arguments thereby correspond to an equidistant grid
\begin{align}
\label{t-grid}
t\in \{jh: j=-N_{\rm ft}/2,...,N_{\rm ft}/2-1\},
\end{align}
and the frequencies represent the dual FFT grid
\begin{align}
\label{w-grid}
\omega \in  \{ \Delta_\omega j: j= -N_{\rm ft}/2,...,N_{\rm ft}/2-1,\Delta_\omega= 2\pi/(hN_{\rm ft})\}.
\end{align}
We require $N_{\rm ft}$  to be even, and use the conventional layout of the arrays where $G^{<|R}(jh)$ is represented by the $j$th element of {\tt G.[les|ret]\_.time\_},  if $0\le j< N_{\rm ft}/2$, and by the  $(j+N_{\rm ft})$th  element if  $-N_{\rm ft}/2\le j< 0$ (analogous for frequency).  For convenience we provide a small helper class {\tt fft\_grid}, with constructor  {\tt fft\_grid grid(Nft,h)}. {\tt grid.time\_at(j)} then returns the time at index $j$, which is $jh$ for $j < N_{\rm ft}/2-1$ and $(j-N_{\rm ft})h$ for $j \ge N_{\rm ft}/2$ (i.e. mapped into the set \eqref{t-grid}), and {\tt grid.freq\_at(j)} returns $j\Delta_\omega$, with $j$ mapped into the set \eqref{w-grid} in the same way. Note, that for Hermitian Green's functions it is in principle redundant to store both negative and positive time data, but we nevertheless do save the whole domain in order to allow for an efficient implementation of the Fourier transform without extraneous operations on the data. 

\begin{table}[tbp]
\centering
\begin{small}
\begin{tabularx}{\textwidth}{|p{0.38\textwidth}|>{\raggedright\arraybackslash}X|}
\hline
\texttt{ herm\_matrix\_ness(Nft, size)} & Construct \texttt{ herm\_matrix\_ness} with Fourier domain size  {\tt Nft\_} and matrix dimension {\tt size}, optional third argument  {\tt FFTW\_FLAG}. \\
\hline
\texttt{C.set\_[les|ret](i,M,domain)} & Set \( C_{\rm time|freq}^{<|R}[i] \) to \texttt{M}.
\\
\hline
\texttt{C.get\_[les|ret](i,M,domain)} & \texttt{M}  is set to \( C_{\rm time|freq}^{<|R}[i] \).  \\
\hline
\texttt{C.retarded()} & Return reference to {\tt C.ret\_}.
\\
\hline
\texttt{C.lesser()} & Return reference to {\tt C.les\_}.
\\
\hline
\end{tabularx}
\end{small}
\caption{
Most important members of the  \texttt{ herm\_matrix\_ness} class. As for {\tt fft\_array}, the argument {\tt domain} can be {\tt fft\_domain::time} or {\tt fft\_domain::freq} to indicate whether the respective operation is done on $C_{\rm time}$ or $C_{\rm freq}$.   In addition, routines \texttt{incr}, \texttt{smul}, \texttt{set\_zero}, \texttt{set\_matrixelement}, \texttt{left\_multiply}, \texttt{right\_multiply}, \texttt{left\_multiply\_hermconj}, \texttt{right\_multiply\_hermconj} extend to both components {\tt les\_} and {\tt ret\_} with the same syntax as for {\tt fft\_array}.
}
\label{tab:hermatroixness}
\end{table}

The constructor of {\tt herm\_matrix\_ness} and important member functions are summarized in Table \ref{tab:hermatroixness}. As for {\tt fft\_array},  in the individual element access routines {\tt [set|get]\_[ret|les]}, time (frequency) arguments $i$ outside the interval $[0,N_{\rm ft}-1]$ are mapped back to this interval by adding an integer multiple of $N_{\rm ft}$. For example  {\tt C.set\_les(j,M,fft\_domain::time)} will set $C^<(jh)$ to $M$ when  $j\in[-N_{\rm ft}/2,...,N_{\rm ft}/2-1]$ corresponding to the grid \eqref{t-grid}.

\subsubsection*{File access}
\vspace*{2mm}

For \texttt{herm\_matrix\_ness} we also provide the member functions \texttt{print\_to\_file} and \texttt{read\_from\_file} to write to and read from human readable text files, while for HDF5 we provide the functions \texttt{print\_to\_hdf5} and \texttt{read\_from\_hdf5}, see the usage in the example program Sec.~\ref{sec:nessexample1}. The Python module \texttt{ReadNESS} contains tools for post-processing in the steady-state code, which allow for reading HDF5 Green's functions into standard arrays (also see Sec.~\ref{sec:nessexample1}). As for \NESSione{} and the truncated code, HDF5 files can be created and opened using the HDF5 interface (see Ref.~\cite{NESSi}, as well as the online documentation) and compilation of \texttt{libcntr} using the \texttt{hdf5=ON} flag (see Sec.~\ref{subsec:installation_libcntr}) is a requirement. 

\begin{table}[tbp]
\centering
\begin{small}
\begin{tabularx}{\textwidth}{|p{0.4\textwidth}|X|}
\hline
\texttt{C.print\_to\_file(filename, precision = 12)} & 
Create a text file named \texttt{filename} and write the content of \texttt{C} into it. \\
\hline
\texttt{C.read\_from\_file(filename, FFTW\_FLAG = FFTW\_ESTIMATE)} & 
Initialize \texttt{C} with data from a text file previously written with \texttt{print\_to\_file}.  \\
\hline
\texttt{C.write\_to\_hdf5(ARGS)} &
Store \texttt{C}  in a HDF5 file defined by {\tt ARGS}. \\
\hline
\texttt{C.read\_from\_hdf5(ARGS)} &
Initialize \texttt{C}  from a HDF5 file defined by {\tt ARGS}. \\
\hline
\end{tabularx}
\end{small}
\caption{Selected member functions of \texttt{herm\_matrix\_ness} for writing to (reading from) text or HDF5 files. In the HDF5 variants, if {\tt ARGS } is {\tt (hid\_t fid}), the data is written to (read from) a HDF5 group with handle {\tt fid}. If {\tt ARGS} is { \tt(hid\_t fid, const char *g\_name}), the data is written to (read from) a sub-group with name {\tt g\_name} within the group {\tt fid} (for  {\tt write}, the sub-group is first created). If {\tt ARGS } is { \tt(const char *fame, const char *g\_name)}, the file {\tt f\_name} is created ({\tt write}) or opened for reading ({\tt read}), and the data is written to (read from) a sub-group with name {\tt g\_name}.
}
\label{tab:nessfileio}
\end{table}

\subsubsection*{Integral transforms of {\tt herm\_matrix\_ness}}
\vspace*{2mm}

For the Fourier transform of Green's functions, we must distinguish between the plain discrete Fourier transforms \eqref{tofreq} and \eqref{totime}, which can be accessed by the call {\tt [ret|les]\_.fft\_to\_[time|freq]()} to the members of  {\tt herm\_matrix\_ness}, and approximations to the integral transforms \eqref{gw2t01} to \eqref{gw2t04}.
The access to discrete integral Fourier transforms of Green's functions is summarized in Table \ref{tab:integrals_ness}.

\begin{table}[tbp]
\centering
\begin{small}
\begin{tabularx}{\textwidth}{|p{0.45\textwidth}|X|}
\hline
\texttt{C.integral\_transform\_to\_freq(h, METHOD)}
&
Compute the frequency-domain Fourier integral of the Green's function \eqref{gw2t02} and \eqref{gw2t04}, assuming a timestep $h$.\\
\hline
\texttt{C.integral\_transform\_to\_time(h, METHOD)} &
Compute the time-domain Fourier integral of the Green's function  \eqref{gw2t01} and \eqref{gw2t03}, assuming a timestep $h$. \\
\hline 
\end{tabularx}
\end{small}
\caption{
Integral transformations of the Green's functions (see main text for explanations).}
\label{tab:integrals_ness}
\end{table}

The transformation {\tt C.integral\_transform\_to\_freq(h, METHOD)} computes an approximation to the integral \eqref{gw2t02} and  \eqref{gw2t04} on the frequency grid \eqref{w-grid}, 
assuming a time grid \eqref{t-grid} with timestep $h$. The integrals are computed with boundaries $[0,t_{\rm c}]$ (for $G^R$) and $[-t_{\rm c},t_{\rm c}]$ (for $G^<$), with $t_{\rm c}=(N_{\rm ft}/2-1)h$. The general discrete representation of the integral is 
\begin{align}
G^{R,<}(\omega_j) =  h\phi (h\omega_j) \mathcal{F}\{G^{R,<}\}[j] + BC(\omega_j),
\end{align}
where $\mathcal{F}\{G^{R,<}\}$ denotes the FFT \eqref{tofreq}, and $ \phi$ and $BC$ are weighting functions and boundary corrections, respectively.  If the argument {\tt METHOD} is omitted or given by the keyword {\tt FFT\_TRAPEZ} (default), we compute the Fourier integral with the trapezoidal rule, where  $\phi=1$, and $BC=(h/2) G^R_{\rm time}[0]$ for the retarded component due to the lower bound ($t=0$) of the integral  \eqref{gw2t02}. As we assume all functions to sufficiently decay for large times, there is no further boundary correction from $t=\pm t_{\rm c}$. The reverse transformation {\tt C.integral\_transform\_to\_time(h, METHOD)} computes an approximation to the integral \eqref{gw2t01} and  \eqref{gw2t03} on the time grid \eqref{t-grid}. For both directions, we also provide an implementation {\tt METHOD=FFT\_CUBIC}, where the integrals are approximated by the exact Fourier transform of a piecewise cubic interpolating function, as discussed in Chapter 13.9 of Ref.~\cite{NumericalRecipes}. There are no boundary corrections, because the integrand typically decays exponentially at the boundaries of the interval \eqref{w-grid}. In many cases, however, this does not provide an improvement over the trapezoidal rule summation.

\subsubsection*{Equilibrium Green's functions}
\vspace*{2mm}

Analogous to real-time Green's functions in {\tt libcntr}, we provide routines that allow to compute equilibrium Green's functions for a given DOS, solving the integrals \eqref{gw2t01} and \eqref{gw2t03} with a DOS $A(\omega)$ and the equilibrium relation \eqref{eq:eq:fdt}, see Table~\ref{tab:diagram_ness_equi}. If the argument {\tt METHOD} is {\tt FFT\_TRAPEZ}, the integral is computed as in {\tt integral\_transform\_to\_time} (Table \ref{tab:integrals_ness}), after initializing the imaginary part of the frequency-dependent components {\tt ret\_.freq\_} and {\tt les\_.freq\_} using the {\tt dos} function on the grid \eqref{w-grid}. Note, that the routine is implemented only for scalar $A(\omega)$; if $G$ is matrix-valued, it is set to a diagonal matrix containing the values obtained with $A(\omega)$. Alternatively, if {\tt METHOD = FFT\_ADAPTIVE}, the Fourier integrals are computed using a cubically corrected Fourier transform of {\tt libcntr}, with a subdivision of the integration domain of $A(\omega)$ in at most {\tt limit} intervals of {\tt nn} points (default is {\tt limit=100} and {\tt nn=20}) in order to reach the desired accuracy. Hence, the adaptive Fourier integration is more accurate but slower, because it does not exploit the FFT algorithm. 

Moreover, we provide a function {\tt force\_equilibrium} (see Table~\ref{tab:diagram_ness_equi}), which does a Fourier transform of $G^R(t)$ to frequency (analogous to {\tt integral\_transform\_to\_freq} with {\tt METHOD = FFT\_TRAPEZ}), uses Eq.~\eqref{eq:eq:fdt} to initialize $G^<(\omega)$ on the grid \eqref{w-grid}, and transforms $G^<(\omega)$ to $G^<(t)$ analogous to {\tt transform\_to\_time}.

\begin{table}[H]
\centering
\begin{small}
\begin{tabularx}{\textwidth}{|p{0.45\textwidth}|X|}
\hline
\texttt{void green\_equilibrium\_ness(sign, G, DOS \&dos, beta, mu,h, METHOD, limit=100, nn=20)} &
Set $G^{<|R}(t)$ to equilibrium Green's function for bosons ({\tt sign=+1}) or fermions  ({\tt sign=-1}).\\
\hline
\texttt{G.force\_equilibrium(sign, beta, mu, h)} &
Set  $G^<(t)$ according to Eqs.~\eqref{gw2t03} and \eqref{eq:eq:fdt} for bosons ({\tt sign=+1}) or fermions  ({\tt sign=-1}), with $A(\omega)$ from $G^R(t)$.
\\
\hline
\end{tabularx}
\end{small}
\caption{Routines to construct an equilibrium NESS Green's function {\tt G} of type {\tt herm\_matrix\_ness} from Eqs.~\eqref{gw2t01}, \eqref{gw2t03}, and \eqref{eq:eq:fdt}, for inverse temperature {\tt beta} and chemical potential {\tt mu}. {\tt DOS} is a class representing $A(\omega)$, which provides an operation {\tt dos(double omega)} to return $A(\omega)$, and the numbers {\tt dos.lo\_} and {\tt dos.hi\_} representing the lower and upper bound of the support of $A(\omega)$.  One may use keywords {\tt BOSON} and {\tt FERMION} for the sign $\pm 1$. {\tt METHOD} can be {\tt FFT\_TRAPEZ} or {\tt FFT\_ADAPTIVE}, see text; for {\tt FFT\_TRAPEZ}, the arguments {\tt nn} and {\tt limit} are ignored.
 }
\label{tab:diagram_ness_equi}
\end{table}

\subsection*{Data exchange with two-time functions}
\vspace*{2mm}

Data exchange between two-time and steady-state functions is enabled by the functions in Table \ref{tab:conversion}. The function {\tt cntr2ness} allows to set the {\tt time\_} data of a steady-state function $G_{\rm ness}$   from a given time slice $t_0$ of a two-time object $G$, such that  $G_{\rm ness}^{R|<}(t)\leftarrow G_{\rm cntr}^{R|<}(t_0,t_0-t)$ for  all $t$. At negative times $t$, $G_{\rm ness}$ is set assuming Hermitian symmetry; if {\tt tstp} is smaller than the maximum time   $N_{\rm ft}/2-1$ in $G_{\rm ness}$, the remaining entries in $G_{\rm ness}$ are left zero. The reverse function {\tt ness2cntr} sets $G_{\rm cntr}^{R|<}(t,t')\leftarrow G_{\rm ness}^{R|<}(t-t')$ for all arguments $(t,t')$ where $t-t'$ is in the domain of $G_{\rm ness}$, and zero otherwise. \\
\begin{table}[H]
\centering
\begin{small}
\begin{tabularx}{\textwidth}{|p{0.4\textwidth}|X|}
\hline
\texttt{cntr2ness(Gness, Gcntr, tstp)} &
Set values  $G^<(t)$ and $G^R(t)$ of a {\tt herm\_matrix\_ness} object {\tt Gness} from time slice  {\tt tstp} of a two-time function {\tt Gcntr}. \\
\hline
\texttt{ness2cntr(Gcntr, Gness)} &
Set $G^{<|R}$ of a two-time function {\tt Gcntr} from a steady-state function {\tt Gness}, assuming time-translational invariance.\\
\hline
\end{tabularx}
\caption{
\label{tab:conversion}
Data exchange between two-time functions {\tt Gcntr} and steady-state functions {\tt Gness}.  {\tt Gcntr} can be  {\tt cntr::herm\_matrix}, {\tt cntr::herm\_matrix\_timestep}, {\tt cntr::herm\_matrix\_moving} or  {\tt cntr::herm\_matrix\_timestep\_moving}. For the latter three types, the argument {\tt tstp} in {\tt cntr2ness} is ignored and can be omitted, because the leading physical timestep is always addressed. If the argument {\tt tstp} is omitted for {\tt Gcntr} of type {\tt cntr::herm\_matrix}, it defaults to the largest physical time ({\tt Gcntr.nt\_}). 
}
 \end{small}
\end{table}
Since in steady-state calculations much finer time grids are possible than in two-time calculations, downsampling and upsampling routines are necessary to read and write Green's functions defined on different grids
in 
the two interfaces. The routines summarized in Tab.~\ref{tab:resampling} resample steady-state {\tt herm\_matrix\_ness} objects, so that they can be brought into the same shape as their {\tt cntr} counterparts and be processed with the {\tt cntr2ness} and {\tt ness2cntr} functions, see Sec.~\ref{sec:nessexample2} for an example usage and the documentation for details.

\begin{table}[H]
\centering
\begin{small}
\begin{tabularx}{\textwidth}{|p{0.4\textwidth}|X|}
\hline
\texttt{upsample(double h\_in,herm\_matrix\_ness in, int factor)} &
Upsample a {\tt herm\_matrix\_ness} object {\tt in} with initial timestep {\tt h\_in} to a finer grid with a factor of {\tt factor} more points. \\
\hline
\texttt{downsample(herm\_matrix\_ness in, int factor)} &
Downsample a {\tt herm\_matrix\_ness} object {\tt in} to a coarser grid with a factor of {\tt factor} fewer points.\\
\hline
\end{tabularx}
\caption{
\label{tab:resampling}
Upsampling and downsampling functions for {\tt ness2::herm\_matrix\_ness} objects (return type {\tt ness2::herm\_matrix\_ness}).
}
 \end{small}
\end{table}

\subsection*{Diagram utilities}
\vspace*{2mm}

Analogous to real-time Green's functions in {\tt libcntr}, we provide two basic functions that compute particle-particle and particle-hole bubbles. These functions are identical to those for the memory truncated Green's functions (Table \ref{tab:diagram}) when the last timestep of the moving window is replaced by the steady-state functions $G^<(t)$ and $G^R(t)$. More precisely, the 
particle-hole bubble ({\tt Bubble1}, Eq.~\eqref{bubble1}), which is defined such that  $C_{c1,c2}(t,t')$ is set to $ i A_{a1,a2}(t,t') \, B_{b2,b1}(t',t)$, becomes 
\begin{align}
C_{c1,c2}^<(t-t')= i A_{a1,a2}^<(t-t') \, B^>_{b2,b1}(t'-t),
\label{bubble1NESSles}
\\
C_{c1,c2}^>(t-t')= i A_{a1,a2}^>(t-t') \, B^<_{b2,b1}(t'-t),
\label{bubble1NESSgtr}
\end{align}
when written for translationally-invariant steady-state functions; the retarded component is then constructed from Eq.~\eqref{eq:retarded_nessi}. Likewise, the  particle-particle bubble {\tt Bubble2} (Eq.~\eqref{bubble2}), defined as  $ i A_{a1,a2}(t,t') \, B_{b1,b2}(t,t')$, becomes 
\begin{align}
C_{c1,c2}^<(t-t')= i A_{a1,a2}^<(t-t') \, B^<_{b1,b2}(t-t'),
\label{bubble2NESSles}
\\
C_{c1,c2}^>(t-t')= i A_{a1,a2}^>(t-t') \, B^>_{b1,b2}(t-t'),
\label{bubble2NESSgtr}
\end{align}
when written for translationally-invariant steady-state functions. The function call is therefore similar to that for the real-time functions (see Table \eqref{tab:diagram_ness}). For {\tt herm\_matrix\_ness}, $A$ and $B$ are assumed to be Hermitian.

\begin{table}[tbp]
\centering
\begin{small}
\begin{tabularx}{\textwidth}{|p{0.5\textwidth}|X|}
\hline
\texttt{void Bubble1\_ness(C, c1, c2, A, a1, a2, B, b1, b2)} &
Particle-hole bubble \eqref{bubble1} in the steady state (Eqs.~\eqref{bubble1NESSles} and \eqref{bubble1NESSgtr}) \\
\hline
\texttt{void Bubble2\_ness(C, c1, c2, A, a1, a2, B, b1, b2)} &
Particle-particle bubble \eqref{bubble2} in the steady state (Eqs.~\eqref{bubble2NESSles} and \eqref{bubble2NESSgtr}). \\
\hline 
\end{tabularx}
\caption{Diagram utilities: {\tt A,B,C} are of type \texttt{herm\_matrix\_ness} and indices {\tt a1, a2, \dots} are integer; if all {\tt A,B,C} are scalar ({\tt size1=1}), indices can be omitted.
 \label{tab:diagram_ness}}
 \end{small}
\end{table}

\subsection*{Further utilities}
An important utility is the evaluation of the density matrix \eqref{density_matrix_A} and the equal-time convolution \eqref{density_matrix_AB}. The density in the steady state is directly evaluated from the lesser Green's function
\begin{align}
\label{ness_density_matrix_A}
\rho_{A} = \xi i A^<(0), 
\end{align}
where $\xi=\pm1$ for bosonic (fermionic) Green's functions. The equal-time convolution \eqref{density_matrix_AB} can be obtained from the Fourier transform \eqref{gw2t03} of $C=A\ast B$, where 
\begin{align}
C^<(\omega)
&= A^<(\omega)\,B^A(\omega)\;+\;A^R(\omega)\,B^<(\omega),
\label{eq:conv_density_matrix_cont}
\end{align}
(using $B^A(\omega)=[B^R(\omega)]^\dagger$ for Hermitian $B$). The corresponding function calls are listed in Table \ref{tab:density_distance_ness}. 

Finally, we provide the function  {\tt distance\_norm2(A,B,domain)} to return the  distance   $||A-B||_2$  ($L_2$ norm) for the data arrays  {\tt A} and {\tt B}  of type {\tt fft\_array} or  {\tt herm\_matrix\_ness} on the time grid ({\tt domain=} {\tt fft\_domain::time}) or frequency grid ({\tt domain=fft\_domain::fft\_freq}), see Table \ref{tab:density_distance_ness}.

\begin{table}[tbp]
\centering
\begin{small}
\begin{tabularx}{\textwidth}{|p{0.4\textwidth}|X|}
\hline
\texttt{convolution\_density\_matrix( result, bosefermi, A, B, double h)} &
{\tt result} (complex matrix) is set to the steady-state $\rho_{AB}$ (Eq.~\eqref{density_matrix_AB}), for the convolution of two functions {\tt A} and {\tt B} of type {\tt herm\_matrix\_ness}. 
\\ 
\hline
\texttt{density\_matrix(result, bosefermi, A)} &
{\tt result} (complex matrix) is set to $\rho_{A}$ (Eq.~\eqref{ness_density_matrix_A}), for a functions {\tt A} of type {\tt herm\_matrix\_ness}. 
\\
\hline
\texttt{double distance\_norm2(A,B,domain)} &
Difference norm $\|A-B\|_2$ like \eqref{distancenorm} for {\tt herm\_matrix\_ness} or {\tt fft\_array} objects, on the time or frequency grid ({\tt domain=fft\_domain::[time|freq]}). \\ 
\hline
\end{tabularx}
\caption{
Some utilities for steady-state Green's functions.
\label{tab:density_distance_ness}
}
\end{small}
\end{table}

\subsection*{OMP parallelization}

In the steady-state code, the main numerical effort is related to the Fourier transforms.  Using the built-in functionalities of the FFTW library, these can be (shared-memory) parallelized. The parallelization should be effective for Fourier transforms with a large domain {\tt Nft} and small orbital dimension (such as {\tt size=1}), as well as for  Fourier transforms with a smaller domain {\tt Nft} but large orbital dimension. In order to make this functionality available, we provide a small helper class {\tt FFT\_OMP\_Manager}. It must be initialized at the beginning of the program:
 \begin{lstlisting}[language=c++,numbers=none]
FFT_OMP_Manager::initialize(); 
\end{lstlisting}
Somewhere before creating an {\tt fft\_array} or {\tt herm\_matrix\_ness} object  (which contains an {\tt fft\_plan} of the FFTW library), call 
 \begin{lstlisting}[language=c++,numbers=none]
FFT_OMP_Manager::set_threads_for_new_plans(nomp); 
\end{lstlisting}
All plans which are created after this call will be generated such that their execution spans over {\tt nomp} threads. 
Note, that new plans are also created when new {\tt fft\_array} or {\tt fft\_herm\_matrix\_ness} are generated via an assignment and or copy assignment. 
At the end end of the program, call 
 \begin{lstlisting}[language=c++,numbers=none]
FFT_OMP_Manager::finalize(); 
\end{lstlisting}
Inside the code, you can access the number of threads currently used in the construction of plans from the variable {\tt FFT\_OMP\_Manager::current\_threads\_}. To test the parallel setup, we provide a notebook {\tt utils/test\_ness2\_omp.ipynb} which calls a small test program (compiled and run just as the longer examples in Sec.~\ref{sec:examples}). 

A few comments are in order: 
\begin{itemize}
\item
This functionality is added automatically when the library is built with {\tt omp=ON} and {\tt ness=ON} (see Sec.~\ref{subsec:installation_libcntr}). In addition it requires the FFTW library to be compiled with {\tt --enable-threads} \cite{FFTW_documentation}.
If only the non-threaded FFTW library is available (or if {\tt omp=OFF}), the calls to the {\tt FFT\_OMP\_Manager}  have simply no effect, and all Fourier transforms are single threaded. 
\item
The generation of plans (i.e., the creation of any {\tt fft\_herm\_matrix\_ness} or  {\tt fft\_array})  is not thread-safe, and should be called from a serial region of the code. 
\item
In general, one should avoid nesting outer parallelization (e.~g. over operations on different Green's functions $G_k$) and parallelization of the Fourier transforms. Using outer parallelization in combination with {\tt fft\_plans} with more than one thread may lead to oversubscription. If outer parallelization is used, plans should therefore simply be created with {\tt nomp=1} threads. 
\end{itemize}

\subsection{Numerical solution of the Dyson equation }\label{sec:steadystate_dyson_num}
\vspace*{2mm}

The call for the solution of the Dyson equation  \eqref{intdif_eq} in the steady state is 
\begin{align}
\text{\begin{minipage}{0.9\textwidth}
 {\tt dyson(G, mu, epsilon, Sigma, h, METHOD, [ETA])}.
\end{minipage}}
\end{align}
Here {\tt G} and {\tt Sigma} are Green's functions of type {\tt herm\_matrix\_ness}, which must be defined on the same grid size  $N_{\rm ft}$ and must have the same matrix dimension; {\tt epsilon} (a complex matrix) is the matrix $\epsilon$ in Eq.~\eqref{dyson-rett}, {\tt mu} is the chemical potential $\mu$, and {\tt h} the timestep. {\tt [ETA]} stands for further optional parameters which introduce a long-time regularization of the Dyson equation as defined below.

At present, for the  solution of the Dyson equation, we provide only one method {\tt METHOD=FFT\_TRAPEZ}, which is based on a straightforward discrete Fourier transform: The self energy is transformed to the frequency grid \eqref{w-grid} using the integral transform {\tt transform\_to\_freq()}, corresponding to a trapezoidal evaluation of Eqs.~\eqref{gw2t02} and \eqref{gw2t04}. Then the frequency-dependent functions $G^{<|R}(\omega)$ are computed on the grid \eqref{w-grid} using Eqs.~\eqref{eq:ness_dyson_1} and \eqref{eq:ness_dyson_2} (optionally adding a regularization $\Sigma_{\rm reg}$ as explained below). For the back-transform, Eqs.~\eqref{gw2t01} and  \eqref{gw2t03}, we use {\tt transform\_to\_time()}. In order to reduce Fourier artifacts from the large frequency region we only use $G^{<|R}(\omega)$  at frequency grid points $-(N_{\rm freq}/2-1),...,N_{\rm freq}/2-1$, with $N_{\rm freq}=N_{\rm ft}/3$, and set $G^{<|R}(\omega)$ to zero out-side this interval. With the step size $\Delta_\omega$, the maximum frequency is therefore $\omega_{\rm max} = \frac{2\pi }{6h}$. An accurate solution of the Dyson equation requires  $h$ to be small enough such that the spectral function is sufficiently decayed outside  $[-\omega_{\rm max},\omega_{\rm max}]$. Moreover, $N_{\rm ft}$ must be sufficiently large such that the functions $G^{R|<}(t)$ have decayed at the largest time $t_{\rm max} = \frac{hN_{\rm ft}}{2}$, and $\Delta_\omega$ can be chosen small enough to resolve the most narrow structures in frequency. The convergence of the Dyson equation for $h\to 0$ and $t_{\rm max}\to\infty$ is analyzed in the demonstration examples.

In case the long-time behavior requires a regularization, we provide the following possibilities to add a bath self-energy $\Sigma_{\rm reg}$ to $\Sigma$:
\begin{itemize}
\item
{\tt [ETA] = REG\_CONST, eta, beta, mu} corresponds to a bath self-energy defined by Eqs.~\eqref{bath-const}  and \eqref{bath-less-fermion}, for a fermionic $G$ and $\Sigma$.
\item
{\tt [ETA] = REG\_GAUSS, eta, beta, mu} corresponds to a bath self-energy defined by Eqs.~\eqref{bath-gauss}  and \eqref{bath-less-fermion}, for fermionic $G$ and $\Sigma$.
The cutoff $\omega_c$ is chosen consistent with the grid \eqref{w-grid} as $\omega_c=\frac{\pi}{10h}$.
\item
{\tt [ETA] = REG\_OHMIC, eta, beta} corresponds to the ohmic bath defined by Eqs.~\eqref{bath-ohmic} with cutoff $\omega_c=\frac{\pi}{10h}$, for a bosonic $G$ and $\Sigma$.
\end{itemize} 
Alternatively, one can construct any user-defined self-energy $\Sigma_{\rm reg}$ for regularization, and add it explicitly to the input $\Sigma$ before calling {\tt dyson}. 

\section{Compilation of \libcntr{}\label{subsec:installation_libcntr}}

\NESSitwo{} is an extension of the previous \NESSi{} implementation, and all additional functionalities are incorporated into the same \libcntr{} library. The compilation of the \libcntr{} library, using the {\tt  cmake} building environment (version 3.1 or higher is required) therefore differs only minimally from the instructions in Ref.~\cite{NESSi}, and we only highlight the main differences here. 

To call {\tt cmake}, set up a script within the {\tt configure.sh} within the {\tt libcntr/}  directory, for which we suggest the following structure:
 \begin{lstlisting}[language=bash,numbers=none]
CC=[C compiler] CXX=[C++ compiler] \
cmake \
    -DCMAKE_INSTALL_PREFIX=[install directory] \
    -DCMAKE_BUILD_TYPE=[Debug|Release] \
    -Domp=[ON|OFF] \
    -Dhdf5=[ON|OFF] \
    -Dmpi=[ON|OFF] \
    -Dness=[ON|OFF] \
    -DBUILD_DOC=[ON|OFF] \
    -DCMAKE_INCLUDE_PATH=[include directory] \
    -DCMAKE_LIBRARY_PATH=[library directory] \
    -DCMAKE_CXX_FLAGS="[compiling flags]" \
    ..
\end{lstlisting}
The only difference with respect to the \NESSione{} implementation is the switch {\tt -Dness}. If the latter is turned {\tt ON}, the library has to be linked against the FFTW library \cite{FFTW}. To enable this, the {\tt CMAKE\_INCLUDE\_PATH} must include the location of the FFTW headers, and {\tt CMAKE\_LIBRARY\_PATH} must include the location of the FFTW library. Apart from this, the other compilation steps and variables remain the same as explained in Ref.~\cite{NESSi}, and in the online documentation. The compilation of the example programs remains unchanged with respect to \NESSione{}.


\section{Example programs and benchmarks}
\label{sec:examples}

\subsection{Running the examples}
\label{sec:examples_ramarks}

The source code for all examples is found  in {\tt nessi/examples/programs/}. To run the programs, one must first compile and install the {\tt libcntr} library with NESS   and HDF5 support, and then compile the examples. After that, one should find an executable {\tt [name].x} for each main program {\tt [name].cpp} in {\tt nessi/examples/exe/}. For details, follow the instructions for installing the \NESSi{} examples  in the first release of the library \cite{NESSi}. The same instructions can also be found in the {\tt README} file on github, as well as in the html manual.  For each example below, we provide a Jupyter notebook to run the code and to do the postprocessing, and a Python script which executes the same commands. Both are located in {\tt nessi/examples/utils/}. To run simulations using the Jupyter notebook, copy the notebook to a working directory and adapt the path of the executables.  The Python scripts depend on the Python utilities for reading HDF5 Green's function files, so that {\tt libcntr/python3} should be part of the {\tt PYTHONPATH}.

\subsection{DMFT with a memory-truncated time propagation}
\label{sec:movingexample}

\subsubsection{Model setup}

As a demonstration program for the  truncated KBEs, we consider a simple Hubbard model with time-dependent interaction $U(t)$, solved within DMFT with a second-order perturbation theory impurity solver, similar to the problem studied in Ref.~\cite{Stahl2022}. The Hubbard model is defined by the Hamiltonian
\begin{equation}
\label{Hubbardmodel}
H(t) = -J \sum_{\langle i,j \rangle, \sigma} c_{i \sigma}^\dagger c_{j \sigma} + U(t) \sum_i \Big(n_{i \uparrow}-\tfrac12\Big)\Big(n_{i \downarrow}-\tfrac12\Big),
\end{equation}
where \( c_{i \sigma}^\dagger \) and \( c_{i \sigma} \) are the creation and annihilation operators for fermions at site \( i \) with spin \( \sigma \) (\( \sigma \in \{\uparrow, \downarrow\} \)), and \( n_{i \sigma} = c_{i \sigma}^\dagger c_{i \sigma} \) is the number operator. The first term represents the hopping of fermions between neighboring sites \( \langle i,j \rangle \), with hopping amplitude \( J \), and \( U \) is the on-site interaction between fermions of opposite spins. We consider an interaction quench, where the system is prepared in an equilibrium state with temperature $T=1/\beta$ and $U=0$ for $t<0$, and the interaction is switched to a nonzero value $U(t)=U$ for $t\ge 0$. The model is studied on a Bethe lattice, where the self-consistency for the DMFT impurity model becomes particularly simple (see, e.g., Ref.~\cite{Stahl2022} for details). Within second order perturbation theory, the equations reduce to  the Dyson equation for the local contour-ordered Green's function $G(t,t') = -i \langle T_\mathcal{C} c_{i\sigma}(t)c_{i\sigma}^\dagger(t')\rangle $,
\begin{equation}
\label{btrega1}
G^{-1} = (i\partial_t +\mu - K),
\end{equation}
with a self-consistent memory kernel
\begin{equation}
\label{btrega2}
K(t,t') = \Delta(t,t') + \Sigma(t,t').
\end{equation}
Here \( \Delta(t,t') \) is the hybridization function and \( \Sigma(t,t') \) is the self-energy. 
For the Bethe lattice, the hybridization function takes the closed form
\begin{equation}
\label{btrega3}
\Delta(t,t') = J_0^2 G(t,t'),
\end{equation}
with a rescaled hopping $J_0$. With this, the DOS of the noninteracting model has a semielliptic form
\begin{equation}
\label{semiel}
D(\epsilon) = \frac{\sqrt{4J_0^2-\epsilon^2}}{2\pi J_0^2} 
\end{equation}
with bandwidth $4J_0$. We take $J_0=1$ as the unit of energy, and $1/J_0$ as the unit of time ($\hbar=1$). The expression for the impurity model self-energy in second order perturbation theory is 
\begin{equation}
\label{btrega4}
\Sigma(t,t') = U(t) \mathcal{G}(t,t') \mathcal{G}(t,t') \mathcal{G}(t',t) U(t'),
\end{equation}
where $\mathcal{G}$ is self-consistently determined by setting $\mathcal{G}=G$.  The chemical potential $\mu$ will be set to $\mu=0$ throughout  this example, which corresponds to the particle-hole symmetric case. We also remark that Eqs.~\eqref{btrega1} to \eqref{btrega4}  are equivalent to the equations of motion for the Green's function in an extended Sachdev-Ye-Kitaev (SYK)  model \cite{Chowdhury2022}, which describes the crossover from a Fermi liquid (at low temperatures and small $U$) to a non-Fermi liquid state at large $T$. 
 
From the converged solution of  Eqs.~\eqref{btrega1} to \eqref{btrega4}, we can calculate the momentum-dependent Green's function \( G_k(t,t') \). Due to the \( k \)-independent self-energy, $G_k$ depends on $k$ only via the single-particle energy \( \epsilon_k \), and is the solution of the Dyson equation with
\begin{equation}
\label{scdyson}
 G_k^{-1} =  \big( i\partial_t +\mu - \epsilon_k - \Sigma\big).
\end{equation}
For the given particle-hole symmetric system ($\mu=0$),  \( \epsilon_k = 0 \) corresponds to the Fermi energy, while states at the edge of the noninteracting bandwidth have \(| \epsilon_k| = 2 \). From $G_k$, we can finally extract the momentum occupation
\begin{align}
\label{rktt}
\rho(\epsilon_k) = \langle c_{k\sigma}^\dagger (t) c_{k\sigma}(t) \rangle = -i G_k^<(t,t).
\end{align}
This quantity is particularly well suited for revealing the two-staged dynamics characterized by a fast prethermalization and a slow thermalization: In the initial state with $U=0$ and inverse temperature $\beta$, one has the Fermi distribution $\rho(\epsilon_k)=\frac{1}{1+e^{\beta\epsilon_k}}$, with a step singularity at $\epsilon_k=0$ in the zero temperature limit (in the numerical results below we will take  a small but nonzero temperature $T=1/\beta=0.01$). Within few inverse hopping times $1/J_0$ after the quench at $t=0$, the system reaches a prethermal state in which the momentum occupations are modified from the initial state, but a jump remains at  $\epsilon_k=0$ \cite{Moeckel2008, Eckstein2009}.  This slowly varying distribution reflects the existence of quasiparticles which are already dressed by the interaction, but not yet thermalized. Thermalization finally leads to a distribution which is smooth across the Fermi energy. This occurs on a  time set by both the interaction and the available scattering phase space, which can be orders of magnitude longer than the inverse hopping. The memory-truncated KBE framework allows to study both the short-time and long-time dynamics of the system with the same numerical formalism.

\subsubsection{Implementation}
\vspace*{4mm}
\label{sec:impl:kbe}

We solve Eqs.~\eqref{btrega1} to \eqref{btrega4} on an equidistant time grid with discretization $h$, up to a maximum number of timesteps {\tt tmax}. The solution is generated with the same memory cutoff of {\tt tc}  timesteps in the kernel $ K $ of Eq.~\eqref{btrega1} and the kernel $\Sigma$ of Eq.~\eqref{btrega4}, and convergence with {\tt tc} is verified at the end. To generate the solution with memory cutoff {\tt tc}, we will first solve the equations on the full contour up to a given number {\tt nt} of timesteps,  where {\tt nt} $\ge$ {\tt tc}. The resulting full Green's functions $G$, $\Sigma$, and $G_k$ are stored as {\tt cntr::herm\_matrix<double>} to a HDF5 file. In a separate program, we read the functions \( G \), \( \Sigma \), \( G_k \) from the file, initialize the corresponding moving Green's function windows of type {\tt cntr::herm\_matrix\_moving<double>}, and perform the truncated time evolution  over the timesteps {\tt tstp = tc+1}, {\tt \dots}, {\tt tmax}. The relevant files can be found in {\tt nessi/examples/} and are listed in Table \ref{tab_ex_1}.
\begin{table}[htb]
    \centering
    \begin{small}
    \begin{tabular}{|l|l|}
        \hline
                   \begin{minipage}{0.4\textwidth}
        \texttt{programs/trunc\_bethe\_start.cpp}          \end{minipage}& 
      \begin{minipage}{0.5\textwidth}
         Source for initial non-truncated evolution
         \end{minipage}
          \\\hline
                   \begin{minipage}{0.4\textwidth}
        \texttt{programs/trunc\_bethe.cpp}          \end{minipage}& 
              \begin{minipage}{0.5\textwidth}
Source for memory-truncated evolution
         \end{minipage}
        \\\hline
                   \begin{minipage}{0.4\textwidth}
        \texttt{utils/demo\_trunc\_bethe.ipynb}          \end{minipage}& 
                      \begin{minipage}{0.5\textwidth}
Jupyter notebook to run the program 
         \end{minipage}
 \\\hline
                   \begin{minipage}{0.4\textwidth}
        \texttt{utils/demo\_trunc\_bethe.py}          \end{minipage}& 
                   \begin{minipage}{0.5\textwidth}
         A Python script; same as the notebook 
         \end{minipage}
\\\hline
    \end{tabular}\caption{Relevant files for running the truncated memory code example.}\label{tab_ex_1}
    \end{small}
\end{table}

For the installation, see Sec.~\ref{sec:examples_ramarks}.  To run simulations, copy the Jupyter notebook to a working directory and adapt the path of the executables. 

\subsubsection*{Implementation: Initial non-truncated simulation}
\vspace*{4mm}

The structure of the program for the non-truncated examples is similar to the examples provided with \NESSione{} \cite{NESSi}.  
The implementation is build on functions in the {\tt cntr} namespace, which is included at the top of the program:
\begin{lstlisting}[language=c++,numbers=none]
using namespace cntr;
\end{lstlisting}
The input parameters are {\tt nt} (the number of real-timesteps), {\tt h} (the time discretization $h$), {\tt beta} (the inverse temperature $\beta$), {\tt ntau} (the number of timesteps on the imaginary axis), {\tt U1} (the final interaction $U$), as well as the numerical parameters {\tt BootstrapMaxIter}, {\tt BootstrapMaxErr} and  {\tt CorrectorSteps}, which will be explained below. After reading the input parameters from the input file and initializing the data structures, the noninteracting equilibrium problem is initialized by
\begin{lstlisting}[language=c++, numbers=none]
cntr::green_equilibrium_mat_bethe(G, beta);
\end{lstlisting}
which replaces the Matsubara component of the Green's function $G$  with the noninteracting Green's function for a semi-elliptic DOS \eqref{semiel}.  The following loop then encompasses the bootstrapping and the actual time propagation step using {\tt cntr::dyson\_timestep}.  To compute the \textit{self-consistent kernel} \( K(t, t') = \Delta(t, t') + \Sigma(t, t') \)  from Eqs.~\eqref{btrega2} and  \eqref{btrega3}, we use the following two functions. The self energy $\Sigma$ at a given timestep {\tt tstp} is obtained from ({\tt GREEN}, {\tt GREEN\_TSTP}  and {\tt CFUNC}  are  synonymous with {\tt cntr::herm\_matrix<double>}, {\tt cntr::herm\_matrix\_timestep<double>} and {\tt cntr::function<double>}, respectively).

\begin{lstlisting}[language=c++, numbers=none]
void get_Sigma_timestep(int tstp,GREEN &Sigma,GREEN &G,CFUNC &U) {
  // temporary variable W to store time slice
  GREEN_TSTP W(tstp, Sigma.ntau(), Sigma.size1(), BOSON); 
  cntr::Bubble1(tstp, W, G, G);   // W(t,t') = ii*G(t,t')G(t',t)
  W.left_multiply(tstp, U);   // W(t,t') <-- U(t)W(t,t')
  W.right_multiply(tstp, U);  // W(t,t') <-- W(t,t')U(t'): 
  Bubble2(tstp, Sigma, G, W); //Sigma(t,t') = ii*G(t,t')W(t',t);
  Sigma.smul(tstp, -1.0); // a final -1 sign.
}
\end{lstlisting}
The Kernel $K$ is then summed up using 
\begin{lstlisting}[language=c++, numbers=none]
void get_K_timestep(int tstp, GREEN &K, GREEN &Sigma, GREEN &G) {
  K.set_timestep(tstp, Sigma); // K=Sigma at timestep tstp
  K.incr_timestep(tstp, G, 1.0);  // K += G at timestep tstp
}
\end{lstlisting}

As described for \NESSione{} \cite{NESSi}, timestepping starts with a bootstrapping phase, which solves the KBEs simultaneously on time slices {\tt 0,\dots,SolveOrder}, where {\tt SolveOrder} is the order $k$ of the Volterra integrator \cite{NESSi} in Sec.~\ref{kbetheory}. We use the maximum value {\tt SolveOrder=MAX\_SOLVE\_Order} $5$ in the present implementation.
\begin{lstlisting}[language=c++, numbers=none]
tstp = SolveOrder;
GREEN_TSTP gtemp(tstp, ntau, 1); // to store last timestep with matrix size 1
set_t0_from_mat(G); // initialize time 0 from Matsubara 
for (iter = 0; iter <= BootstrapMaxIter; iter++) {
  gtemp.set_timestep(tstp, G); // store last iteration
  // K-=Sigma[G]+G on all timesteps 0...SolveOrder:
  for (int n = 0; n <= SolveOrder; n++) {
    get_Sigma_timestep(n, Sigma, G, U);
    get_K_timestep(n, K, Sigma, G);
  }
  // Solve Dyson [idt + mu - eloc]G - K*G =1 for G 
  // on all timesteps 0...SolveOrder (here eloc=0):
  dyson_start(G, mu, eloc, K, beta, h, SolveOrder);      
  double err = distance_norm2(tstp, gtemp, G); // convergence?
  if (err < BootstrapMaxErr && iter > 3) break;
}
\end{lstlisting}
The iteration does a maximum number of {\tt BootstrapMaxIter} iterations, where  Eqs.~\eqref{btrega2} and  \eqref{btrega3}  are solved iteratively to determine $K$ from $G$, and Eq.~\eqref{btrega1} is solved to determine $G$ from $K$. The error measure {\tt cntr::distance\_norm2} returns a sum of the 2-norm of the difference between the previous iteration of $G$ (stored in {\tt gtemp}) and the updated $G$ on the last time slice {\tt tstp}. After the bootstrapping, a similar iteration of Eqs.~\eqref{btrega2},  \eqref{btrega3} and \eqref{btrega1} is performed for each timestep {\tt tstp=}{\tt SolveOrder+1,\dots,nt}:
\begin{lstlisting}[language=c++, numbers=none]
// extrapolate G from tstp-1,...tstp-SolveOrder to timestep tstp: 
cntr::extrapolate_timestep(tstp - 1, G, SolveOrder);
// self-consistent iteration:
for (iter = 0; iter <= CorrectorSteps; iter++) {
    G.get_timestep(tstp, gtemp); // store last iteration
    get_Sigma_timestep(tstp, Sigma, G, U); // get Sigma on tstp
    get_K_timestep(tstp, K, Sigma, G); // K<--Sigma+G
    // solve [idt + mu - eloc]G - K*G =1 for G on timestep tstp
    cntr::dyson_timestep(tstp,G,mu,eloc,K,beta,h,SolveOrder);
    ... // convergence error to previous iteration,  etc.  ...
}
\end{lstlisting}
 Since an initial guess for $G$ on the timestep can be obtained by extrapolation, the iterations converge quickly, and we keep a fixed number {\tt CorrectorSteps} of iterations at each time (typically {\tt CorrectorSteps=3} is sufficient). 
 
After the solution of the self-consistent equation for $G$, we solve the Dyson equation \eqref{scdyson} with a single call to {\tt cntr::dyson} (no self-consistency is needed for the Kernel in this case). At the end, all Green's functions are stored into a single HDF5 file using the file i/o routine described in \NESSione{} \cite{NESSi}.
\begin{lstlisting}[language=c++, numbers=none]
hid_t file_id = open_hdf5_file(flout); // create hdf5 file flout
G.write_to_hdf5(file_id, "G"); // write G to group /G in the file 
... // ... similar writing of Sigma, and other params nt, h, etc.
close_hdf5_file(file_id);
\end{lstlisting}
 The file can be read using the Python utilities in {\tt ReadCNTR} and {\tt ReadCNTRhdf5}.

\subsubsection*{Implementation: Memory-truncated simulation}
\vspace*{4mm}

The memory-truncated simulation over the timesteps {\tt tc+1,\dots,tmax}  is performed in a separate program, which reads the previously computed Green's functions from file;  {\tt tc} $\le$ {\tt nt} is required to be able to initialize the moving Green's functions from the data, and {\tt tc} $\ge$ {\tt SolveOrder} is needed such that Volterra Integrators of order {\tt SolveOrder} can be used. After reading the  input parameters ({\tt tc}, {\tt tmax}, {\tt CorrectorSteps}), the program reads the HDF5 input and initializes the moving Green's functions
({\tt GTRUNC}, {\tt GTRUNC\_TSTP} and {\tt CTRUNC} are short for {\tt cntr::herm\_matrix\_moving<double>}, \texttt{cntr::herm\_matrix\_timestep\_moving\\<double>} and {\tt cntr::function\_moving<double>}, respectively):
\begin{lstlisting}[language=c++, numbers=none]
hid_t file_id = read_hdf5_file(fldata); // open HDF5 file fldata
nt = read_primitive_type<int>(file_id, "nt"); // read nt from file
// allocate moving window with memory depth tc and matrix size 1:
GTRUNC G_t(tc, 1, FERMION);
CTRUNC U_t(tc, 1); // allocate moving contour function 
... // similar allocation for other functions
GREEN Gtmp; // temporary herm_matrix<double> 
Gtmp.read_from_hdf5(file_id, "G"); // read G and store into Gtmp
// Initialize G_t from timesteps  tc, tc-1,...,0 measured from nt
// (the two arguments are G and its Hermitian conjugate)
G_t.set_from_G_backward(Gtmp, Gtmp, tc);
\end{lstlisting}
The functions to compute the kernel from \eqref{btrega2} and \eqref{btrega3} on a given time slice are very similar to the corresponding functions for the full {\tt herm\_matrix<double>} objects explained above. They only differ in the referencing of the time slices: For the memory-truncated Green's functions, the functions below always act on the leading time slice {\tt 0} of the moving window:
\begin{lstlisting}[language=c++, numbers=none]
void get_Sigma_timestep(GTRUNC &Sigma, GTRUNC &G, CTRUNC &U) {
  // temporary W to store time slice of memory depth tc:
  GTRUNC_TSTP W(Sigma.tc(), Sigma.size1(), BOSON);    
  Bubble1(W,G,G); // W(t,t') <-- ii*G(t,t')G(t',t) on leading time t
  W.left_multiply(U); // W(t,t') <-- U(t)W(t,t') on leading time t
  W.right_multiply(U); // W(t,t') <-- W(t,t')U(t'):
  Bubble2(Sigma, G, W); // Sigma(t,t') <-- ii*G(t,t')W(t',t):
  Sigma.smul(0, -1.0); // *=-1 on timestep 0 of moving window (first argument)
}
\end{lstlisting}
\begin{lstlisting}[language=c++, numbers=none]
void get_K_timestep(GTRUNC &K, GTRUNC &Sigma, GTRUNC &G)  {
  K.set_timestep(0, Sigma, 0); // K <-- Sigma on step 0
  K.incr_timestep(0, G, 0, 1.0); // K += G on step 0
}
\end{lstlisting}

Finally, we present the implementation of the timestepping, which, similar to the timestepping of the full KBE, 
performs a self-consistent iteration of  Eqs.~\eqref{btrega2} and \eqref{btrega3} for the Kernel and  Eq.~\eqref{btrega1} 
with a fixed number {\tt CorrectorSteps} of iterations for each {\tt tstp = tc + 1,\dots,tmax}:
\begin{lstlisting}[language=c++, numbers=none]
GTRUNC_TSTP gtmp(tc, 1, FERMION);
// move windows forward by one step
Sigma_t.forward();
K_t.forward();
G_t.forward();
// extrapolate G from tstp-1,...tstp-SolveOrder to timestep tstp: 
extrapolate_timestep(G_t, SolveOrder);
// self-consistent iteration
for (iter = 0; iter <= CorrectorSteps; iter++) { 
  gtmp.set_timestep(G_t, 0); // store leading time of G into gtmp
  get_Sigma_timestep(Sigma_t,G_t,U_t); //get Sigma on leading time
  get_K_timestep(K_t, Sigma_t, G_t); // K=G+Sigma  on leading time
  // solve Dyson G^{-1} = (idt + mu - eloc - K) on leading time
  dyson_timestep(G_t, K_t, eloc_t, mu, SolveOrder, h);
}
\end{lstlisting}
Within the loop over timesteps from {\tt tc+1,\dots,tmax}, the first operation is to move the windows forward by one step, using the {\tt forward()} method. If the leading timestep of the window initially corresponds to the physical timestep {\tt tstp-1}, then after the action of {\tt forward()} the leading timestep of the window  corresponds to the physical timestep {\tt tstp}. Next, an estimate for $G$ on its leading timestep is obtained by extrapolating from the sub-leading timesteps {\tt 1,...,SolveOrder}. The self-consistent iteration itself, inside the loop over {\tt iter}, then proceeds in the same way as the standard KBE timestepping described above. 

For the computation of $G_k$ it is important to note that $\Sigma$ is not stored as the window is moved forward. Hence, the timestepping for the solution for $G_k$ based on Eq.~\eqref{btrega4} must be computed within the same timestepping loop as $\Sigma$, following the convergence of the DMFT iteration above. This is in contrast to the non-truncated simulation, where one can compute $G_k$ outside the timestepping loop for $\Sigma$. For each {\tt k = 0,\dots,nk}, $G_k$ is obtained as
\begin{lstlisting}[language=c++, numbers=none]
Gk_t[k].forward();
// solve Dyson [idt + mu - esp_k]G_k - K*G_k = 1 for each k 
dyson_timestep(Gk_t[k],Sigma_t,epsk_t[k],mu,SolveOrder,h);
Gk_t[k].density_matrix(0, mtmp); // rho_k(t)= -i Gk^les(t,t)
densk[k][tstp] = mtmp(0, 0).real(); // save rho_k(t)
\end{lstlisting}
The last lines extract the values for the momentum occupation \eqref{rktt}.
 
 \subsubsection*{HDF5 output of time slices} 
 \vspace*{2mm}
 
 Finally, since in the memory truncated Green's functions are not stored as the window is moved forward, one must actively save the intermediate Green's function data if needed. As an example, in {\tt trunc\_bethe.cpp} we have implemented the possibility to write selected time slices to a HDF5 file during the timestepping.  For this, we create a HDF5 file for writing before entering the propagation loop
\begin{lstlisting}[language=c++, numbers=none]
hid_t  file_id=open_hdf5_file(flout1); // flout1 is filename
\end{lstlisting}
using the function {\tt open\_hdf5\_file} from the \NESSione{} HDF5 interface (see Ref.~\cite{NESSi}, or the online documentation). During  step {\tt tstp} of the evolution, we can write the current leading time slice of the moving window to a new group  t[{\tt tstp}]/G within this file by calling (assuming the group t[{\tt tstp}] does not yet exist)
\begin{lstlisting}[language=c++, numbers=none]
hid_t sub_group = create_group(file_id,"t"+std::to_string(tstp));
G_t.write_timestep_to_hdf5(0,sub_group,"G");  
// save timestep 0 of moving window
close_group(sub_group); 
\end{lstlisting}
In the Jupyter notebook, we can use the helper functions of {\tt readCNTRhdf5} to extract the data (timestep, size, retarded and lesser component) at a timestep {\tt tstp} as simple arrays:
\begin{lstlisting}[language=Python, numbers=none]
with h5py.File(out_file_name, 'r') as fd:
  key=f"t{tstp}/G" # the key under which the timestep is stored
  G=read_herm_matrix_timestep_moving_group(fd[key])
  # now G.ret[s,a,b]=G^R(t,t-s)_{ab}, G.les[s,a,b]=G^<(t,t-s)_{ab}.
\end{lstlisting}
A similar routine  {\tt read\_herm\_matrix\_moving\_group} can be used to read a full moving Green's function from a HDF5 group.

\subsubsection{Results}
\vspace*{4mm}
\label{sec:impl:res}

The results shown below have been obtained for an interaction quench to $U=1$, with an initial temperature $T=0.01$ ({\tt beta=100}), {\tt ntau=2000} steps on the imaginary contour, and a time discretization {\tt dt=0.04}. The initial time evolution is performed up to {\tt nt=800} (corresponding to physical time $32$). The memory truncated simulation is performed for {\tt tmax=30000} timesteps ($t_\text{max}=1200$) with memory depth of {\tt tc=200,400,600} steps ($t_c=8,16,24$). The simulation runs in roughly $10$ minutes on a MacBook with an Apple M2 processor, consuming $130$MB of memory to simultaneously store  the temporary {\tt cntr::herm\_matrix<double>} object for initialization as well as the moving windows for  $\Sigma$, $G$, $K$, and $G_k$ for $7$ values of $k$. In contrast, a single full {\tt cntr::herm\_matrix<double>} with {\tt tmax=30000} real-timesteps and {\tt ntau=2000} imaginary timesteps would require roughly $14$GB of memory.

\begin{figure}[tbp]
\centerline{\includegraphics[width=0.8\textwidth]{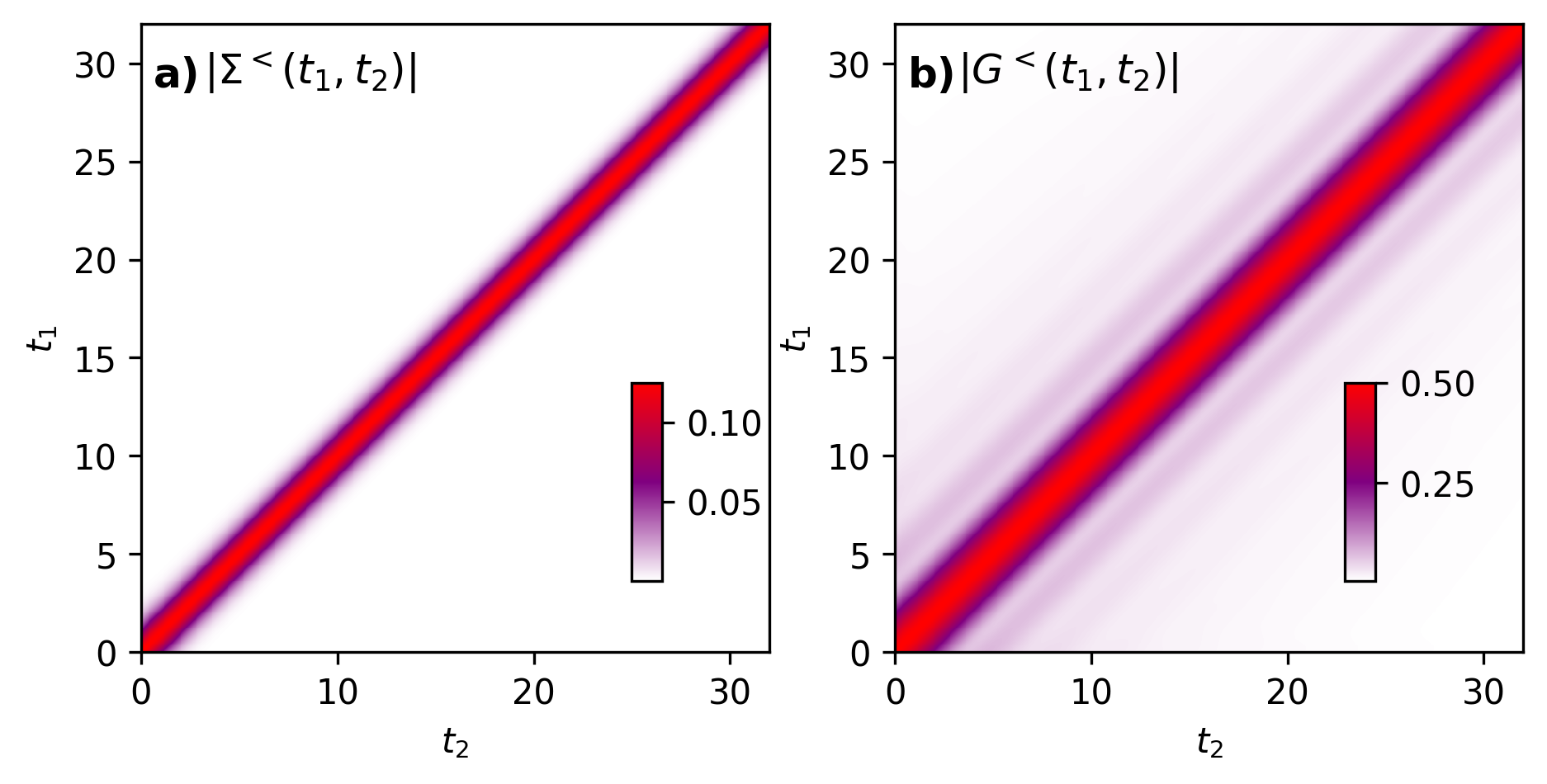}}
\caption{The functions $|\Sigma^{<}(t,t')|$ (a) and $|G^{<}(t,t')|$ (b) in the initial time window $0\le t,t'\le 32$.}
\label{figtrunc1}
\end{figure}

\begin{figure}[tbp]
\centerline{\includegraphics[width=0.99\textwidth]{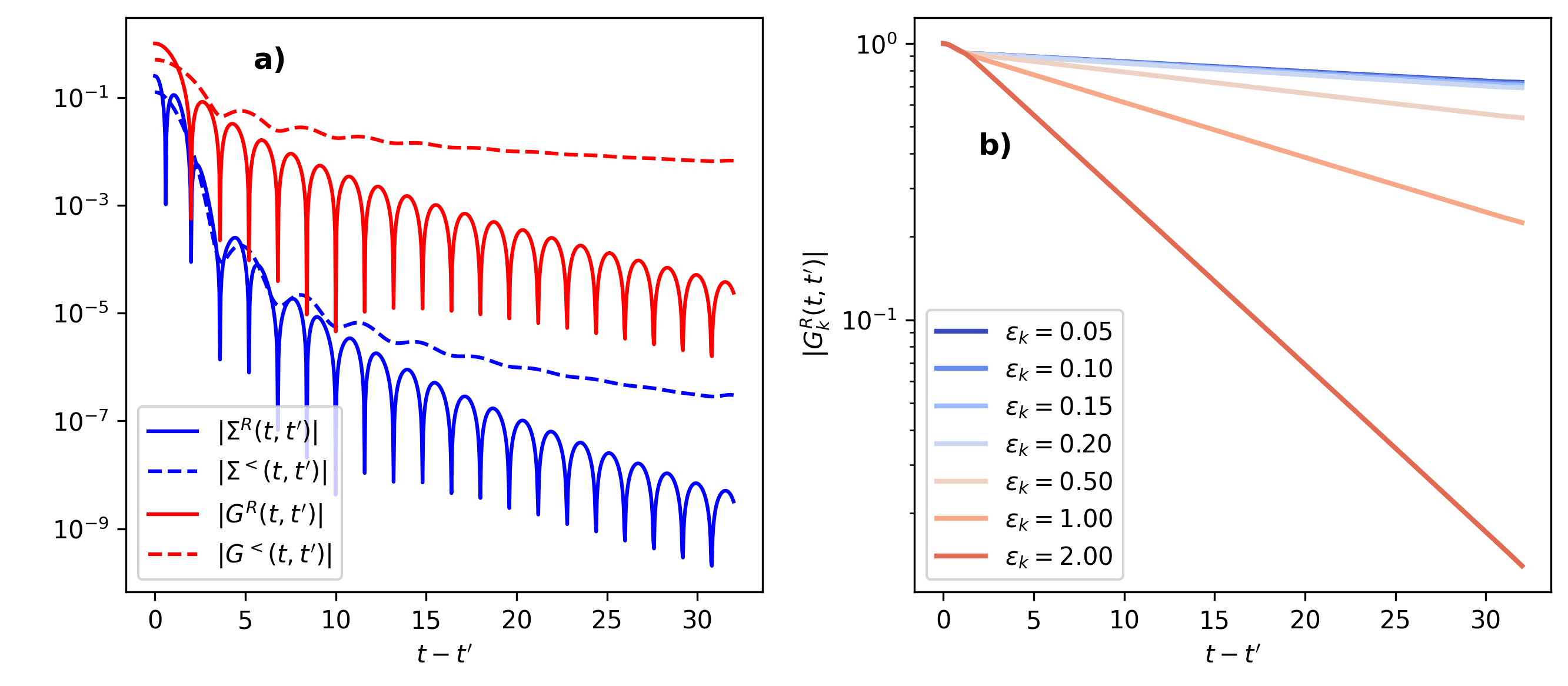}}
\caption{ (a) The functions $|\Sigma^{R,<}(t,t')|$ and $|G^{R,<}(t,t')|$ as a function of the time difference $t-t'$ at a given time slice  ($t=32$) on a logarithmic scale. (b) The function $|G_k^{R}(t,t')|$ as a function of the time difference $t-t'$ at a given time slice  ($t=32$) on a logarithmic scale, for different values of $\epsilon_k$, where $\epsilon_k=0$ corresponds to the Fermi edge. Note the different vertical scale in the two plots.}
\label{figtrunc2}
\end{figure}

Figure~\ref{figtrunc1} shows the functions $|\Sigma^{<}(t,t')|$ and $|G^{<}(t,t')|$ in the initial time window $0\le t,t'\le 32$.   One can see that all functions decay rapidly as a function of the time difference $|t-t'|$, which justifies the truncation approach (the retarded components would show a similar or even faster decay with $|t-t'|$). In particular, for the self-energy, we find an exponentially decaying envelope over several orders of magnitude (see the line plots in the left panels of Fig.~\ref{figtrunc2}. Because $\Sigma$ is a point-wise product of Green's functions, it decays faster than $G$. The decay of the Kernel  $K=\Sigma+G$ for the self-consistent equation \eqref{btrega1} is therefore dominated by $G$, and one can expect that the memory-truncated evolution for $G_k$ (Eq.~\eqref{btrega4}, with memory kernel $\Sigma$) is faster convergent than the determination of $G$. However, because the memory integrals are convolutions of the Green's functions and the Kernel, it is not easy to estimate a priori the required memory depth {\tt tc}, and the parameter {\tt tc}  will instead be used as a numerical convergence parameter. 

It is important to note, that for the memory-truncated time propagation to work, it is sufficient that the Kernel decays, while the Green's function (which is the solution of the Dyson equation) can still be large outside the memory truncated window. This becomes evident for the momentum-resolved Green's functions $G_k$, for which the kernel $\Sigma$  decays quickly (left panel of Fig.~\ref{figtrunc2}), while the decay of the Green's functions is much slower (right panel of Fig.~\ref{figtrunc2}).   The decay of $G_k$ reflects the quasi-particle lifetime, which becomes long in particular close to the Fermi energy  $\epsilon_k=0$.

\begin{figure}[tbp]
\includegraphics[width=0.99\textwidth]{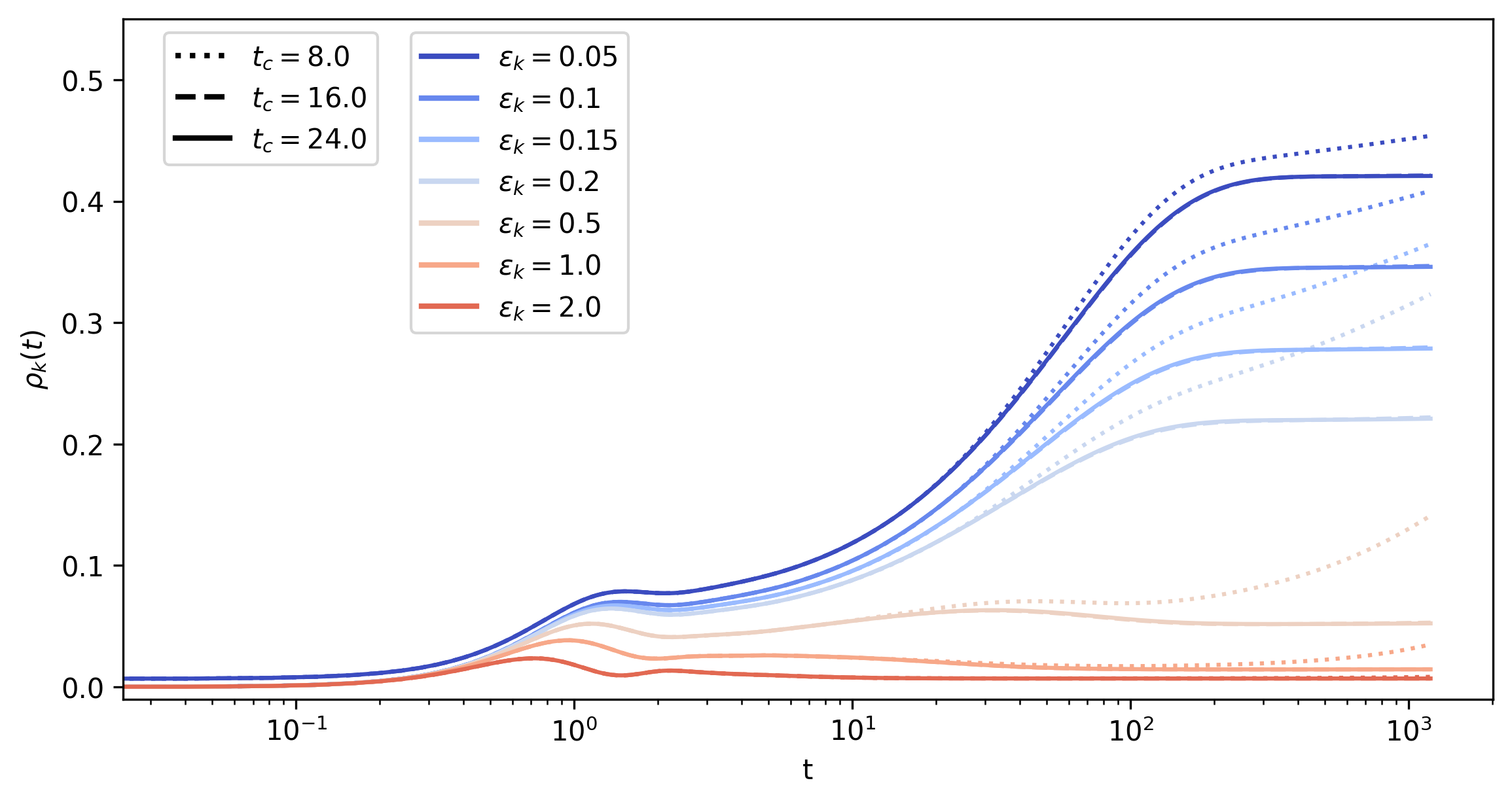}
\caption{Momentum occupation $\rho_k(t)$ for selected values of $\epsilon_k$, obtained with the memory-truncated time evolution with different {\tt tc}. Different line colors represent different values of $\epsilon_k$, close to the Fermi energy ($\epsilon_k=0$), in the middle of the band ($\epsilon_k=1$), and close the band edge ($\epsilon_k=2$). Different line-styles distinguish different values of {\tt tc}.
}
\label{figtrunc3}
\end{figure}

Finally, in Fig.~\ref{figtrunc3} we show the momentum occupation $\rho_k(t)$ for selected values of $\epsilon_k$ close to the Fermi energy ($\epsilon_k=0$), in the middle of the band ($\epsilon_k=1$) and at the band edge ($\epsilon_k=2$). By increasing {\tt tc} (compare the different linestyles), one can see that a relatively short memory window of $t_c=24$ is sufficient to reach a converged solution over the full interval. On the other hand, if the truncation window is too small, the results strongly deviate. For even shorter $t_c=4$ (not shown here) the solution of the memory-truncated KBEs becomes unstable within the simulated time range of Fig.~\ref{figtrunc3}. 

In the converged results one can clearly see the  two-stage dynamics: The prethermal state  is reached after times of order $1$ (few inverse hoppings). The prethermal momentum distribution still has a pronounced step at the Fermi energy, which is evident by comparing $\rho_k$ at the smallest value of $\epsilon_k$ ($\epsilon_k=0.05$) to $\rho_{k_f}=0.5$ at the Fermi energy $\epsilon_{k_F}$ (not shown in the plot). The thermalization time is of the order of a few $100$ hopping times, after which the momentum distribution takes the smooth form corresponding to the equilibrium distribution $\rho_{k}(T_f,U)$ at the final interaction $U$ and a final temperature $T_f$ which is set by the total energy of the system.

\subsection{Steady-state Dyson equation: Anderson impurity model}
\label{sec:nessexample1}

\subsubsection{Model setup}
\vspace*{2mm}

Here we demonstrate the use of the steady-state implementation for a simulation of transport through a single impurity Anderson model, corresponding to the perturbative solution of the setup studied in Ref.~\cite{Werner2010}. The  on-site impurity Hamiltonian is 
\begin{align}
H_{\mathrm{loc}} = U n_\uparrow n_\downarrow + \left(\mu-\frac{U}{2} + \epsilon_d\right)(n_\uparrow + n_\downarrow),
\end{align}
where $U$ is the local interaction, and $\epsilon_d$ the bare level energy; $\mu=0$ will be set to zero in the following, such that $\epsilon_d=0$ corresponds to a particle-hole symmetric case.  The impurity is coupled to two infinite metallic leads, the left ($L$) and right ($R$) bath, which are kept at a voltage bias $V$. Integrating out the leads gives rise to an embedding self-energy $\Sigma_{\rm bath}=\Sigma_{L}+\Sigma_{R}$, which is defined via a spectral representation as in Eqs.~\eqref{gw2t01} and \eqref{gw2t03}. Following Ref.~\cite{Werner2010}, we use a smooth box DOS
\begin{align}
\label{abox}
A_\Sigma(\omega) = \frac{\Gamma/\pi}{(1+e^{\nu(\omega-\omega_c)})(1+e^{-\nu(\omega_c+\omega)})},
\end{align} 
where $\Gamma$ is the hybridization strength, $\omega_c$ is the half bandwidth, and $\nu$ a smoothening parameter. 
While the DOS $A_\Sigma(\omega)$ is identical for the left and right baths, $\Sigma^<_{L,R}$ is given by the equilibrium distribution \eqref{eq:eq:fdt} with different chemical potentials $\mu_{L/R}=\pm V/2$. In the following, we use $\omega_c=10\Gamma$ and $\nu=3/\Gamma$; $\Gamma=1$ defines the energy unit, and $1/\Gamma$ is the unit of time.  Note, that in Ref.~\cite{Werner2010}, steady-state Green's functions were computed by simulating a real-time evolution into the steady state, which can be avoided by the direct steady-state simulation.

The noninteracting impurity Green's function $G_{0}$ is determined through the steady-state variant of the Dyson equation $G_0 =( i\partial_t + \mu - \Sigma_{\rm bath})^{-1}$, while the interacting Green's function $G$ is obtained from a Dyson equation  $G =( i\partial_t + \mu - \Sigma_{\rm bath}+\Sigma_U)^{-1}$, with an additional self-energy $\Sigma_U$ due to interactions. In the example below, we approximate  $\Sigma_U$ by a (non self-consistent) 2nd order perturbation theory, i.e., $\Sigma_U$  is given by Eq.~\eqref{btrega4} with $\mathcal{G}$ replaced by $G_0$, evaluated in the steady state.

Finally, the current is defined by the rate of particle transfer from the left to the right reservoir, $J= \frac{dN_L}{dt}= -\frac{dN_R}{dt}$. An exact expression can be derived using equations of motion for the Green's functions, as in the appendix of Ref.~\cite{Werner2010}, and is given by the equal-time convolution 
\begin{align}
\label{currentsiam}
J = 2\left(\Sigma_L\ast G -  G\ast\Sigma_L\right)^<(t,t),
\end{align}
where the factor $2$ is due to spin. This relation is in general not satisfied away from half-filling in bare (non-conserving) second order perturbation theory, which we make use of for calculating the self-energy.  In the steady state, equation \eqref{currentsiam} can be evaluated using the {\tt convolution\_density\_matrix} method, see 
Eq.~\eqref{density_matrix_AB},
\begin{align}
\label{currentsiam-1}
J =  2i\left( \rho_{\Sigma_L,G} -  \rho_{G,\Sigma_L} \right) =-4\text{Im} \left( \rho_{\Sigma_L,G}   \right).
\end{align}
Here the second equation uses the Hermitian symmetry $ \rho_{\Sigma_L,G} = \rho_{G,\Sigma_L} ^\dagger$.

\subsubsection{Implementation}
\vspace*{2mm}
\label{sec:nessexample1-1}

The relevant files for the implementation, found in {\tt nessi/examples/}, are listed in Table \ref{tab_ex_2}.
\begin{table}[htb]
    \centering
    \begin{small}
    \begin{tabular}{|l|l|}
        \hline
                   \begin{minipage}{0.4\textwidth}
        \texttt{programs/ness2\_siam.cpp}          \end{minipage}& 
      \begin{minipage}{0.5\textwidth}
         Source code.
         \end{minipage}
          \\
        \hline
                   \begin{minipage}{0.4\textwidth}
        \texttt{utils/demo\_ness2\_siam.ipynb}          \end{minipage}& 
                      \begin{minipage}{0.5\textwidth}
Jupyter notebook to run the program.
         \end{minipage}
 \\\hline
                   \begin{minipage}{0.4\textwidth}
        \texttt{utils/demo\_ness2\_siam.py}          \end{minipage}& 
                   \begin{minipage}{0.5\textwidth}
         Python script;  same as the notebook.
         \end{minipage}
\\\hline
    \end{tabular}\caption{Relevant files for running the steady-state Anderson impurity model example.}\label{tab_ex_2}
    \end{small}  
\end{table}

The implementation is built on the functions in the {\tt ness2} namespace, which is included at the top of the source code:
\begin{lstlisting}[language=c++,numbers=none]
using namespace ness2;
\end{lstlisting}
The input parameters for the main program  are the physical parameters {\tt U} (interaction $U$), {\tt beta} (inverse temperature $\beta$), {\tt V} (voltage bias $V$), and {\tt epsd} (on-site energy $\epsilon_d$), as well as the numerical parameters {\tt Nft} (number of time/frequency points) and {\tt h} (timestep $h$). Moreover, we allow for a nonzero {\tt eta} for the regularization of the Dyson equation (which can however be set to zero in the example below, and would only be relevant if the level $\epsilon_d$ has no spectral overlap with the baths).  After reading the input parameters from the input file we initialize the {\tt herm\_matrix\_ness} object
\begin{lstlisting}[language=c++,numbers=none]
herm_matrix_ness G0(Nft,size); // size is 1 here
\end{lstlisting}
for the noninteracting Green's functions $G_0$, and similarly for 
{\tt G} ($G$), 
{\tt SL} ($\Sigma_L$), 
{\tt SR} ($\Sigma_R$), 
{\tt Sbath} ($\Sigma_L+\Sigma_R$), 
{\tt SU} ($\Sigma_U$), and 
{\tt S} ($\Sigma_L+\Sigma_R+\Sigma_U$). Next, the left and right baths are initialized with the given DOS  \eqref{abox}. For this we define a class to provide the DOS
\begin{lstlisting}[language=c++,numbers=none]
class box_dos{
public:
  double hi_,lo_,wc_,nu_;
  box_dos() wc_(10.0), nu_(3.0), hi_(15.0), lo_(-15.0) {}
  
  double operator()(double w){
    return 1/((1+exp(nu_*(w-wc_))*(1+exp(-nu_*(wc_+w)));
  }
};
\end{lstlisting}
Here {\tt hi\_} and {\tt lo\_} define the bounds of the integrals \eqref{gw2t01} and \eqref{gw2t03} in the spectral representation. The bath self-energies are then initialized using (c.f.~Table~\ref{tab:diagram_ness_equi})
\begin{lstlisting}[language=c++,numbers=none]
box_dos dos();
green_equilibrium_ness(FERMION,SL,dos,beta,+0.5*V,h,FFT_TRAPEZ);
green_equilibrium_ness(FERMION,SR,dos,beta,-0.5*V,h,FFT_TRAPEZ);
Sbath=SL;
Sbath.incr(SR,1.0,fft_domain::time); // Sbath+= SR on time-data
\end{lstlisting}
Next, we solve the noninteracting dyson equation using 
\begin{lstlisting}[language=c++,numbers=none]
cdmatrix eps_matrix(size,size); 
// complex-valued double precision variable-size eigen matrix
eps_matrix(0,0) = epsd; 
dyson(G0,mu,eps_matrix,Sbath,h,FFT_TRAPEZ,BATH_GAUSS,eta,beta,mu);
\end{lstlisting}
Here, we allow for the regularization using the Gaussian bath if {\tt eta} is nonzero.
With the resulting {\tt G0} one can determine the 2nd order self-energy. The structure of the diagram is analogous to the previous real-time example, and hence also the implementation is similar:
\begin{lstlisting}[language=c++,numbers=none]
herm_matrix_ness W(Nft, size);  // temporary W
Bubble1_ness(W, G0, G0);  //W(t,t') = ii*G0(t,t')G0(t',t)
W.smul(U*U,fft_domain::time); // W(t,t') = U^2 W(t,t')
Bubble2_ness(SU,G0,W); //Sigma_U(t,t')=ii*G0(t,t')W(t',t)
SU.smul(-1.0,fft_domain::time);
\end{lstlisting}
Finally, we  solve the interacting Dyson equation:
\begin{lstlisting}[language=c++,numbers=none]
S=Sbath;
S.incr(SU,1.0,fft_domain::time); // S += SU
dyson(G,mu,eps_matrix,S,h,FFT_TRAPEZ,BATH_GAUSS,eta,beta,0);
\end{lstlisting}
For postprocessing we compute the convolutions  (c.f.~Eq.~\eqref{currentsiam-1})
\begin{lstlisting}[language=c++,numbers=none]
cdmatrix SUG,SLG,rho;
convolution_density_matrix(SLG,FERMION,SL,G,h); // ii*[SL*G]^<(t=0)
convolution_density_matrix(SUG,FERMION,SU,G,h); // ii*[SU*G]^<(t=0)
density_matrix(rho,FERMION,G); //   = ii*G^<(t=0)
double Current = 4.0*SLG.trace().imag(); 
double Eint = (SUG).trace().real(); // interaction energy
double dens = 2*rho.trace().real(); // <n_up + n_do>
\end{lstlisting}
 At the end, all output is  stored into a single HDF5 file:
\begin{lstlisting}[language=c++,numbers=none]
// create the hdf5 file (char *flout points to the filename)
hid_t file_id = open_hdf5_file(flout);
// write Nft to group "/Nft" in the file :
store_int_attribute_to_hid(file_id, "Nft", Nft); 
[...] // similar for dens, Eint, Current
G0.write_to_hdf5(file_id, "G0"); // new group "/G0" in file  
[...]  // same for G,S,...
close_hdf5_file(file_id);
\end{lstlisting}
The HDF5 output is conveniently interpreted  using the Python utilities provided with the {\tt ReadNESS} modules.

\subsubsection{Results}
\vspace*{2mm}

\begin{figure}[tbp]
\centerline{\includegraphics[width=0.99\textwidth]{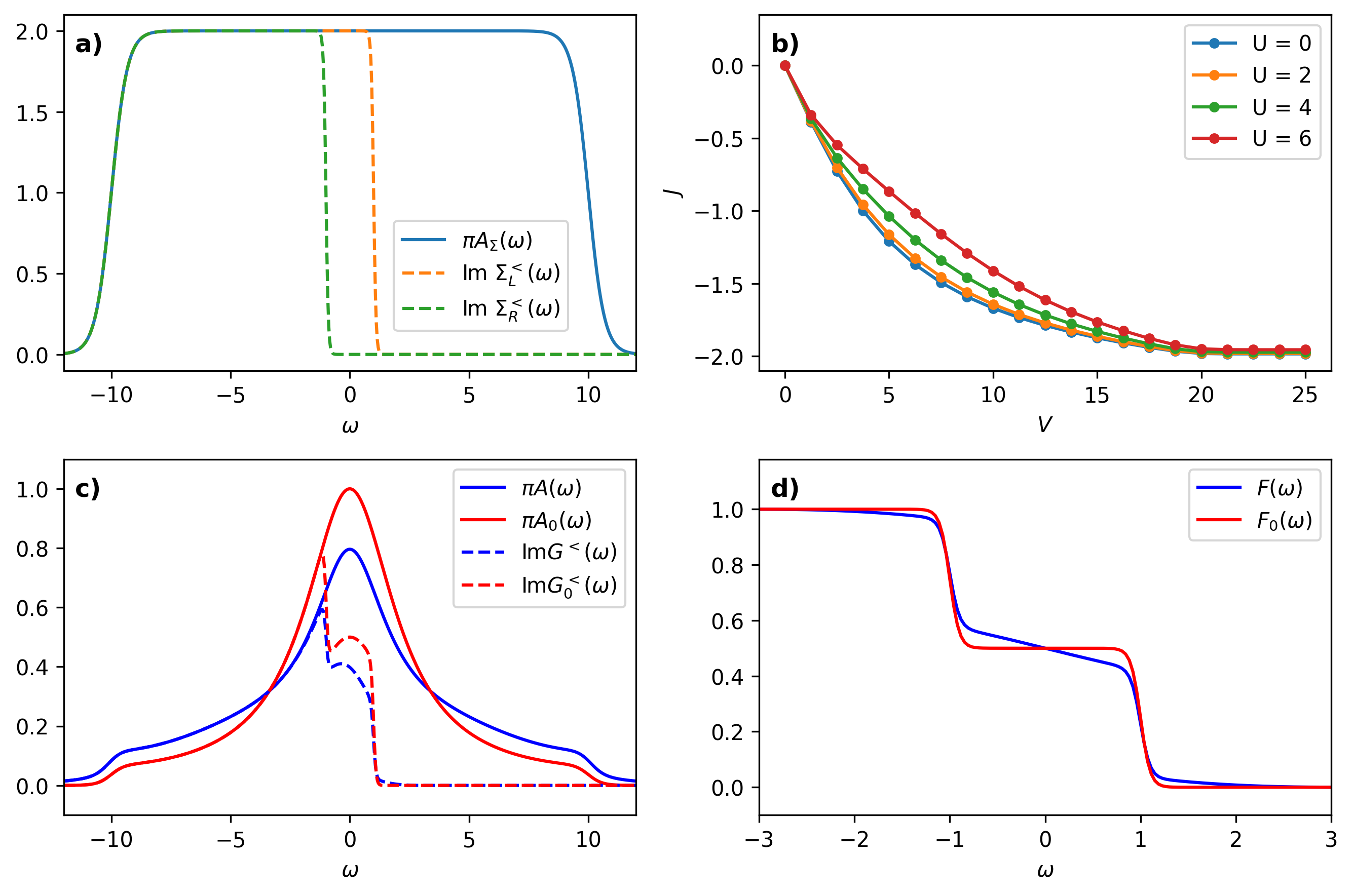}}
\caption{
Anderson model for  $U=6$, and $\beta=20$ ($\epsilon_d=0$). (a) Bath spectral function \eqref{abox} and occupation functions (dashed) for voltage $V=2$.
(b) Current as a function of voltage $V$, for $\beta=20$.
(c) Interacting spectral function $A(\omega)$ (full blue line) and noninteracting spectral function $A_0(\omega)$ (full red line) for $V=2$. The dashed lines show the  imaginary part of the corresponding lesser Green's functions. (d) Occupation functions \eqref{Fratio} for the interacting case ($F(\omega)$) and noninteracting case ($F_0(\omega)$) for $V=2$. }
\label{fig:test_siam_1}
\end{figure}

Fig.~{\ref{fig:test_siam_1}}(c) shows the converged results for the interacting  impurity spectral function $A(\omega)$ and the noninteracting  impurity spectral function $A_0(\omega)$, for the baths illustrated in Fig.~{\ref{fig:test_siam_1}}(a). The spectral function is essentially a Lorentzian peak, which becomes slightly more broadened for nonzero interaction $U$. The nonequilibrium nature of the state is evident from the distribution function
\begin{align}
\label{Fratio}
F(\omega) = \frac{G^<(\omega)}{2\pi i A(\omega)}.
\end{align}
The latter becomes clearly non-thermal, simultaneously reflecting the Fermi edges in the left and right bath (see Fig.~{\ref{fig:test_siam_1}}(d)). Interactions support thermalization and therefore slightly reduce the sharp edges, compare the blue and red curves in  Fig.~{\ref{fig:test_siam_1}}(c) for $F(\omega)=G^<(\omega)/2\pi i A(\omega)$ and $F_0(\omega)=G_0^<(\omega)/2\pi i A_0(\omega)$, respectively. 

Fig.~\ref{fig:test_siam_1}(b) shows the current \eqref{currentsiam} as function of the voltage, which evolves from the linear response regime at small $V$ to a saturated value of $J=2$ (corresponding to one transport channel for each spin) when $V$ becomes comparable to the bandwidth $2\omega_c$, such that the left (right) bath is full (empty). In the interacting case, the current is reduced, consistent with the broadening of the steps in the  distribution function.  Of course, the bare second order perturbation theory cannot correctly describe the Kondo effect at low temperatures and large $U$, which is beyond the scope of the present code. The steady-state code can however be easily  combined with more accurate diagrammatic computations of real-time Green's functions and self-energies in the steady state, which are nowadays becoming feasible with the help of various techniques \cite{Arrigoni2013,Profumo2015, Erpenbeck2023,Eckstein2024, Kim2024}. 

\begin{figure}[tbp]
\centerline{\includegraphics[width=0.99\textwidth]{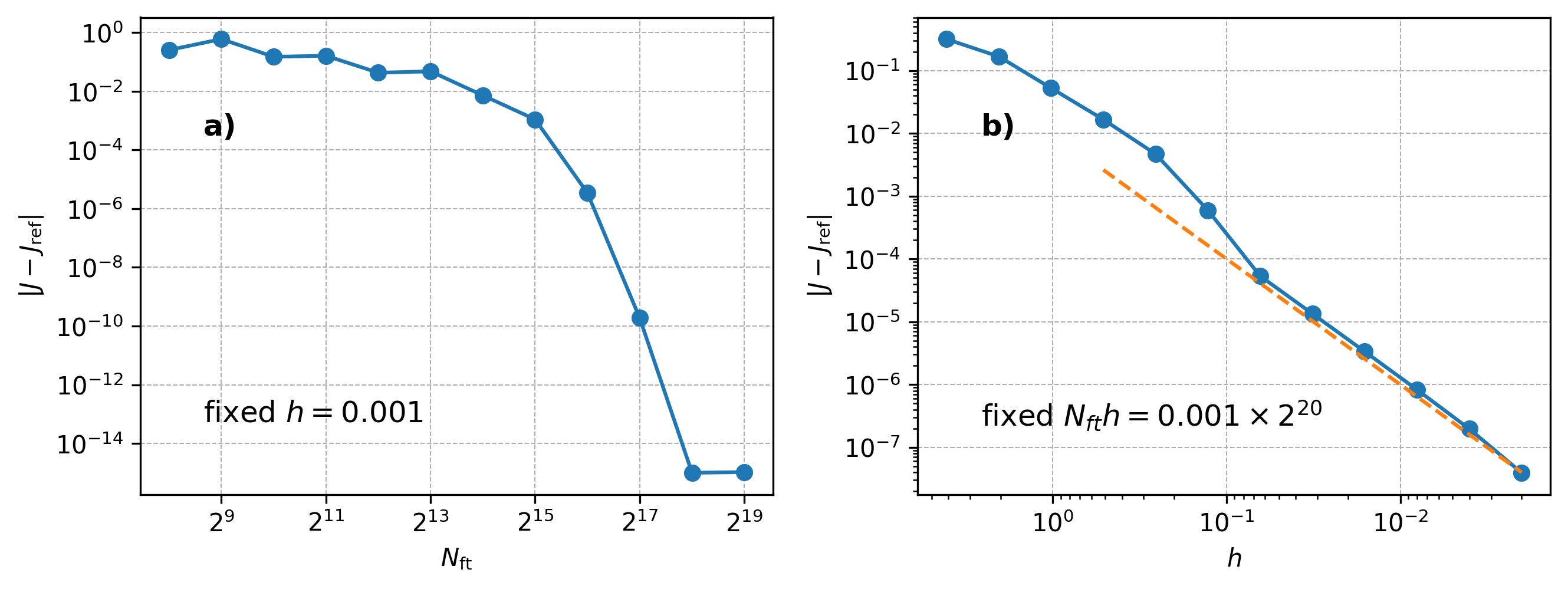}}
\caption{
a) Convergence of the steady-state solution for fixed timestep $h=0.001$ with $N_{\rm ft}$ (maximal time $t_{\rm c}\sim hN_{\rm ft}/2$), for $U=6$, $V=2$, $\beta=20$. b) Convergence of the steady-state solution with the timestep $h$, for fixed $h N_{\rm ft}=0.001\times 2^{20}$ (fixed $t_{\rm c}$). In both cases, the difference $\varepsilon=|J-J_{\rm ref}|$ of the current  to a reference $J_{\rm ref}$ is analyzed, where $J_{\rm ref}$ corresponds to the largest grid $N_{\rm ft}=2^{20}$ and smallest $h=0.001$. The dashed line in b) indicates the scaling behavior $\varepsilon\sim h^2$.
}
\label{fig:test_siam_2}
\end{figure}

In Fig.~\ref{fig:test_siam_2}, we demonstrate the numerical convergence of the steady-state approach (for $U=6$, $V=2$, $\beta=20$). We first fix a small value $h=0.001$ of the timestep, and perform simulations with different length $N_{\rm  ft}$ of the Fourier domain.  This tests the convergence with the maximal real time $t_{\rm c}= h N_{\rm ft}/2$ which is represented by the grid \eqref{t-grid}. For accurate results, it is necessary that all functions $G^{R,<}(t)$ and $\Sigma^{R,<}(t)$ essentially decay to zero within the domain $|t|<t_{\rm c}$. In Fig.~\ref{fig:test_siam_2}(a), we plot the difference $|J(N_{\rm ft}) - J_{\rm ref}|$, where the reference result $J_{\rm ref}$ is simply the result for the largest grid ($N_{\rm ft}=2^{20}$). The sharp drop in the error $|J(N_{\rm ft}) - J_{\rm ref}|$ is consistent with an exponential decay of $|G^{R,<}(t)|$ and $|\Sigma^{R,<}(t)|$, such that there is no dependence on $t_{\rm c}$ for sufficiently large $t_{\rm c}$. 

To analyze the convergence with the timestep $h$, we perform simulations with different $N_{\rm ft}$ and $h$, keeping the product $N_{\rm ft}h$ fixed. This corresponds to varying $h$ at fixed cutoff $t_c$; the latter is chosen as the largest value in Fig.~\ref{fig:test_siam_2}(a), i.e., $hN_{\rm ft} = 0.001 \times 2^{20}$. The error $|J(N_{\rm ft}) - J_{\rm ref}|$ with respect to the reference result $J_{\rm ref}$ at the largest $N_{\rm ft}$ (smallest $h$) is shown in Fig.~\ref{fig:test_siam_2}(b). The plot demonstrates an error of order $\mathcal{O}(h^2)$ (compare to the dashed line in Fig.~\ref{fig:test_siam_2}(b)), consistent with the trapezoidal evaluation of the convolution integrals (Sec.~\ref{sec:ness:impl}).

\subsection{DMFT in the steady state}
\label{sec:nessexample2}
\vspace*{2mm}

\subsubsection{Model setup}
\vspace*{2mm}

In order to benchmark the steady-state code with respect to the real-time propagation, we use a similar physical setup as in Sec.~\ref{sec:movingexample}. We again consider the Hubbard model \eqref{Hubbardmodel} on the Bethe lattice at half-filling, now with a time-independent interaction \(U\). This requires the self-consistent solution of Eqs.~\eqref{btrega1}, \eqref{btrega2}, and \eqref{btrega3} in the steady state. The self-energy is approximated by the second order diagram \eqref{btrega4}, but here we allow for two variations: (i) Self-consistent perturbation theory, where the self-energy is expanded in the fully interacting Green's function, such that $\mathcal{G}=G$, and (ii), iterated perturbation theory (IPT) \cite{Georges1996}, where $\Sigma_U$ is expanded in the bare Green's function of the impurity model. The latter is obtained via another Dyson equation,
\begin{align} \label{gweiss_ipt}
\mathcal{G}^{-1} = i\partial_t + \mu - \Delta,
\end{align}
with the self-consistent $\Delta$ given by Eq.~\eqref{btrega3}. 

We will solve the problem in thermal equilibrium at temperature $T=1/\beta$ in three ways: (i) First, we will use the two-time implementation with a timestep $h_{\rm cntr}$ up to a given time $t_{\rm max}$ (referred to as real-time or ``cntr'' simulation in the following). Here the equilibrium state is prepared through the imaginary time branch with a given number of timesteps $n_{\tau}$. The resulting Green's function should be translationally invariant in time, $G_{\rm cntr}^{R,<}(t_1,t_2)\equiv G^{R,<}(t_1-t_2)$. (ii) In a second calculation, the same problem is solved using the steady-state implementation with a given Fourier domain size $N_{\rm ft}$ and a timestep $h_{\rm ness}$ (referred to as NESS simulation in the following). The resulting NESS solution should match the real-time solution, $G_{\rm cntr}^{R,<}(t_1,t_2)= G_{\rm ness}^{R,<}(t_1-t_2)$ up to numerical accuracy, so that the real-time result can be used as a benchmark for the steady-state result. (iii)  Finally, we will demonstrate how the steady-state result can be used to prepare an initial equilibrium solution for the real-time evolution in the memory-truncated KBE, thereby avoiding the need for the imaginary-time simulation. 

\subsubsection{Implementation}
\vspace*{4mm}

The relevant files for the implementation, found in {\tt nessi/examples/}, are listed in Table \ref{tab_ex_3}. For installation instructions, see Sec.~\ref{sec:examples_ramarks}.
\begin{table}[htb]
    \centering
    \begin{small}
    \label{tab:nessex2}
    \begin{tabular}{|l|l|}
        \hline
                   \begin{minipage}{0.4\textwidth}
        \texttt{programs/ness2\_bethe\_prop.cpp}          \end{minipage}& 
      \begin{minipage}{0.45\textwidth}
         Source code for the real-time evolution.
         \end{minipage}
          \\\hline
                   \begin{minipage}{0.4\textwidth}
        \texttt{programs/ness2\_bethe.cpp}          \end{minipage}& 
              \begin{minipage}{0.5\textwidth}
Source code for the NESS solution.
         \end{minipage}
          \\\hline
                   \begin{minipage}{0.4\textwidth}
        \texttt{programs/ness2\_bethe\_trunc.cpp}          \end{minipage}& 
              \begin{minipage}{0.5\textwidth}
Memory-truncated~KBEs starting from NESS.
         \end{minipage}
\\\hline
                   \begin{minipage}{0.4\textwidth}
        \texttt{utils/demo\_ness2\_bethe.ipynb}          \end{minipage}& 
                   \begin{minipage}{0.5\textwidth}
         Jupyter notebook to run the program.
         \end{minipage}
 \\\hline
                   \begin{minipage}{0.4\textwidth}
        \texttt{utils/demo\_ness2\_bethe.py}          \end{minipage}& 
                   \begin{minipage}{0.5\textwidth}
         Python script;  same as the notebook.
         \end{minipage}
\\\hline
    \end{tabular}\caption{Relevant files for running the DMFT in the steady-state example.}\label{tab_ex_3}
        \end{small}
\end{table}

\subsubsection*{Implementation: Real-time simulation}
\vspace*{4mm}

The real-time benchmark  {\tt ness2\_bethe\_prop.cpp} is almost identical to the startup routine  {\tt trunc\_bethe\_} {\tt start.cpp} in the example for the memory truncated KBEs (sec.~\ref{sec:movingexample}), and will therefore not be discussed in detail here. The  difference is that the evolution computed for a time-independent $U$ (such that a DMFT iteration is also needed on the imaginary time branch), there is no determination of momentum-dependent Green's functions $G_k$, and there is an input flag {\tt ipt\_flag} to choose between an IPT self-energy ({\tt ipt\_flag=1}) and self-consistent perturbation theory ({\tt ipt\_flag=0}). The IPT solution involves one more call at each timestep to solve the Dyson equation \eqref{gweiss_ipt}. 

\subsubsection*{Implementation: NESS simulation}
\vspace*{4mm}
\label{sec:impl:ness:bethe}

The input parameters for the main program are the flag {\tt ipt\_flag} to choose the self-energy, the physical parameters {\tt U} (interaction $U$), {\tt beta} (inverse temperature $\beta$), {\tt mu} (chemical potential $\mu$), as well as the numerical parameters {\tt Nft} (number of time/frequency points) and {\tt h} (timestep $h_{\rm ness}$). In addition, the parameters {\tt N\_it} (maximum number of iterations), {\tt errmax} (error cutoff) and {\tt mix} (linear mixing) are used to control the DMFT iteration (see below). Finally, the flag {\tt out\_every} allows to save the Green's functions at every DMFT iteration.  The implementation is built on the functions in the {\tt ness2} namespace, which is included at the top of the source code:
\begin{lstlisting}[language=c++,numbers=none]
using namespace cntr;
using namespace ness2;
\end{lstlisting}
After reading the input from a file, we allocate a {\tt herm\_matrix\_ness} object with {\tt size=1} for the local Green's function $G$,
\begin{lstlisting}[language=c++,numbers=none]
herm_matrix_ness G_ness(Nft,size);
\end{lstlisting}
and similarly for {\tt SU\_ness} ($\Sigma$),  {\tt Gweiss\_ness} ($\mathcal{G}$) and some temporary variables. Next, $G$ is initialized with the noninteracting Green's function
\begin{lstlisting}[language=c++,numbers=none]
ness2::bethedos dos(-2,2); // bethe dos: dos(w)=sqrt(4-w**2)/(2pi)
green_equilibrium_ness(FERMION,G_ness,dos,beta,mu,h,FFT_TRAPEZ);    
\end{lstlisting}
Because the accuracy of the initialization is not too important, we can use the faster {\tt FFT\_TRAPEZ} method instead of the slower {\tt FFT\_ADAPTIVE}.  Following this, we enter a loop over at most {\tt N\_it} iterations for the DMFT self-consistency. At the beginning of each loop, we copy the current Green's function into a new {\tt herm\_matrix\_ness} object {\tt G\_old}. We can then compute $\mathcal{G}$ either by copying $\mathcal{G}=G$ (self-consistent perturbation theory), or by solving Eq.~\eqref{gweiss_ipt} (IPT):
\begin{lstlisting}[language=c++,numbers=none]
if(ipt_flag) dyson(Gweiss_ness,mu,ham,G_ness,h,FFT_TRAPEZ); // IPT
else  Gweiss_ness=G_ness;
\end{lstlisting}
Here {\tt ham} is a zero matrix of dimension {\tt size=1}. The computation of the self-energy is then identical to the corresponding section in the steady-state example of Sec.~\ref{sec:nessexample1} ({\tt W\_ness} is a temporary {\tt herm\_matrix\_ness}):
\begin{lstlisting}[language=c++,numbers=none]
Bubble1_ness(W_ness, Gweiss_ness, Gweiss_ness);  // W = ii * G * G
W_ness.smul(U*U,fft_domain::time);  // W <-- W*U**2
Bubble2_ness(SU_ness, Gweiss_ness, W_ness);  // SU = ii * G * W
SU_ness.smul(-1.0, fft_domain::time);       // SU *= -1
K_ness.set_zero(fft_domain::time);          // kernel K=0
K_ness.incr(SU_ness,1.0,fft_domain::time);  // K += SU
K_ness.incr(G_ness,1.0,fft_domain::time);   // K += G (using Delta=G)
\end{lstlisting}
Finally, we solve the Dyson equation \eqref{btrega1}, and compute the $L_2$ norm difference to the previous iteration  (stored in {\tt G\_old}) on the time grid:
 \begin{lstlisting}[language=c++,numbers=none]           
dyson(G_ness, mu, ham, K_ness, h, FFT_TRAPEZ); // update G_ness
err = distance_norm2(G_ness,G_old,fft_domain::time);
\end{lstlisting}
The convergence of the DMFT iteration is typically improved if the Green's function at the new iteration is not taken as the updated result $G'$, but as a linear combination $G_{\rm new} = \alpha G' + (1-\alpha) G_{\rm old} $ ($\alpha$ is the mixing parameter $\tt mix$). 
 \begin{lstlisting}[language=c++,numbers=none]                 
G_ness.smul(mix,fft_domain::time); // linear mixing for convergence
G_ness.incr(G_old,1.0-mix,fft_domain::time);
\end{lstlisting}
The DMFT iteration loop is stopped if the maximum number of iteration is reached, or if the $L_2$ error {\tt err} falls below the cutoff  provided by {\tt errmax}. Finally, we write all relevant Green's functions to a HDF5 file, just as in  the example of Sec.~\ref{sec:nessexample1}. To read the file one can again use the Python utilities provided with the {\tt ReadNESS} modules. 

\begin{figure}[tbp]
\centerline{\includegraphics[width=0.99\textwidth]{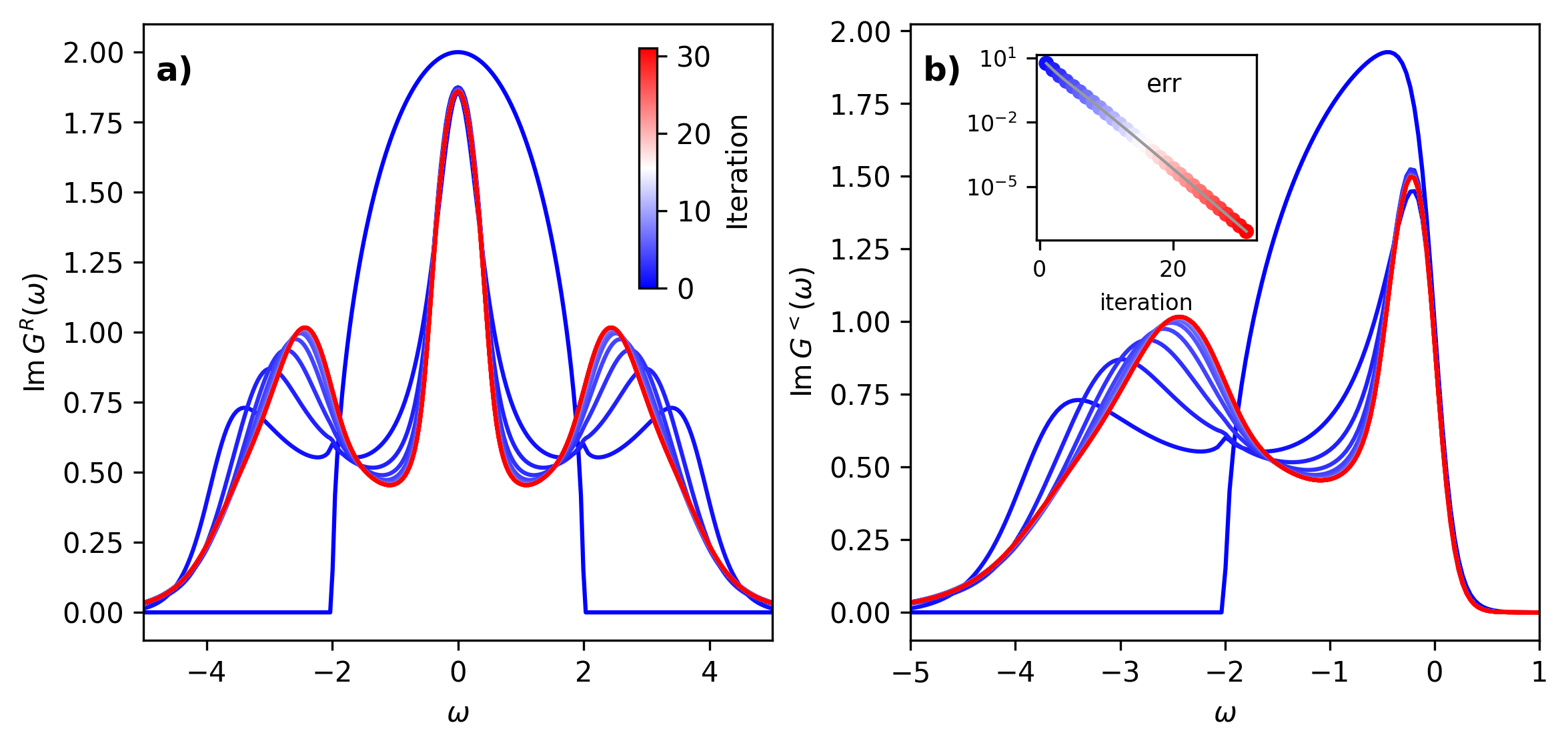}}
\caption{
Steady-state DMFT loop ($U=4$, $\beta=10$, IPT), with $N_{\rm ft}=2^{13}$ and a timestep $h_{\rm ness}=0.02$. (a) Convergence of the spectral function
with iteration iter=$0,...,30$ (see color bar). (b) Same as (a), but for  $\text{Im}G^<(\omega)$. The inset in (b) shows the convergence error $\epsilon_{\rm iter}= |G_{\rm iter+1}-G_{\rm iter}|_2$ as a function of iteration; the difference is evaluated in the time-domain, as explained in Sec.~\ref{sec:impl:ness:bethe}.}
\label{fig:test_bethe_1}
\end{figure}

\subsubsection{Results}
\vspace*{2mm}

We first show in Fig.~\ref{fig:test_bethe_1} the convergence of the spectrum $A(\omega)$ and the occupied DOS $G^<(\omega)$ during the DMFT loop. In general, we observe that the linear mixing is more relevant for the DMFT loop in the steady-state implementation than for the imaginary-time evolution within the real-time code. (In the example, we use a mixing factor $\alpha=0.5$.) Moreover, if the time cutoff $\sim hN_{\rm ft}$ in the simulation is not sufficient for the Dyson equation to be accurately solved, the decrease of the DMFT error with iteration slows down around a value that is determined by the accuracy of the Dyson equation.

\begin{figure}[tbp]
\centerline{\includegraphics[width=0.99\textwidth]{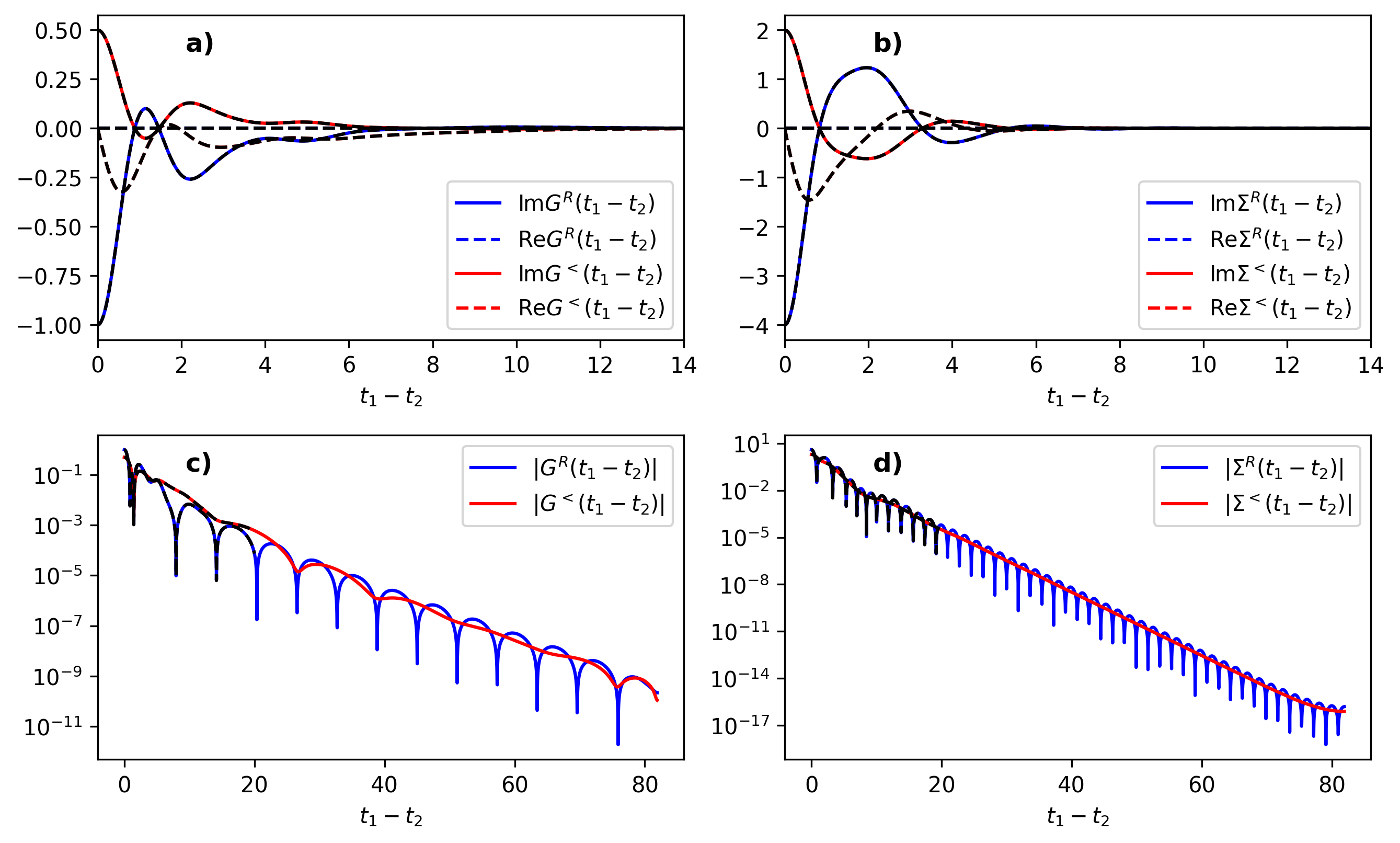}}
\caption{
Converged Green's function $G$ (a) and self-energy $\Sigma$ (b) for the same parameters as in Fig.~\ref{fig:test_bethe_1}. Colored lines correspond to the results for the NESS simulation, while the dashed black lines, which lie on top of the colored lines, correspond to the results obtained from the last timestep of the real-time simulation (see main text). The real-time simulation has been performed with $h_{\rm cntr}=h_{\rm ness}=0.02$, {\tt nt=1000}, and {\tt ntau=1000} (the black dashed real part curves are exactly on top of the colored ones).  (c) and (d) Same data as in panels  (a) and (b), but on a larger time window and on a logarithmic scale.
}
\label{fig:test_bethe_2}
\end{figure}

In Fig.~\ref{fig:test_bethe_2}, we demonstrate that the NESS simulation and the real-time simulation converge to the same result.   In equilibrium, the real-time solution produces Green's functions which are translationally invariant in time. A slice of such a function at a given timestep  $t_1$ should therefore coincide with the corresponding NESS result,
\begin{align}
\label{ness-cntr}
X_{\rm cntr}(t_1,t_2) = X_{\rm ness}(t_1-t_2),
\end{align}
for $X=G^{R,<},\Sigma^{R,<}$. The colored lines in Figs.~\ref{fig:test_bethe_2}(a) and (b) show the results for the NESS simulation, while the dashed black lines correspond to the last time slice ($t_1=h_{\rm cntr}\cdot n_t$) of the real-time simulation.
Both real-time and NESS calculations take less that a minute on a MacBook with an Apple M1 processor.  The results indeed coincide within the line-width of the plots. A quantitative analysis will be given below. Moreover, Figs.~\ref{fig:test_bethe_2}(c) and (d) show $G^{R,<}(t)$ and $\Sigma^{R,<}(t)$ on a logarithmic scale over the full time domain $|t|<t_{\rm c}=h_{\rm ness}(N_{\rm ft}/2-1)$ of the NESS simulation. This demonstrates the decay of the functions at the boundary of the domain, which is needed for an accurate solution of the Dyson equation in the steady-state formalism. 

\begin{figure}[tbp]
\centerline{\includegraphics[width=0.99\textwidth]{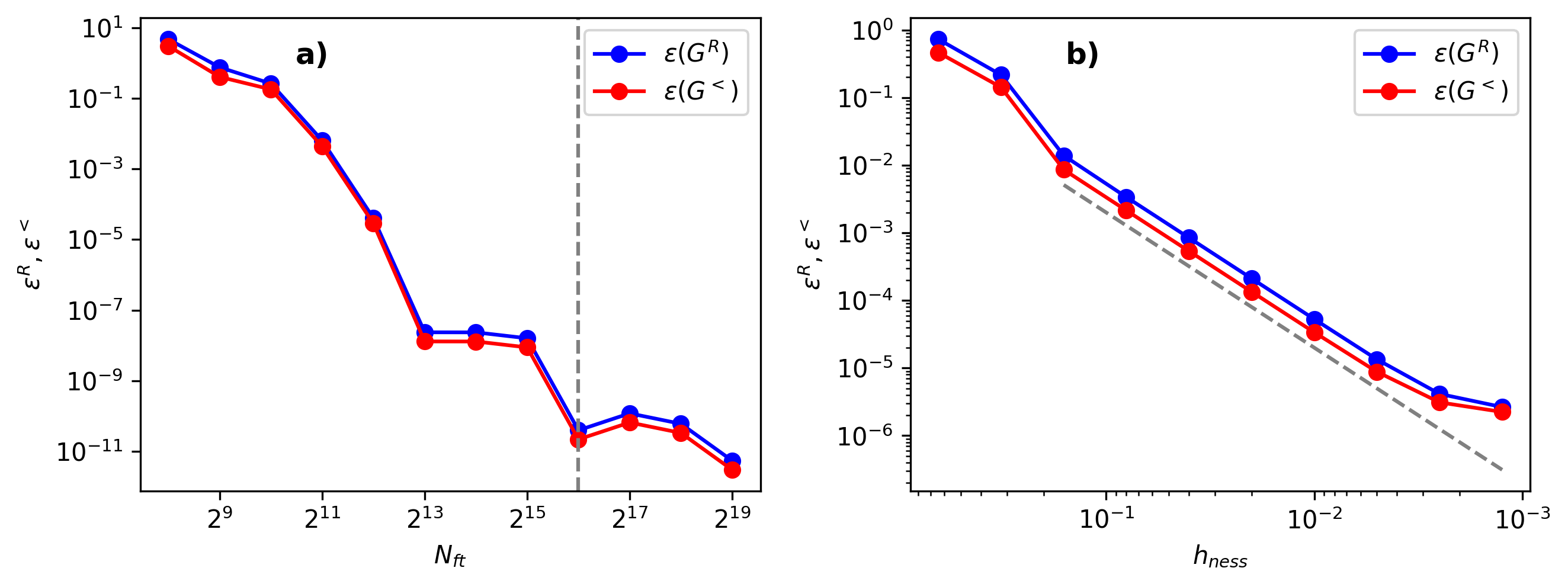}}
\caption{
Convergence of the NESS simulation with increasing $N_{\rm ft}$ and decreasing $h_{\rm ness}$ ($U=4$, $\beta=10$, IPT). (a) Difference $\epsilon(G^{R,<}) = |G^{R,<}_{N}-G^{R,<}_{N^*}|_2$ between a steady-state DMFT solution with domain size $N_{\rm ft}=N$ and a reference simulation with large $N_{\rm ft}=N^*=2^{20}$, 
at fixed timestep $h_{\rm ness}=0.02$.  The difference is computed on the smallest common grid. (b) Difference $\epsilon^{R,<}\equiv |G^{R,<}_{\rm ness}(t_1-t_2)-G^{R,<}_{\rm cntr}(t_1,t_2)|_2$ between the steady-state result and the real-time benchmark $G_{\rm cntr}$, as a function of $h_{\rm ness}$ (see main text). Results are obtained for fixed $N_{\rm ft}\times h_{\rm ness}=2^{16}\times 0.02$; corresponding to a cutoff $t_{\rm c}\approx 650$, well beyond the convergence threshold found in (a). The dashed line in (b) indicates an error scaling $\epsilon \sim h^2$.
}
\label{fig:test_bethe_3}
\end{figure}

Finally, we can quantitatively demonstrate the convergence of the NESS and real-time calculations to the same results (for the same parameters $U=4$, $\beta=10$, and IPT).  Because of the high-order accurate quadrature used in the \NESSi{} real-time simulation, the real-time result $G_{\rm cntr}$ with $h_{\rm cntr}=0.02$ and {\tt ntau=1000} can be taken as an accurate benchmark. In Fig.~\ref{fig:test_bethe_3}(a), we first show the convergence of the NESS result at a fixed timestep $h_{\rm ness}=0.02$ for increasing $N_{\rm ft}$, corresponding to an increasing time cutoff $t_{\rm c}\sim h_{\rm ness}N_{\rm ft}/2$. One can see that the results are converged with the cutoff for $N_{\rm ft}\gtrsim 2^{13}$, corresponding to a cutoff $t_{\rm c}\approx 80$. This is consistent with the exponential decay of $G$ and $\Sigma$ shown in Fig.~\ref{fig:test_bethe_2}(c) and (d). Next we perform a series of simulations with different $N_{\rm ft}$ and fixed $h_{\rm ness}N_{\rm ft}$ by varying the timestep $h_{\rm ness}$ at fixed cutoff $t_{\rm c}$. The convergence is analyzed in terms of the difference $\epsilon^{R,<} = |G^{R,<}_{\rm ness}(t_1-t_2)-G^{R,<}_{\rm cntr}(t_1,t_2)|_2$ between the steady-state result and the real-time benchmark $G_{\rm cntr}$, on the largest timestep $t_1=n_th_{\rm cntr}$ of the real-time simulation (Fig.~\ref{fig:test_bethe_3}(b)).  The cutoff $t_{\rm c}$ for the simulations in Fig.~\ref{fig:test_bethe_3}(b) is well  beyond the threshold found in Fig.~\ref{fig:test_bethe_3}(a), with $t_{\rm c}\sim 0.02\cdot 2^{16}/2\approx 650$. The comparison again confirms the decrease of the error like $\mathcal{O}(h^2)$, similar to Fig.~\ref{fig:test_siam_2}(b), consistent with the trapezoidal evaluation of the convolution integrals. The saturation of the error at small $h_{\rm ness}$ can be related to numerical errors in the benchmark itself, or in the convergence of the DMFT iteration. 

\subsubsection{Interface with the memory-truncated KBE}
\vspace*{2mm}

Because both the memory-truncated and the NESS Green's functions are entirely defined in terms of their real-time components $G^R$ and $G^<$, one can straightforwardly exchange data between the two objects (see Table~\ref{tab:conversion}). Here we provide an example that illustrates how a NESS simulation can be used to initialize the time evolution with the truncated KBE. This circumvents the initialization via the imaginary time propagation, as in the startup routine of the example in Sec.~\ref{sec:movingexample}. Specifically, we will initialize memory-truncated Green's functions and self-energies for the Hubbard model on a Bethe lattice using the result of the steady-state simulation in {\tt ness2\_bethe.x}, and then use the truncated KBEs to further propagate the solution in time. While the expected result should simply maintain the time-translationally invariant solution \eqref{ness-cntr}  for all times, this is still a numerically nontrivial test which can be used to check the accuracy of the approach. 

Running the test is also part of {\tt demo\_ness2\_bethe.ipynb}. We first use the NESS simulation with {\tt ness2\_bethe.x} to prepare a steady-state equilibrium solution with a given timestep $h_{\rm ness}$ and domain size $N_{\rm ft}$.  The output file of the NESS simulation is then read by the executable {\tt ness2\_bethe\_trunc.x}, to perform the truncated KBE simulation. Because the numerical error in the NESS simulation decreases with the timestep only like $\mathcal{O}(h^2)$, in contrast to the higher order accurate real-time KBE implementation, it can be beneficial to perform the NESS simulation with a smaller timestep than the one used in the truncated evolution. We will set $h_{\rm cntr}=d\cdot h_{\rm ness}$, with an integer factor $d$.  

The truncated simulation thus takes as input the parameters {\tt downsampling} (the factor $d$), {\tt tc} (memory cutoff in the KBE simulation), {\tt tmax} (number of timesteps over which the KBE is propagated), {\tt out\_every} (frequency at which time slices of the KBE are written to file), and {\tt CorrectorSteps} (number of DMFT iterations at each timestep, see explanation in Sec.~\ref{sec:impl:kbe}).  Further parameters ($U$, $\beta$, $h_{\rm ness}$, $N_{\rm ft}$, and the {\tt ipt\_flag}) are read, together with the Green's functions, from the HDF5 file which is written as output of the NESS  simulation. 

After reading the input, the main step is the initialization of the memory-truncated functions, which we explain exemplarily for the Green's function. (All routines use the namespaces {\tt ness2} and {\tt cntr}.) First, we allocate the memory truncated function with a cutoff {\tt tc} and orbital dimension {\tt size=1}:
\begin{lstlisting}[language=c++,numbers=none]
herm_matrix_moving<double> Gcntr(tc,1,FERMION);
\end{lstlisting}
We then read the NESS Green's functions from the HDF5 output of the NESS simulation (with name {\tt ness\_filename}),
where it is stored under a group {\tt G}:
\begin{lstlisting}[language=c++,numbers=none]
herm_matrix_ness Gness;
Gness.read_from_hdf5(ness_filename,"G"); 
\end{lstlisting}
The object {\tt Gness} is thereby also resized to the correct dimension {\tt Nft}. Since the real-time evolution will be run on a grid that is coarser by a factor $d$, we must downsample the steady-state Green's function:
\begin{lstlisting}[language=c++,numbers=none]
herm_matrix_ness Gness1;
Gness1=downsample(Gness,downsampling);
\end{lstlisting}
This will automatically resize {\tt Gness1} to the correct dimension {\tt Nft1}, where {\tt Nft1=Nft/downsampling}. In the time domain, downsampling simply copies every $d$th value of {\tt Gness} (on the fine grid) to ${\tt Gness1}$ (coarse grid). In turn, the frequency grid \eqref{w-grid} is restricted to the lowest frequencies {\tt \{-Nft1/2, ..., Nft1/2-1\}}. The routine therefore requires {\tt Nft} to be an integer multiple of $d$, and {\tt Nft1} to be a multiple of $2$. Here we  choose both {\tt Nft} and $d$ to be  a power of $2$. Finally, the memory truncated Green's function is initialized using 
\begin{lstlisting}[language=c++,numbers=none]
ness2cntr(Gcntr, Gness1);
\end{lstlisting}
Here {\tt ness2cntr} will  initialize {\tt G\_{\rm cntr}} such that 
\begin{align}
G_{\rm cntr}^{<,R}(t_1,t_2)=G^{<,R}_{\rm ness}(t_1-t_2)
\end{align}
is translationally invariant in time in the full moving domain $\mathcal{M}[G]$ (Eq.~\eqref{time slice-moving}).
The cutoff {\tt tc} of $G_{\rm cntr}$ should be smaller or equal to the maximum number of timesteps in {\tt Gness1}, which is {\tt Nft1/2-1}; otherwise part of  $G_{\rm cntr}$ would be left zero. The initialization is repeated for the self-energy and for $\mathcal{G}$, which are both stored in the NESS output file. 

After this point, the truncated time evolution in {\tt ness2\_bethe\_trunc.x} is identical to the example of Sec.~\ref{sec:impl:kbe}, where the truncated Green's functions are initialized from a full real-time simulation. Selected time slices {\tt t} of the KBE simulation will be stored in the HDF5 output file under a group with key t[{\tt t}]/G. As an alternative to writing the slices directly, as in the example {\tt trunc\_bethe} above, we can use the reverse data exchange {\tt cntr2ness} to initialize a {\tt herm\_matrix\_ness} from the real-time Green's function, and store the latter:
\begin{lstlisting}[language=c++,numbers=none]
void write_slice(hid_t fl_id,int t,herm_mat_moving<double> &Gcntr){
  // fl_id is a handle to an open and writable HDF5 file
  [...] // create herm_matrix_ness Gness with Nft/2-1>=G.tc_
  cntr2ness(Gness,Gcntr); // read last timestep of G into Gness 
  // write Gness to new group t[t]/G in file fl_id:
  hid_t sub_group_id = create_group(file_id,"t"+std::to_string(t));
  Gness.write_to_hdf5(sub_group_id, "G");
  close_group(sub_group_id);
}
\end{lstlisting}

\begin{figure}[tbp]
\centerline{\includegraphics[width=0.99\textwidth]{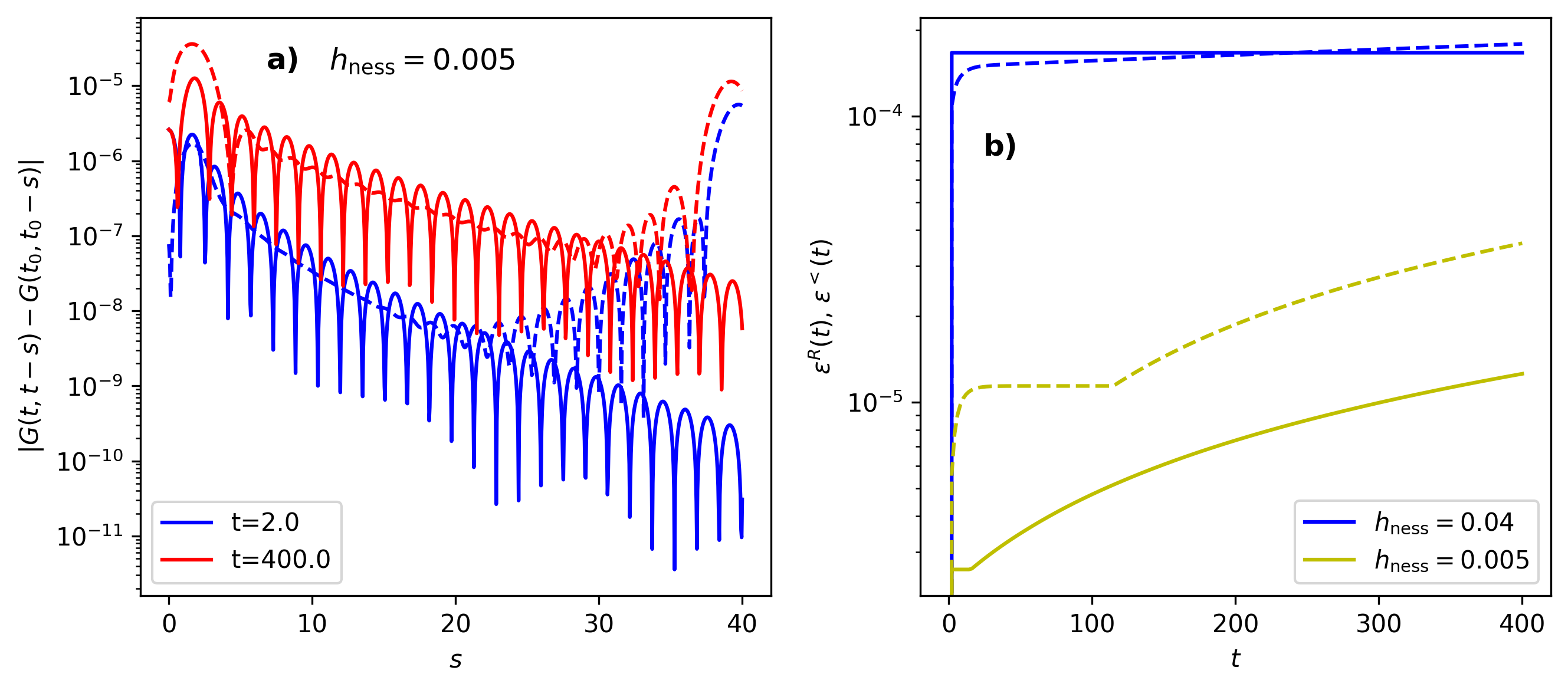}}
\caption{
Memory-truncated time evolution of an equilibrium state, initialized with a NESS simulation.
(Self-consistent perturbation theory, $U=1$, $\beta=2$.) The truncated KBE evolution is performed with $h_{\rm cntr}=0.04$ and a cutoff {\tt tc=1000}, and the initial NESS simulation uses $h_{\rm ness}=0.005$.  (a) Difference $\epsilon^{R,<}(t,s)=|G_{\rm cntr}^{R,<}(t,t-s)-G_{\rm cntr}^{R,<}(t_0,t_0-s)|$ between the Green's functions on time slice $t$ and the first time slice $t_0=0$. Full (dashed) lines show $\epsilon^R$ ($\epsilon^<$). (b) Maximum difference $\epsilon^{R,<}(t)=\text{max}_s (\epsilon^{R,<}(t,s))$ as function of $t$, for different $h_{\rm ness}$.}
\label{fig:test_bethe_4}
\end{figure}

Some results are shown in Fig.~\ref{fig:test_bethe_4}. We performed the test for similar parameters as in the example in Sec.~\ref{sec:impl:res}, using a self-consistent perturbation theory for $U=1$ and $\beta=2$. Note that this high temperature is of the order of magnitude of the final thermalized temperature in the previous example, rather than the initial temperature before the quench. Lower temperatures typically require a longer memory cutoff, because a sharp Fermi edge in the distribution function $G^<(\omega)$ implies a slower decay of $G^<(t)$. We also note that the IPT simulation is less stable under long-time evolution, which may be related to its non-conserving nature \cite{Eckstein2010}.

Figure~\ref{fig:test_bethe_4}(a) shows the comparison of the real-time result $G_{\rm cntr}^{R,<}(t,t-s)$ to the initial NESS Green's function $G_{\rm ness}^{R,<}(s)$, which matches by construction on the initial time slice $t=0$. The two time slices $t=2$ and $t=400$ correspond to $50$ and $10000$ timesteps $h_{\rm cntr}=0.04$, respectively. One can see that very early a difference is building up in the lesser component at large relative times $s$, which is related to the finite cutoff $t_c$ in the memory-truncated evolution (which is smaller than the maximum time $h_{\rm ness} N_{\rm ft}/2$ in the NESS simulation). At later times, the difference $|G_{\rm cntr}^{R,<}(t,t-s)-G_{\rm ness}^{R,<}(s)|$ grows slowly due to a linear error accumulation, while remaining small on the absolute scale. In Fig.~\ref{fig:test_bethe_4}(b), we analyze the maximum difference 
\begin{align}
\epsilon^{R,<}(t)=\text{max}_s\left(|G_{\rm cntr}^{R,<}(t,t-s)-G_{\rm ness}^{R,<}(s)|\right)
\end{align}
 over a time slice $t$ as a function of $t$. If the NESS simulation is performed with the same timestep as the real-time simulation, the NESS simulation is less accurate and therefore slightly inconsistent with a translationally invariant real-time solution (see the results for $h_{\rm ness}=h_{\rm cntr}=0.04$ in Fig.~\ref{fig:test_bethe_4}(b)). The time evolution then leads to the  build-up of a nonnegligible difference $\epsilon^{R,<}(t)$ over a few timesteps, so that the linear error accumulation  starts from a higher level. This initial error buildup can be simply reduced by performing the NESS simulation with a smaller timestep (see the results for $h_{\rm ness}=0.005$ in Fig.~\ref{fig:test_bethe_4}(b)). Since the numerical  cost of the NESS simulation scales only with $\mathcal{O}(N_{\rm ft}\log N_{\rm ft})$ due to the use of FFT, the preparation of the equilibrium state via the NESS simulation is still cheaper than the preparation via a full real-time evolution.

\section{Conclusions}
We presented the new version \NESSitwo{} of the open-source computational physics library \NESSi{}. \NESSitwo{} extends the Green's function based nonequilibrium dynamics simulation package by the memory-truncated KBE formalism as well as nonequilibrium steady-state functionalities. We described the newly added classes and routines in detail and showcased their usage and intuitive integration into the existing framework with illustrative example programs. 
This should facilitate the implementation of 
custom programs and usage of this library for a 
wide range of applications.

Also for the extension \NESSitwo{} we appreciate feedback as well as contributions from the user community. For this we suggest direct contact with the main \NESSi{} authors and for any issues with the library we refer the user to the contact address on the website. We continuously keep working on future additions to this library, especially a software package for nonequilibrium dynamical mean-field theory calculations based on strong-coupling impurity solvers. A renewed web page {\tt https://nessi.readthedocs.io/en/latest/} is available and will be kept up to date. It contains a link to the repository of the \NESSitwo{} extension as well as updated installation instructions, a detailed manual of all relevant classes and routines of the memory-truncated and steady-state code and the new example programs.

\section*{CRediT authorship contribution statement}
\textbf{Fabian K\"unzel}: Software, Validation, Writing - original draft, Writing - review \& editing, Visualization, Project administration. 
\textbf{Michael Sch\"uler}: Software, Validation, Writing - review \& editing, Visualization, Supervision. 
\textbf{Denis Gole\v{z}}: Software, Validation, Writing - original draft, Writing - review \& editing, Visualization. 
\textbf{Yuta Murakami}: Software, Validation, Writing - original draft, Writing - review \& editing, Visualization. 
\textbf{Sujay Ray}: Software, Validation, Writing - review \& editing, Visualization. 
\textbf{Christopher Stahl}: Conceptualization, Software, Validation, Writing - review \& editing, Visualization. 
\textbf{Jiajun Li}: Software, Validation, Writing - review \& editing. 
\textbf{Hugo U.R. Strand}: Software, Writing - review \& editing, Supervision. 
\textbf{Philipp Werner}: Conceptualization, Validation, Writing - original draft, Writing - review \& editing, Supervision, Project administration, Funding acquisition. 
\textbf{Martin Eckstein}: Software, Conceptualization, Methodology, Writing - original draft, Writing - review \& editing, Supervision, Project administration, Funding acquisition.

\section*{Declaration of competing interest}
The authors declare that they have no known competing financial interests or personal relationships that could have appeared to influence the work reported in this paper.

\section*{Acknowledgements}
We thank Paul Fadler and Bastian Schindler for valuable feedback while developing the library. F.K. and M.E. were funded by the Deutsche Forschungsgemeinschaft through QUAST- FOR5249-449872909 (Project P6),  and through the Cluster of Excellence ``CUI: Advanced Imaging of Matter'' of the Deutsche Forschungsgemeinschaft (DFG) - EXC 2056 - project ID 390715994. M.S. acknowledges support from the
NCCR MARVEL, a National Centre of Competence in Research, funded by the Swiss National Science Foundation
(Grant Number 205602). D.G. acknowledges support from No. P1-0044, No. J1-2455, No. J1-2458, and No. MN-0016-106 of the Slovenian Research Agency (ARIS). Y.M. is  supported by a Grant-in-Aid for Scientific Research from the Japan Society for the Promotion of Science (JSPS) (Project Numbers: JP21H05017, JP24H00191, JP25K07235). S.R. and P.W. acknowledge support from the Swiss National Science Foundation via Grant No.~200021-196966, 2000-1-240023, and NCCR Marvel.

\bibliographystyle{elsarticle-num} 
\bibliography{bib.bib}

\begin{thebibliography}{10}
\expandafter\ifx\csname url\endcsname\relax
  \def\url#1{\texttt{#1}}\fi
\expandafter\ifx\csname urlprefix\endcsname\relax\def\urlprefix{URL }\fi
\expandafter\ifx\csname href\endcsname\relax
  \def\href#1#2{#2} \def\path#1{#1}\fi

\bibitem{KadanoffBaym_1962}
L.~Kadanoff, G.~Baym, Quantum Statistical Mechanics: Green's Function Methods
  in Equilibrium and Nonequilibrium Problems, Frontiers in physics, W.A.
  Benjamin, 1962.

\bibitem{Keldysh_1964}
L.~V. Keldysh, Diagram technique for nonequilibrium processes, Sov.~Phys.~JETP
  20 (1965) 1018.

\bibitem{KamenevBook}
A.~Kamenev, Field Theory of Non-Equilibrium Systems, Cambridge University
  Press, Cambridge, 2011.

\bibitem{stefanucci_nonequilibrium_2013}
G.~Stefanucci, R.~van Leeuwen, Nonequilibrium Many-Body Theory of Quantum
  Systems: A Modern Introduction, Cambridge University Press, Cambridge, 2013.

\bibitem{Sentef2021}
A.~de~la Torre, D.~M. Kennes, M.~Claassen, S.~Gerber, J.~W. McIver, M.~A.
  Sentef, Colloquium: Nonthermal pathways to ultrafast control in quantum
  materials, Rev. Mod. Phys. 93 (2021) 041002.
\newblock \href {https://doi.org/10.1103/RevModPhys.93.041002}
  {\path{doi:10.1103/RevModPhys.93.041002}}.

\bibitem{Giannetti2016}
C.~Giannetti, M.~Capone, D.~Fausti, M.~Fabrizio, F.~Parmigiani, D.~Mihailovic,
  Ultrafast optical spectroscopy of strongly correlated materials and
  high-temperature superconductors: a non-equilibrium approach, Advances in
  Physics 65~(2) (2016) 58--238.
\newblock \href {https://doi.org/10.1080/00018732.2016.1194044}
  {\path{doi:10.1080/00018732.2016.1194044}}.

\bibitem{Murakami2025}
Y.~Murakami, D.~Gole\ifmmode~\check{z}\else \v{z}\fi{}, M.~Eckstein, P.~Werner,
  Photoinduced nonequilibrium states in mott insulators, Rev. Mod. Phys. 97
  (2025) 035001.
\newblock \href {https://doi.org/10.1103/tkjh-lr83}
  {\path{doi:10.1103/tkjh-lr83}}.

\bibitem{NESSi}
M.~Sch\"uler, D.~Golez, Y.~Murakami, N.~Bittner, A.~Herrmann, H.~U. Strand,
  P.~Werner, M.~Eckstein, Nessi: The non-equilibrium systems simulation
  package, Comput. Phys. Commun. 257 (2020) 107484.
\newblock \href {https://doi.org/https://doi.org/10.1016/j.cpc.2020.107484}
  {\path{doi:https://doi.org/10.1016/j.cpc.2020.107484}}.

\bibitem{Aryasetiawan_1998}
F.~Aryasetiawan, O.~Gunnarsson, The gw method, Reports on Progress in Physics
  61~(3) (1998) 237.
\newblock \href {https://doi.org/10.1088/0034-4885/61/3/002}
  {\path{doi:10.1088/0034-4885/61/3/002}}.

\bibitem{Golez2016}
D.~Gole\ifmmode~\check{z}\else \v{z}\fi{}, P.~Werner, M.~Eckstein, Photoinduced
  gap closure in an excitonic insulator, Phys. Rev. B 94 (2016) 035121.
\newblock \href {https://doi.org/10.1103/PhysRevB.94.035121}
  {\path{doi:10.1103/PhysRevB.94.035121}}.

\bibitem{Bickers1989}
N.~E. Bickers, D.~J. Scalapino, S.~R. White, Conserving approximations for
  strongly correlated electron systems: Bethe-salpeter equation and dynamics
  for the two-dimensional hubbard model, Phys. Rev. Lett. 62 (1989) 961--964.
\newblock \href {https://doi.org/10.1103/PhysRevLett.62.961}
  {\path{doi:10.1103/PhysRevLett.62.961}}.

\bibitem{Sayyad2019}
S.~Sayyad, N.~Tsuji, A.~Vaezi, M.~Capone, M.~Eckstein, H.~Aoki,
  Momentum-dependent relaxation dynamics of the doped repulsive hubbard model,
  Phys. Rev. B 99 (2019) 165132.
\newblock \href {https://doi.org/10.1103/PhysRevB.99.165132}
  {\path{doi:10.1103/PhysRevB.99.165132}}.

\bibitem{Stahl2021}
C.~Stahl, M.~Eckstein, Electronic and fluctuation dynamics following a quench
  to the superconducting phase, Phys. Rev. B 103 (2021) 035116.
\newblock \href {https://doi.org/10.1103/PhysRevB.103.035116}
  {\path{doi:10.1103/PhysRevB.103.035116}}.

\bibitem{Aoki_2014}
H.~Aoki, N.~Tsuji, M.~Eckstein, M.~Kollar, T.~Oka, P.~Werner, Nonequilibrium
  dynamical mean-field theory and its applications, Rev. Mod. Phys. 86 (2014)
  779--837.
\newblock \href {https://doi.org/10.1103/RevModPhys.86.779}
  {\path{doi:10.1103/RevModPhys.86.779}}.

\bibitem{stan_time_2009}
A.~Stan, N.~E. Dahlen, R.~van Leeuwen, Time propagation of the kadanoff-baym
  equations for inhomogeneous systems, J. Chem. Phys. 130~(22) (2009) 224101.
\newblock \href {https://doi.org/10.1063/1.3127247}
  {\path{doi:10.1063/1.3127247}}.

\bibitem{Freericks2006}
J.~K. Freericks, V.~M. Turkowski, V.~Zlati\ifmmode~\acute{c}\else \'{c}\fi{},
  Nonequilibrium dynamical mean-field theory, Phys. Rev. Lett. 97 (2006)
  266408.
\newblock \href {https://doi.org/10.1103/PhysRevLett.97.266408}
  {\path{doi:10.1103/PhysRevLett.97.266408}}.

\bibitem{balzer_nonequilibrium_2012}
K.~Balzer, M.~Bonitz, Nonequilibrium Green's Functions Approach to
  Inhomogeneous Systems, Lecture Notes in Physics, Springer Berlin Heidelberg,
  2012.

\bibitem{Kaye2021}
J.~Kaye, D.~Gole\v{z}, {Low rank compression in the numerical solution of the
  nonequilibrium Dyson equation}, SciPost Phys. 10 (2021) 091.
\newblock \href {https://doi.org/10.21468/SciPostPhys.10.4.091}
  {\path{doi:10.21468/SciPostPhys.10.4.091}}.

\bibitem{Shinaoka2023}
H.~Shinaoka, M.~Wallerberger, Y.~Murakami, K.~Nogaki, R.~Sakurai, P.~Werner,
  A.~Kauch, Multiscale space-time ansatz for correlation functions of quantum
  systems based on quantics tensor trains, Phys. Rev. X 13 (2023) 021015.
\newblock \href {https://doi.org/10.1103/PhysRevX.13.021015}
  {\path{doi:10.1103/PhysRevX.13.021015}}.

\bibitem{Sroda2024}
M.~\ifmmode~\acute{S}\else \'{S}\fi{}roda, K.~Inayoshi, H.~Shinaoka, P.~Werner,
  \href{https://link.aps.org/doi/10.1103/dxfb-b3l5}{Memory-efficient
  nonequilibrium green's function framework built on quantics tensor trains},
  Phys. Rev. Lett. 135 (2025) 226501.
\newblock \href {https://doi.org/10.1103/dxfb-b3l5}
  {\path{doi:10.1103/dxfb-b3l5}}.
\newline\urlprefix\url{https://link.aps.org/doi/10.1103/dxfb-b3l5}

\bibitem{Meirinhos2022}
F.~Meirinhos, M.~Kajan, J.~Kroha, T.~Bode, {Adaptive numerical solution of
  Kadanoff-Baym equations}, SciPost Phys. Core 5 (2022) 030.
\newblock \href {https://doi.org/10.21468/SciPostPhysCore.5.2.030}
  {\path{doi:10.21468/SciPostPhysCore.5.2.030}}.

\bibitem{Lang2025}
J.~Lang, S.~Sachdev, S.~Diehl, Numerical renormalization of glassy dynamics,
  Phys. Rev. Lett. 135 (2025) 247101.
\newblock \href {https://doi.org/10.1103/z64g-nqs6}
  {\path{doi:10.1103/z64g-nqs6}}.

\bibitem{Yin2022}
J.~Yin, Y.~hao Chan, F.~H. da~Jornada, D.~Y. Qiu, S.~G. Louie, C.~Yang, Using
  dynamic mode decomposition to predict the dynamics of a two-time
  non-equilibrium green's function, Journal of Computational Science 64 (2022)
  101843.
\newblock \href {https://doi.org/https://doi.org/10.1016/j.jocs.2022.101843}
  {\path{doi:https://doi.org/10.1016/j.jocs.2022.101843}}.

\bibitem{Zhu_2025}
Y.~Zhu, J.~Yin, C.~C. Reeves, C.~Yang, V.~Vlček, Predicting nonequilibrium
  green’s function dynamics and photoemission spectra via nonlinear integral
  operator learning, Machine Learning: Science and Technology 6~(1) (2025)
  015027.
\newblock \href {https://doi.org/10.1088/2632-2153/ada99d}
  {\path{doi:10.1088/2632-2153/ada99d}}.

\bibitem{Lipavski1986}
P.~Lipavsk\'y, V.~\ifmmode \check{S}\else \v{S}\fi{}pi\ifmmode~\check{c}\else
  \v{c}\fi{}ka, B.~Velick\'y, Generalized kadanoff-baym ansatz for deriving
  quantum transport equations, Phys. Rev. B 34 (1986) 6933--6942.
\newblock \href {https://doi.org/10.1103/PhysRevB.34.6933}
  {\path{doi:10.1103/PhysRevB.34.6933}}.

\bibitem{Schluenzen2020}
N.~Schl\"unzen, J.-P. Joost, M.~Bonitz, Achieving the scaling limit for
  nonequilibrium green functions simulations, Phys. Rev. Lett. 124 (2020)
  076601.
\newblock \href {https://doi.org/10.1103/PhysRevLett.124.076601}
  {\path{doi:10.1103/PhysRevLett.124.076601}}.

\bibitem{Schueler2018}
M.~Sch\"uler, M.~Eckstein, P.~Werner, Truncating the memory time in
  nonequilibrium dynamical mean field theory calculations, Phys. Rev. B 97
  (2018) 245129.
\newblock \href {https://doi.org/10.1103/PhysRevB.97.245129}
  {\path{doi:10.1103/PhysRevB.97.245129}}.

\bibitem{Stahl2022}
C.~Stahl, N.~Dasari, J.~Li, A.~Picano, P.~Werner, M.~Eckstein, Memory truncated
  kadanoff-baym equations, Phys. Rev. B 105 (2022) 115146.
\newblock \href {https://doi.org/10.1103/PhysRevB.105.115146}
  {\path{doi:10.1103/PhysRevB.105.115146}}.

\bibitem{Picano2021b}
A.~Picano, M.~Eckstein, Accelerated gap collapse in a slater antiferromagnet,
  Phys. Rev. B 103 (2021) 165118.
\newblock \href {https://doi.org/10.1103/PhysRevB.103.165118}
  {\path{doi:10.1103/PhysRevB.103.165118}}.

\bibitem{Dasari2020}
N.~Dasari, J.~Li, P.~Werner, M.~Eckstein, Photoinduced strange metal with
  electron and hole quasiparticles, Phys. Rev. B 103 (2021) L201116.
\newblock \href {https://doi.org/10.1103/PhysRevB.103.L201116}
  {\path{doi:10.1103/PhysRevB.103.L201116}}.

\bibitem{lange2017}
F.~Lange, Z.~Lenar{\v{c}}i{\v{c}}, A.~Rosch, Pumping approximately integrable
  systems, Nature Communications 8~(1) (Jun. 2017).
\newblock \href {https://doi.org/10.1038/ncomms15767}
  {\path{doi:10.1038/ncomms15767}}.

\bibitem{Jiajun2021PRB}
J.~Li, M.~Eckstein, Nonequilibrium steady-state theory of photodoped mott
  insulators, Phys. Rev. B 103 (2021) 045133.
\newblock \href {https://doi.org/10.1103/PhysRevB.103.045133}
  {\path{doi:10.1103/PhysRevB.103.045133}}.

\bibitem{Kuenzel2024}
F.~K\"unzel, A.~Erpenbeck, D.~Werner, E.~Arrigoni, E.~Gull, G.~Cohen,
  M.~Eckstein, Numerically exact simulation of photodoped mott insulators,
  Phys. Rev. Lett. 132 (2024) 176501.
\newblock \href {https://doi.org/10.1103/PhysRevLett.132.176501}
  {\path{doi:10.1103/PhysRevLett.132.176501}}.

\bibitem{Profumo2015}
R.~E.~V. Profumo, C.~Groth, L.~Messio, O.~Parcollet, X.~Waintal, Quantum monte
  carlo for correlated out-of-equilibrium nanoelectronic devices, Phys. Rev. B
  91 (2015) 245154.
\newblock \href {https://doi.org/10.1103/PhysRevB.91.245154}
  {\path{doi:10.1103/PhysRevB.91.245154}}.

\bibitem{Erpenbeck2023}
A.~Erpenbeck, E.~Gull, G.~Cohen, Quantum monte carlo method in the steady
  state, Phys. Rev. Lett. 130 (2023) 186301.
\newblock \href {https://doi.org/10.1103/PhysRevLett.130.186301}
  {\path{doi:10.1103/PhysRevLett.130.186301}}.

\bibitem{Eckstein2024}
M.~Eckstein, Solving quantum impurity models in the non-equilibrium steady
  state with tensor trains (2024).
\newblock \href {http://arxiv.org/abs/2410.19707} {\path{arXiv:2410.19707}}.

\bibitem{Kim2024}
A.~J. Kim, P.~Werner, Strong coupling impurity solver based on quantics tensor
  cross interpolation, Phys. Rev. B 111 (2025) 125120.
\newblock \href {https://doi.org/10.1103/PhysRevB.111.125120}
  {\path{doi:10.1103/PhysRevB.111.125120}}.

\bibitem{Arrigoni2013}
E.~Arrigoni, M.~Knap, W.~von~der Linden, Nonequilibrium dynamical mean-field
  theory: An auxiliary quantum master equation approach, Phys. Rev. Lett. 110
  (2013) 086403.
\newblock \href {https://doi.org/10.1103/PhysRevLett.110.086403}
  {\path{doi:10.1103/PhysRevLett.110.086403}}.

\bibitem{FFTW}
M.~Frigo, S.~Johnson, The design and implementation of fftw3, Proceedings of
  the IEEE 93~(2) (2005) 216--231.
\newblock \href {https://doi.org/10.1109/JPROC.2004.840301}
  {\path{doi:10.1109/JPROC.2004.840301}}.

\bibitem{NumericalRecipes}
W.~Press, Numerical Recipes 3rd Edition: The Art of Scientific Computing,
  Cambridge University Press, 2007.

\bibitem{FFTW_documentation}
S.~G.~J. Matteo~Frigo, \href{https://www.fftw.org/fftw3.pdf}{FFTW online
  manual}, Massachusetts Institute of Technology (2020) [cited 30.09.2025].
\newline\urlprefix\url{https://www.fftw.org/fftw3.pdf}

\bibitem{Chowdhury2022}
D.~Chowdhury, A.~Georges, O.~Parcollet, S.~Sachdev, Sachdev-ye-kitaev models
  and beyond: Window into non-fermi liquids, Rev. Mod. Phys. 94 (2022) 035004.
\newblock \href {https://doi.org/10.1103/RevModPhys.94.035004}
  {\path{doi:10.1103/RevModPhys.94.035004}}.

\bibitem{Moeckel2008}
M.~Moeckel, S.~Kehrein, Interaction quench in the hubbard model, Phys. Rev.
  Lett. 100 (2008) 175702.
\newblock \href {https://doi.org/10.1103/PhysRevLett.100.175702}
  {\path{doi:10.1103/PhysRevLett.100.175702}}.

\bibitem{Eckstein2009}
M.~Eckstein, M.~Kollar, P.~Werner, Thermalization after an interaction quench
  in the hubbard model, Phys. Rev. Lett. 103 (2009) 056403.
\newblock \href {https://doi.org/10.1103/PhysRevLett.103.056403}
  {\path{doi:10.1103/PhysRevLett.103.056403}}.

\bibitem{Werner2010}
P.~Werner, T.~Oka, M.~Eckstein, A.~J. Millis, Weak-coupling quantum monte carlo
  calculations on the keldysh contour: Theory and application to the
  current-voltage characteristics of the anderson model, Phys. Rev. B 81 (2010)
  035108.
\newblock \href {https://doi.org/10.1103/PhysRevB.81.035108}
  {\path{doi:10.1103/PhysRevB.81.035108}}.

\bibitem{Georges1996}
A.~Georges, G.~Kotliar, W.~Krauth, M.~J. Rozenberg, Dynamical mean-field theory
  of strongly correlated fermion systems and the limit of infinite dimensions,
  Rev. Mod. Phys. 68 (1996) 13--125.
\newblock \href {https://doi.org/10.1103/RevModPhys.68.13}
  {\path{doi:10.1103/RevModPhys.68.13}}.

\bibitem{Eckstein2010}
M.~Eckstein, M.~Kollar, P.~Werner, Interaction quench in the hubbard model:
  Relaxation of the spectral function and the optical conductivity, Phys. Rev.
  B 81 (2010) 115131.
\newblock \href {https://doi.org/10.1103/PhysRevB.81.115131}
  {\path{doi:10.1103/PhysRevB.81.115131}}.

\end{thebibliography}


\begin{thebibliography}{0}
\bibitem{1}
M. Sch\"uler, D. Golez, Y. Murakami, N. Bittner, A. Herrmann, H. U. Strand, P. Werner, M. Eckstein, Nessi: The non- equilibrium systems simulation package, Comput. Phys. Commun. 257 (2020) 107484. doi:https://doi.org/10.1016/j.cpc.2020.107484.
\end{thebibliography}

\end{document}